\documentclass[ALICE,manyauthors]{cernphprep}%alice_public_notes}
\usepackage[english]{babel}
\usepackage{graphicx}
\usepackage{verbatim}
\usepackage{amsfonts}
\usepackage{makeidx}
\usepackage{color}
\usepackage{amsmath}
\usepackage{lineno}
\usepackage{epsfig}
\usepackage{epstopdf}
\usepackage{hyperref}
\usepackage{cite}
\usepackage{multirow}
\usepackage{subfigure}
\usepackage[T1]{fontenc}
\usepackage{placeins}

\usepackage[comma,square,numbers,sort&compress]{natbib}

\newcommand{\pp}{pp~}

\newcommand{\sqrts}{\sqrt{s}}
\newcommand{\sqrtsNN}{\sqrt{s_{\scriptscriptstyle \rm NN}}}

\newcommand{\GeV}{\mathrm{GeV}}
\newcommand{\TeV}{\mathrm{TeV}}
\newcommand{\tev}{\mathrm{TeV}}

\newcommand{\gev}{\mathrm{GeV}}
\newcommand{\gevc}{\mathrm{GeV}/c}
\newcommand {\massStyle}[1] {\mbox{\ensuremath{\text{#1}\kern-0.1em /\kern-0.12em c^2}}}

\newcommand{\mb}{\mathrm{mb}}

\newcommand{\pPb}{\mbox{p--Pb}}

\newcommand{\pt}{p_{\rm T}}

\newcommand{\DtoKpi}{{\rm D}^0 \to {\rm K}^-\pi^+}
\newcommand{\DtoKpipi}{{\rm D}^+\to {\rm K}^-\pi^+\pi^+}
\newcommand{\DstartoDpi}{{\rm D}^{*+} \to {\rm D}^0 \pi^+}

\newcommand{\Dzero}{{\rm D^0}}
\newcommand{\Ds}{{\rm D^+_s}}
\newcommand{\Dzerobar}{\overline{\rm D}{}^{0}}
\newcommand{\Dstar}{{\rm D^{*+}}}

\newcommand{\Dplus}{{\rm D^+}}

\newcommand{\Npart}{N_{\rm part}}
\newcommand{\Ncoll}{N_{\rm coll}}

\newcommand{\RpPb}{R_{\rm pPb}}

\newcommand{\meanpt}{{\langle p_{ {\mathrm T} } \rangle}}
\newcommand{\QpPb}{\ensuremath{Q_\mathrm{pPb}}}
\newcommand{\Qcp}{\ensuremath{Q_\mathrm{CP}}}
\newcommand{\TpPb}{\ensuremath{T_\mathrm{pPb}}}

\newcommand{\Lint}{{L}_{\mathrm{int}}}
\newcommand{\fprompt}{\ensuremath{f_{\rm prompt}}}

\newcommand{\Ntrk}{N_{\rm tracklets}}
\newcommand{\Nch}{N_{\rm ch}}

\newcommand{\dNdEta}{{\rm d}N_{\rm ch}/{\rm d}\eta}

\begin{document}

\begin{titlepage}

\PHyear{2019}
\PHnumber{125}                 % required, obtained from PH
\PHdate{08 June}              % required
% ALICE-PUBLIC-2017-008

\title{ Measurement of prompt $\Dzero$, $\Dplus$, $\Dstar$, and $\Ds$ production in p--Pb collisions 
        at $\mathbf{\sqrt{{\textit s}_{\rm NN}}~=~5.02~TeV}$}
\Collaboration{ALICE Collaboration}

\ShortTitle{D-meson production in p--Pb collisions at $\sqrtsNN=5.02~\tev$}

\begin{abstract} 

%\linenumbers

The measurement of the production of prompt $\Dzero$, $\Dplus$, $\Dstar$, and $\Ds$ mesons in proton--lead (p--Pb) collisions at the centre-of-mass energy per nucleon pair of $\sqrtsNN=5.02~\tev$, 
with an integrated luminosity of $292\pm 11$ $\mu$b$^{-1}$,
are reported. 
Differential production cross sections are measured at mid-rapidity
\mbox{($-0.96<y_{\rm cms}<0.04$)} as a function of transverse momentum ($\pt$) in the intervals \mbox{$0<\pt<36~\gev/c$} for $\Dzero$, 
\mbox{$1<\pt<36~\gev/c$} for $\Dplus$ and $\Dstar$, 
and $2<\pt<24~\gev/c$ for $\Ds$ mesons. 
For each species, the nuclear modification factor $\RpPb$ is calculated as a function of $\pt$ using a proton-proton (pp) reference measured
at the same collision energy. The results are compatible with unity in the whole $\pt$ range. The average of the non-strange D mesons $\RpPb$ is compared with theoretical model predictions that include initial-state effects and parton transport model predictions. 
The $\pt$ dependence of the $\Dzero$,  $\Dplus$, and $\Dstar$ nuclear modification factors is also reported in the interval \mbox{$1<\pt<36~\gev/c$} as a function of the collision centrality,
and the central-to-peripheral ratios are computed from the D-meson yields measured in different centrality classes. The results are further compared with charged-particle measurements and a similar trend is observed in all the centrality classes.
The ratios of the $\pt$-differential cross sections of $\Dzero$, $\Dplus$, $\Dstar$, and $\Ds$ mesons are also reported. 
The $\Ds$ and $\Dplus$ yields are compared as a function of the charged-particle multiplicity for several $\pt$ intervals. No modification in the relative abundances of the four species is observed with respect to pp collisions within the statistical and systematic uncertainties.

\end{abstract}

\end{titlepage}
\setcounter{page}{2}
%\tableofcontents

%\linenumbers
\newpage 

%%----------------------------------------------------------
\section{Introduction}
\label{sec:intro}
Measurements of heavy-flavour hadron production in proton--nucleus collisions allow for an assessment of the various
effects related to the presence of nuclei in the colliding system, denoted as cold-nuclear-matter (CNM) effects.
Heavy quarks (charm and beauty) are primarily produced in hard-scattering processes
with large momentum transfer ($Q^2$) due to their large masses.
Their inclusive production cross sections can therefore be calculated perturbatively in Quantum Chromodynamics (QCD) utilizing the factorisation approach.
In this scheme, the $\pt$ differential production cross sections of hadrons containing charm or beauty quarks are calculated as a convolution of three terms: (i) the parton distribution functions (PDFs) of the incoming nucleons, (ii) the partonic scattering cross section, calculated as a perturbative series in powers of the strong coupling constant $\alpha_{\rm s}$, and (iii) the fragmentation function, which parametrises the non-perturbative evolution of a heavy quark into a given heavy-flavour hadron species.
Theoretical predictions based on perturbative QCD (pQCD) calculations at next-to-leading order accuracy with all-order resummation of next-to-leading logarithms,
such as FONLL~\cite{Cacciari:1998it,Cacciari:2012ny} and GM-VFNS~\cite{Kniehl:2004fy,Kniehl:2005mk,Kniehl:2012ti,Helenius:2018uul}, can describe within uncertainties the production cross sections of D and B mesons measured in pp and ${\rm p\overline{p}}$ collisions in different kinematic regions at centre-of-mass energies from 0.2 to 13~TeV (see e.g. Ref.~\cite{Andronic:2015wma,Prino:2016cni} and references therein).
In proton--nucleus collisions, various effects in the initial and final state could modify the D-meson production cross sections per nucleon--nucleon collision as compared to pp interactions.
In the initial state, the production is affected by the modification of the PDFs in bound nucleons compared to those of free nucleons, depending on the parton momentum fraction $x$, the momentum transfer $Q^2$ in the hard scattering process, and the  nucleus mass number $A$~\cite{Arneodo:1992wf,Malace:2014uea}.
At LHC energies and at mid-rapidity, the most relevant effect on the PDFs is shadowing: a reduction of the parton densities at low $x$ (below $10^{-2}$), which becomes stronger when $Q^2$ decreases and the nucleus mass number $A$ increases.
This effect can be described by means of phenomenological parametrisations of the PDF modifications, denoted as nuclear PDFs (nPDFs)~\cite{Eskola:2009uj,Hirai:2007sx,deFlorian:2003qf,Eskola:2016oht}.
As demonstrated in Refs.~\cite{Kusina:2017gkz,Eskola:2019bgf}, measurements of heavy-flavour and quarkonium production at the LHC can significantly reduce the uncertainties on the gluon nPDFs at small $x$.
If the parton phase-space reaches saturation, the appropriate theoretical description is the Colour Glass Condensate effective theory (CGC)~\cite{Gelis:2010nm,Tribedy:2011aa,Albacete:2012xq,Rezaeian:2012ye,Fujii:2013yja}.
The modification of the small-$x$ parton dynamics can significantly reduce the D-meson yield at low $\pt$. 
Furthermore, the multiple scattering of partons in the nucleus, before and/or
after the hard scattering, can modify the kinematic distribution of the
produced hadrons. Partons can lose energy in the initial stages of the 
collision via initial-state radiation~\cite{Vitev:2007ve}, or experience 
transverse momentum broadening due to multiple soft collisions before the 
heavy-quark pair is produced \cite{Lev:1983hh, Wang:1998ww, Kopeliovich:2002yh}.
These effects can also induce a significant modification of D-meson production at low $\pt$.
In addition, final-state effects may also be responsible for a modification of heavy-flavour hadron yields and momentum distributions.
The presence of significant final-state effects in p--Pb collisions with large multiplicities of produced particles is suggested by different observations,
e.g.\ the presence of long-range structures in two-particle angular correlations of charged hadrons~\cite{CMS:2012qk, Abelev:2012ola, ABELEV:2013wsa, Aad:2012gla,Adam:2015bka,Aaij:2015qcq}, the studies of azimuthal anisotropies in multi-particle correlations~\cite{Khachatryan:2015waa,Aaboud:2017blb}, the evolution with multiplicity of the identified-hadron transverse-momentum distributions~\cite{Abelev:2013haa,Chatrchyan:2013eya}, and the suppression of the $\psi{\rm (2S)}$ production with respect to that of ${\rm J}/\psi$ mesons~\cite{Abelev:2014zpa,Aaij:2016eyl,Adam:2016ohd}.
In particular, the angular correlations in high-multiplicity p--Pb collisions were found to have similar properties (e.g.\ particle mass and $\pt$ dependence~\cite{Abelev:2013haa,Chatrchyan:2013eya}) as those observed in Pb--Pb collisions, where they are commonly interpreted as indications of a collective particle flow produced during the hydrodynamic evolution of the Quark-Gluon Plasma (QGP)~\cite{Borsanyi:2010bp,Bazavov:2011nk,Jaiswal:2016hex,Busza:2018rrf}.
The interpretation of the aforementioned results is highly debated, with the outstanding open question being whether small droplets of a fluid-like QGP are created in small collision systems (see e.g.\,\cite{Nagle:2018nvi} for a recent review).
Hydrodynamic calculations, that assume the formation of a medium with some degree
of collectivity (see e.g.\,\cite{Bozek:2012gr, Bozek:2013uha,Weller:2017tsr}),
can describe the angular correlations measured in p--Pb collisions, which suggests
a common hydrodynamic origin of the experimental observations from small to
large collision systems.
However, alternative explanations exist, based on gluon saturation (CGC) in
the initial state~\cite{Dusling:2012cg,Dusling:2015gta}, the anisotropic
escape probability of partons from the collision zone~\cite{He:2015hfa}, or interactions between string-like colour fields in dense configurations of confined QCD flux tubes~\cite{Bierlich:2014xba,Bierlich:2017vhg}.
If a collective expansion in the final state of the collision occurs, 
the medium could also impart a flow to heavy-flavour quarks or hadrons, and 
modify the hadronisation dynamics of heavy quarks.
Detailed calculations were performed in the framework of transport models, assuming that in p--Pb collisions at LHC energies a QGP is formed, which affects the propagation and hadronisation of heavy quarks~\cite{Xu:2015iha,Beraudo:2015wsd}. These models predict a significant modification of the $\pt$ distributions of heavy-flavour hadrons in high-multiplicity p--Pb collisions as compared to pp interactions, accompanied by the presence of anisotropies in their azimuthal distributions.
Recent measurements of angular correlations in p--Pb collisions involving
J/$\psi$ mesons~\cite{Acharya:2017tfn}, $\Dzero$ mesons~\cite{Sirunyan:2018toe}, and heavy-flavour decay electrons~\cite{Acharya:2018dxy} provided a clear indication that long-range anisotropies are present also in the heavy-flavour sector.

In the presence of a QGP, a modification of the hadronisation is predicted: hadrons can be produced not 
only via the fragmentation mechanism, but also via (re)combination of charm 
quarks with other quarks from the medium during the deconfined phase or at the 
phase boundary ~\cite{Greco:2003mm,Greco:2003vf, Andronic:2007zu, He:2012df}.
%A modification of the hadronisation mecahnisms is expected in presence of a medium 
%composed of deconfined  quarks and gluons, due to the possible formation of hadrons via
%(re)combination of charm quarks with other quarks from the 
%medium during the deconfined phase or at the phase boundary
%~\cite{Greco:2003mm,Greco:2003vf, Andronic:2007zu, He:2012df} in addition to
%fragmentation.
Given the observed increase of strangeness production with increasing particle multiplicity
in p--Pb and pp collisions~\cite{Abelev:2013haa,Adam:2015vsf,ALICE:2017jyt}, 
the modified hadronisation could result in an enhancement of the relative yield of $\Ds$ mesons 
with respect to non-strange charmed mesons in high-multiplicity p--Pb collisions.

In this paper, we report the measurements of the $\pt$-differential 
production cross sections and nuclear modification factors 
of prompt $\Dzero$, $\Dplus$, $\Dstar$, and $\Ds$ mesons in p--Pb collisions 
at $\sqrtsNN=5.02~\TeV$ recorded with the ALICE detector in 2016.
The sample used for these analyses is larger by a factor of about six with respect 
to the sample collected in 2013, which was used in previous publications of these 
observables~\cite{Abelev:2014hha,Adam:2016mkz,Adam:2016ich}. Therefore, it is possible to obtain 
lower statistical and systematic uncertainties by a factor 1.5--2 and extend the $\pt$ reach of the measurements.   
The ratios of the production cross sections of the different D-meson species are also reported and are compared with those measured in pp collisions at the same centre-of-mass energy.
The nuclear modification factor, $\RpPb$, is defined as the ratio of 
the cross section in p--Pb collisions to that in pp interactions scaled by the 
mass number of the Pb nucleus. This ratio is sensitive to cold-nuclear-matter 
and hot-medium effects on D-meson production in p--Pb collisions.
In addition, the measurement of the nuclear modification factor for non-strange D mesons is carried out in intervals of collision centrality, called in the following as $\QpPb$.
The $\QpPb$ is calculated as the ratio of the D-meson yield in p--Pb collisions to the cross section in pp interactions scaled by the nuclear overlap function $ \langle \TpPb \rangle$, which accounts for the average number of nucleon-nucleon interactions in the considered centrality class.
%defined using the energy deposited in the zero-degree neutron calorimeter in the Pb-going side, as described in Ref.~\cite{Adam:2014qja}. 
The $\QpPb$ measurements are performed in finer intervals of collision centrality, enabling in particular the measurements of D-meson production in the 10\% most central collisions, in which possible final-state effects are expected to be stronger.
Further insight into the centrality dependence of prompt D-meson $\pt$ 
distributions is provided by the measurements of the 
ratios of D-meson yields in various centrality classes.
Finally,  the ratio of $\Ds$-meson yield to that of non-strange $\Dplus$ is presented as a function of the multiplicity of 
charged particles produced in p--Pb collisions and is compared with results measured in pp and Pb--Pb collisions at the same centre-of-mass energy.

%%------------------------------------\includegraphics[]{DpPb2016paper.pdf}

\section{Experimental apparatus and data sample}
\label{sec:detector}
The ALICE apparatus~\cite{Aamodt:2008zz} is composed of a central barrel comprising various detectors 
for particle reconstruction and identification at mid-rapidity ($|\eta|<0.9$),
a forward muon spectrometer ($-4<\eta<-2.5$),
and a set of forward-backward detectors for triggering and event characterisation.
Typical detector performance in pp, p--Pb, and Pb--Pb collisions is presented in~\cite{Abelev:2014ffa}. 
The main detector components used in this analysis are the V0 detector, the Inner Tracking System (ITS),
the Time Projection Chamber (TPC), and the Time-Of-Flight (TOF) detector,
which are located inside a large solenoidal magnet providing a maximum uniform magnetic field of 0.5~T
parallel to the LHC beam direction ($z$-axis in the ALICE reference system),
and the Zero-Degree Calorimeter (ZDC), located at $\pm 112.5$~m from the interaction point.

Proton--lead collisions at $\sqrtsNN=5.02~\tev$
were recorded with a minimum-bias (MB) interaction trigger that required coincident signals in both scintillator arrays of  the V0 detector, which
cover the full azimuth in the pseudorapidity intervals $-3.7< \eta <-1.7$
and $2.8< \eta <5.1$.  The V0 timing information was used
together with that from the ZDCs for offline rejection
of beam--beam or beam--gas interactions happening outside of the nominal colliding bunches.

The MB trigger was sensitive to about 96.4\% of the p--Pb inelastic cross section~\cite{Abelev:2014epa}.
Only collision events with a primary vertex reconstructed within $\pm 10$~cm from the
centre of the detector along the beam axis were considered.
Events with several interactions per bunch crossing,
whose probability was below 0.5\%,
were rejected using an algorithm based on track segments,
defined within the Silicon Pixel Detector (SPD, the two innermost ITS layers),
to detect multiple interaction vertices.

The number of events passing these selection criteria was about $6\times 10^8$.
The corresponding integrated luminosity, $\Lint=N_{\rm MB}/\sigma_{\rm MB}$,
is equal to $292 \pm 11~{\mathrm{\mu b}}^{-1}$,
$\sigma_{\rm MB}=2.09$~b being the MB-trigger (i.e.\ visible) cross section
measured via a van der Meer scan,
with negligible statistical uncertainty and a systematic uncertainty of 3.7\%~\cite{Abelev:2014epa}.
During the p--Pb data-taking period, the beam energies were 4~TeV for protons and 1.58~TeV per nucleon for lead nuclei.
With this beam configuration,
the nucleon--nucleon centre-of-mass system moves in rapidity by $\Delta y_{\mathrm{cms}}=0.465$ in the direction
of the proton beam. The D-meson analyses were performed in the laboratory-frame interval $|y_{\mathrm{lab}}|<0.5$,
which leads to a shifted centre-of-mass rapidity coverage of $-0.96 < y_{\mathrm{cms}} < 0.04$.
Additionally, the \pPb~data sample was divided into centrality classes defined as percentiles of the visible cross section. The events were classified according to the energy deposited in the ZDC positioned in the Pb-going side by the neutrons produced in the interaction by nuclear de-excitation processes, or knocked out by wounded nucleons. The multiplicity of these neutrons is expected to grow monotonically with the number of nucleon--nucleon binary collisions, $\Ncoll$.
It was demonstrated in Ref.~\cite{Adam:2014qja} that this is the least-biased centrality estimator for \pPb~interactions.
The description of the average nuclear overlap function,
as well as the values corresponding to the measured centrality classes,
will be given in section~\ref{subsec:QpPb}.

The pp reference for the $\RpPb$ and $\QpPb$ calculation was taken from the
measurements performed on a data sample of about 990 million minimum-bias pp
collisions ($L_{\rm int} = (19.3 \pm 0.4)$~nb$^{-1}$) at $\sqrts = 5.02$~TeV collected
with ALICE in 2017, and published in Ref.~\cite{pp:2019}.

%%----------------------------------------------------------
%\section{Multiplicity determination}
%\label{sec:Multiplicity}
%\input{multiplicity.tex}

%%----------------------------------------------------------
\section{Data analysis}
\label{sec:analysis}
The D-meson yields were extracted using two different analysis methods. 
The first method, described in Section \ref{sec:topol}, is based on the reconstruction of decay vertices displaced from the primary vertex. 
The second method, described in Section \ref{sec:lowpt}, is used only for the $\Dzero$ measurement and is based on the estimation and subtraction of the combinatorial 
background, without any selection criteria on the displaced decay-vertex topology.
The first method allows the D-meson yield to be extracted in a $\pt$-interval of 1--36 GeV/$c$ for $\Dzero$, $\Dplus$, and $\Dstar$ and 2--24 GeV/$c$ for $\Ds$.
The second method allows the  $\Dzero$-meson production to be measured down to $\pt=0$.\\

%%----------------------------------------------------------
\subsection{Analysis with D-meson decay vertex reconstruction}
\label{sec:topol}
The D mesons and their charge conjugates were reconstructed in the decay channels
$\DtoKpi$ (with a branching ratio, BR, of $ 3.89 \pm 0.04\%$), $\DtoKpipi$ (BR of $8.98 \pm 0.28\%$), $\DstartoDpi$ (BR of $67.7\pm 0.5\%$),
and $\Ds\to\phi\pi^+$ (with $\phi\to\rm K^+K^-$) (BR of $2.27 \pm 0.08\%$)~\cite{Tanabashi:2018xmw}.
The analyses were based on the reconstruction of decay vertices displaced from the interaction vertex, exploiting the separation of a few hundred 
microns induced by the weak decays of the $\Dzero$, $\Dplus$, and $\Ds$ mesons. The displacement of the $\Dzero$-meson candidate decay vertex was used to
select the $\Dstar$ meson
which decays strongly at the primary vertex. This is performed by combining the $\Dzero$ candidates with a soft pion in an invariant-mass analysis.

The $\Dzero$, $\Dplus$, and $\Ds$ candidates were defined using pairs or triplets of tracks with proper charge sign combinations with $|\eta| < 0.8$,
$\pt > 0.3~\rm{GeV}/\it{c}$,
at least 70  associated space points in the TPC,
and at least two space points in the ITS, with at least one in the SPD. The $\Dstar$ candidates were formed by combining $\Dzero$ candidates with tracks satisfying $|\eta| < 0.8$, $\pt > 0.1~\gevc$ and at least two space points in the ITS, including at least one in the SPD. The selection of tracks with $|\eta|<0.8$ limits the D-meson acceptance in rapidity, which, depending on $\pt$, varies from $|y_{\rm lab}|<0.5$ at low $\pt$ to $|y_{\rm lab}|<0.8$ at $\pt > 5~\gevc$~\cite{Adam:2015ota}.
A $\pt$-dependent fiducial acceptance region was therefore defined as $y_{\mathrm{fid}}(\pt)>|y_{\rm lab}|$, 
with $y_{\mathrm{fid}}(\pt)$ increasing from 0.5 to 0.8 in the transverse momentum range $0 < \pt < 5~\gevc$ according to a second-order polynomial function, and $y_{\mathrm{fid}}=0.8$ for $\pt > 5~\gevc$.
The selection strategy is the same as in previous analyses~\cite{Adam:2016ich}. 
The main variables used to select the D-meson candidates are the separation between primary and secondary vertex, the displacement of the tracks from the primary
vertex, and the pointing of the reconstructed D-meson momentum to the primary vertex.
For the $\Dplus$, a selection on the impact parameter of the candidate with respect to the primary vertex in the transverse plane was also applied.
For the $\Ds$-candidate selection, one of the two pairs of opposite-sign tracks is required to have a reconstructed ${\rm K^+K^-}$ invariant mass compatible with the PDG world average of the $\phi$ meson mass~\cite{Tanabashi:2018xmw}. 
Further background reduction is achieved by applying particle identification to select charged pions and kaons using information of the TPC and TOF detectors.
The track particle identification (PID) is obtained using a 3$\sigma$ window around the expected mean values of the specific ionisation energy loss (${\rm d}E/{\rm d} x$) in the TPC gas
and of the time of flight from the interaction point to the TOF detector. A 2$\sigma$ window around the expected mean values of the ${\rm d}E/{\rm d} x$ was applied, except for the lowest $\pt$ interval, $1.5< \pt <2~\gevc$,  $\Dstar$ meson, and for the $\Ds$ meson in those cases in which no time-of-flight information was available.

\begin{figure}[!t]
\begin{center}
 	\label{fig:D0InvMass_2_3}
	\includegraphics[width=0.475\columnwidth]{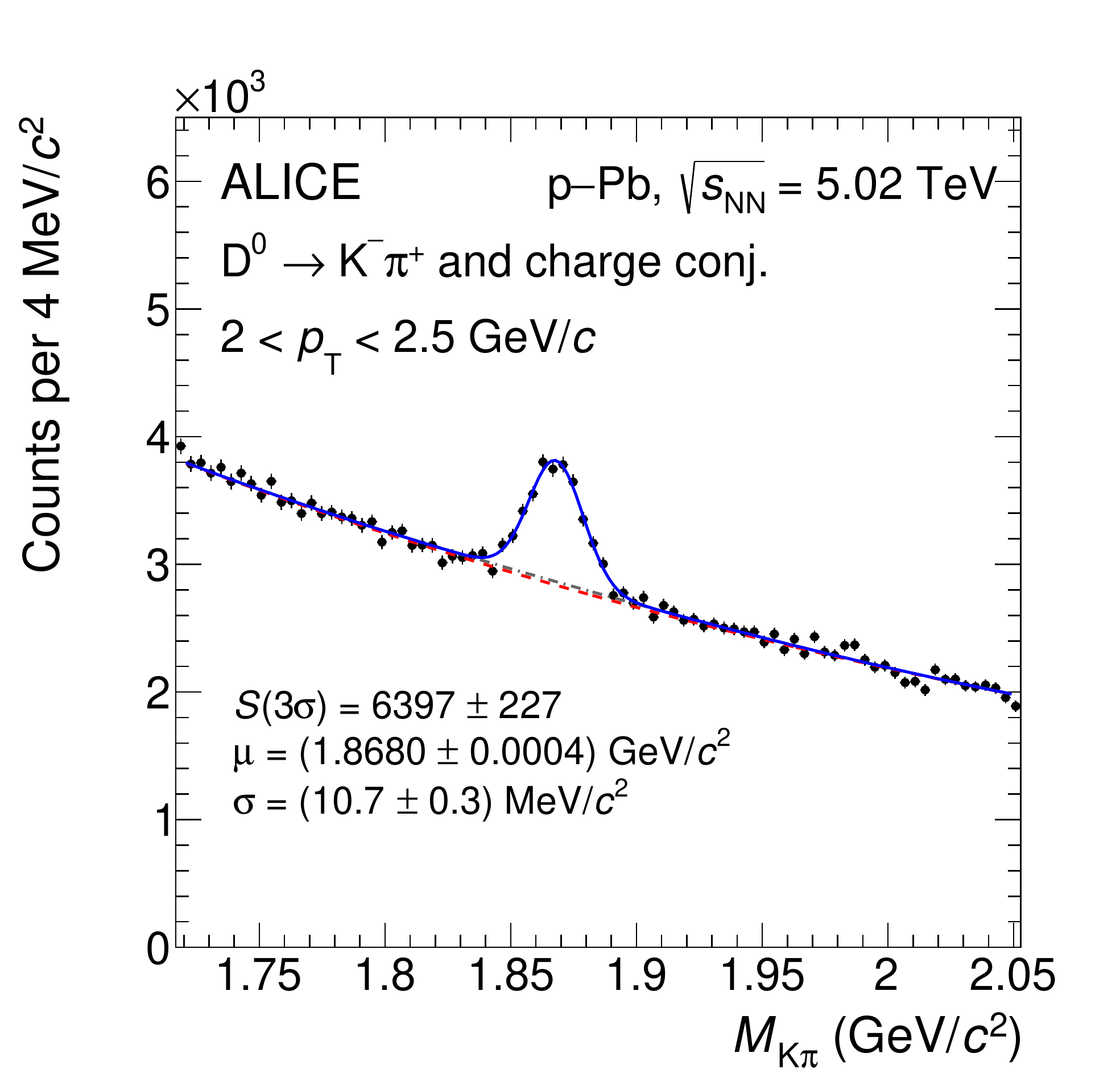}
	\label{fig:DplusInvMass_24_36}
	\includegraphics[width=0.475\columnwidth]{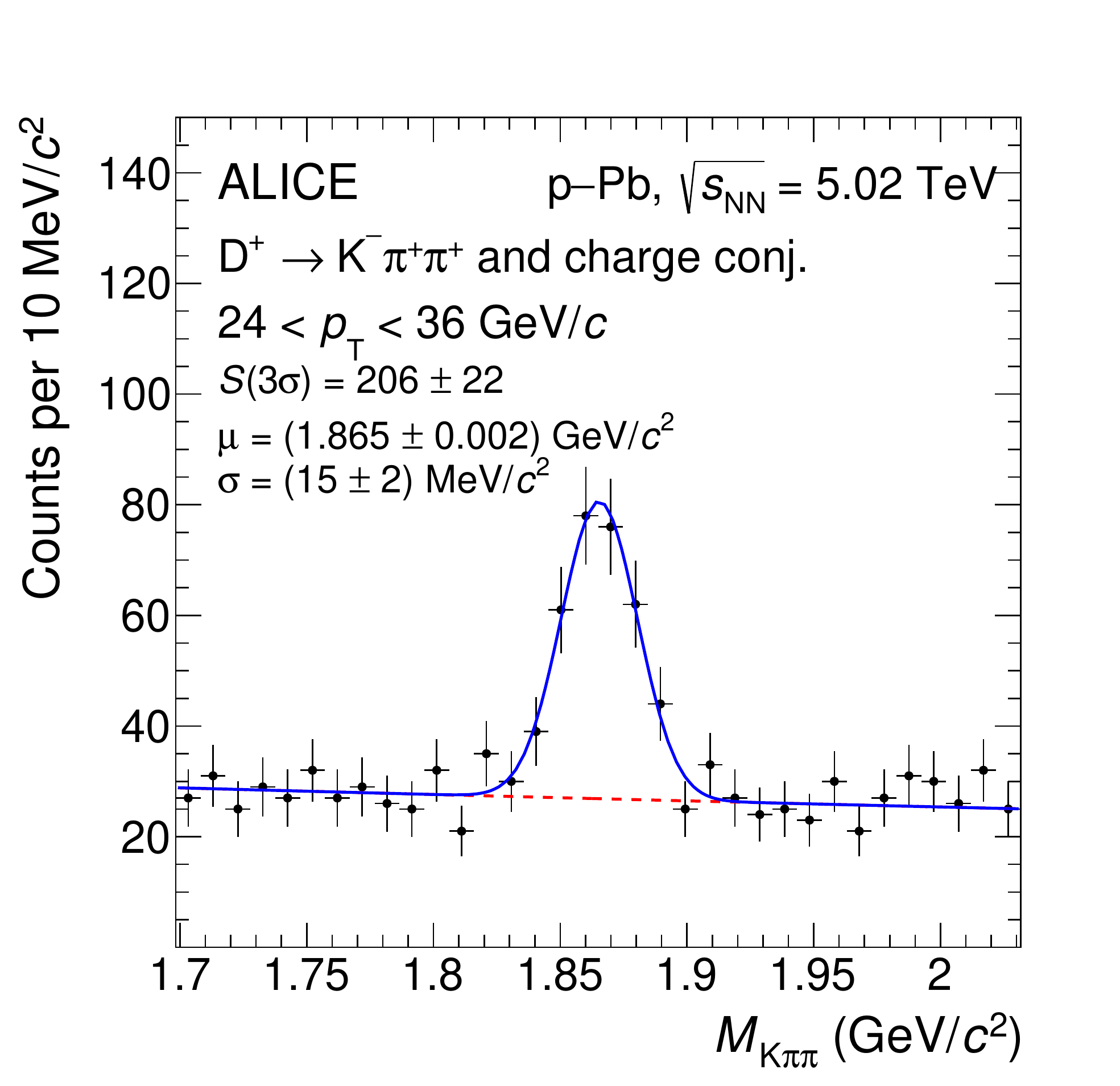}
	\\
	\label{fig:DstarInvMass_9_10}
	\includegraphics[width=0.475\columnwidth]{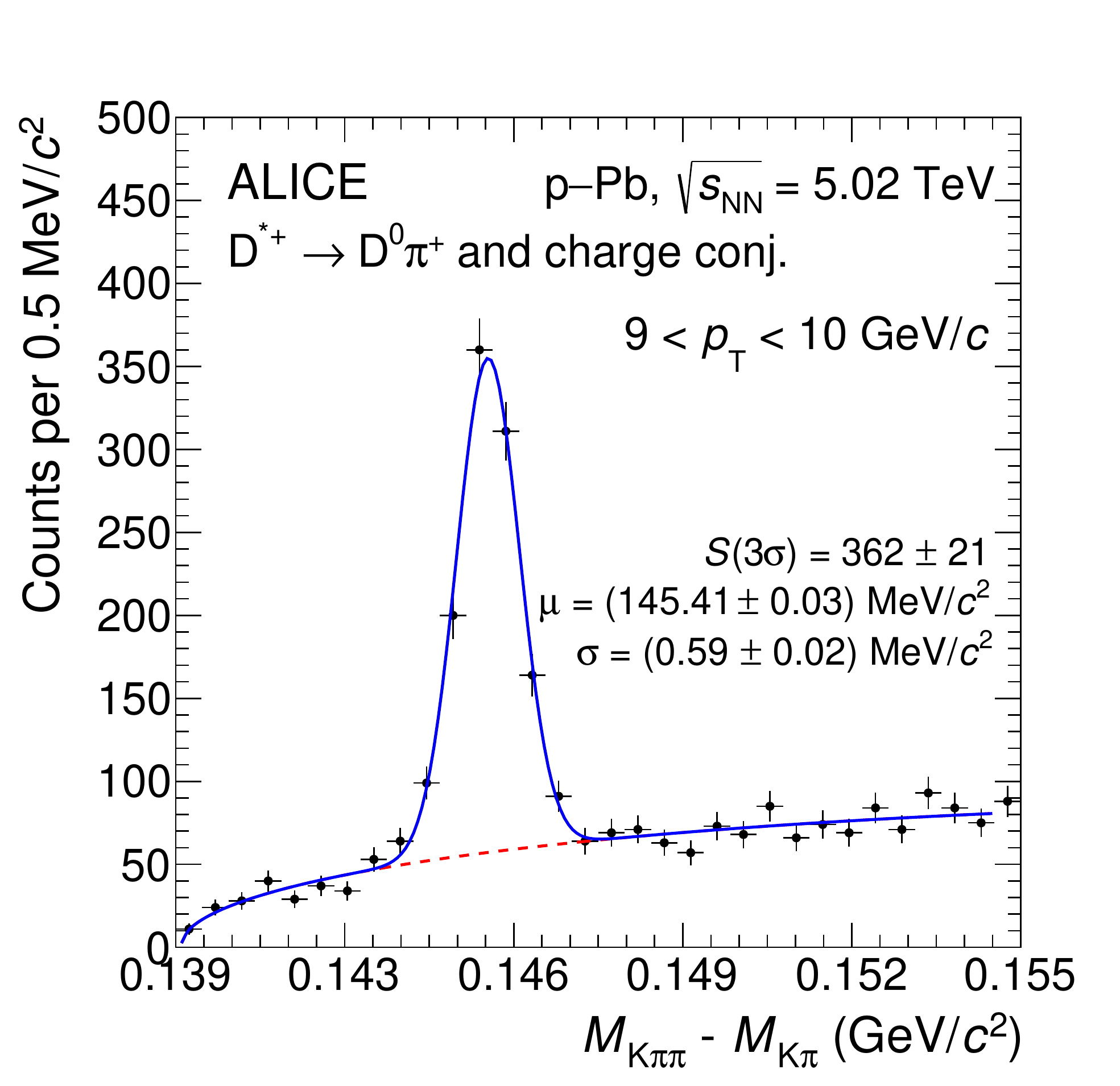}
	\label{fig:DsInvMass_2_3}
	\includegraphics[width=0.475\columnwidth]{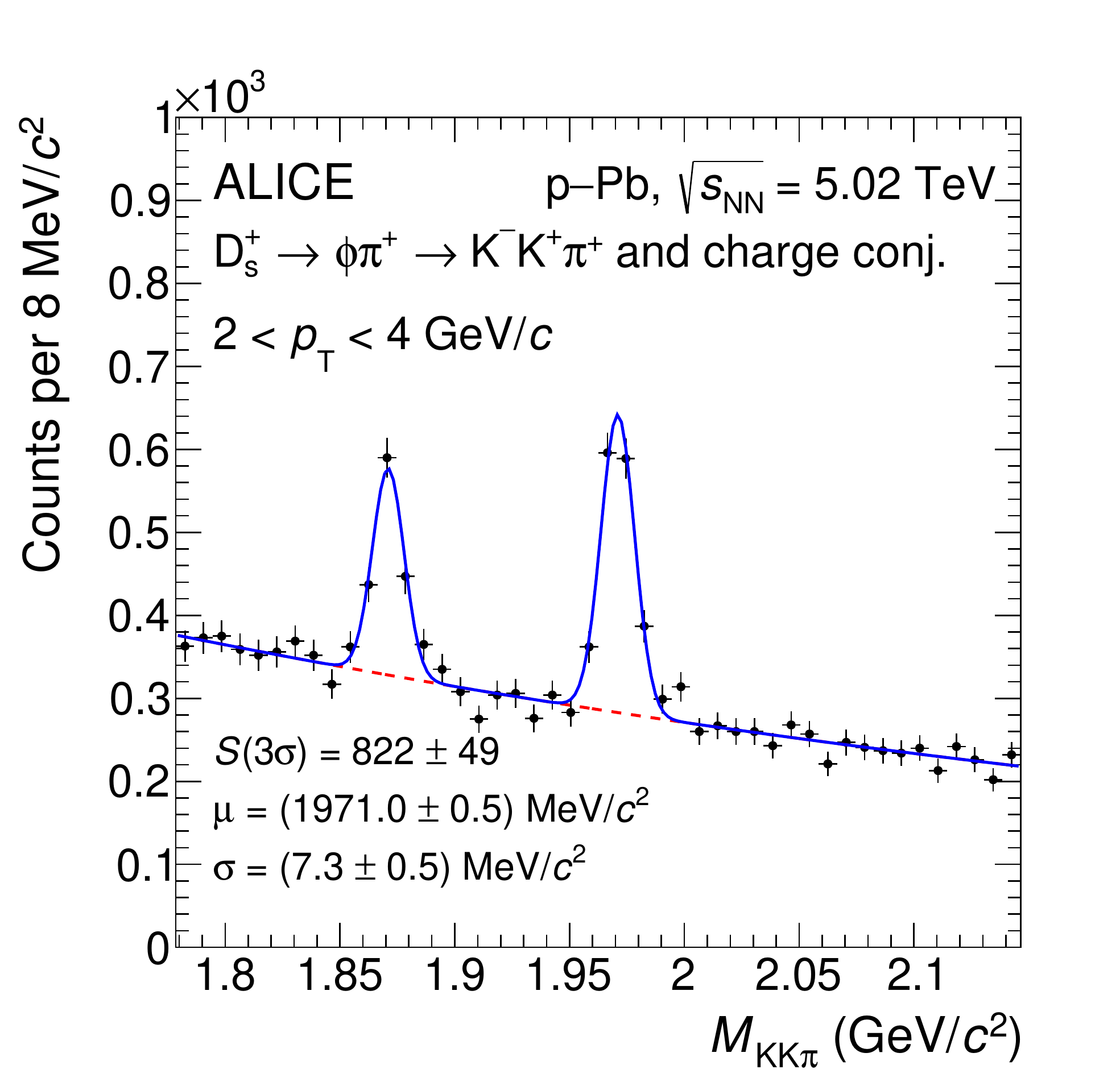}
\caption{Examplary invariant-mass distributions for $\Dzero$, $\Dplus$, and $\Ds$ candidates
        (plus charge conjugates) and the mass difference  $\Delta M = M_{{\rm K} \pi\pi} -M_{{\rm K} \pi}$ for $\Dstar$ candidates (and charge conjugates) 
        in minimum-bias p--Pb collisions at $\sqrtsNN=5.02~\tev$.
        The dashed curves represent the fit applied to the background, while the solid lines represent the total fit function.
        In the case of the $\Dzero$ meson, the contribution of signal reflections in the invariant-mass distribution is shown using a gray dot-dashed line. In the case of the $\Ds$ invariant-mass distribution, an additional Gaussian is used in the fit function in order to describe the $\Dplus$ signal peak on the left side of the $\Ds$ signal. 
        \label{fig:invmass4specpPb}
        }
\end{center}
\end{figure}

The D-meson raw yields were obtained by fitting the candidate invariant-mass distributions
for each D-meson species and the mass difference $\Delta M = M_{{\rm K} \pi\pi} -M_{{\rm K} \pi}$ for $\Dstar$.
Examples of these distributions are shown in Fig.~\ref{fig:invmass4specpPb} for $\Dzero$, $\Dplus$, $\Dstar$, and $\Ds$ mesons in different $\pt$ intervals.
The $\Dzero$, $\Dplus$, and $\Ds$ candidate invariant-mass distributions were fit with a function composed of a Gaussian for the signal shape and an
exponential term to describe the background shape.
The $\Delta M$
distribution of the $\Dstar$ candidates was fit with a Gaussian function for the signal shape and a threshold
function multiplied by an exponential for the background: $a\,\sqrt{\Delta M - m_{\pi} } \cdot {\rm e}^{b (\Delta M-m_{\pi})}$, where $a$ and $b$ are free parameters.
To account for the contribution of signal candidates that are present in the invariant-mass distribution of the $\Dzero$ meson but were assigned the wrong decay-particle mass (reflections) an additional term was included in the fit function.
The contribution of the reflections was modelled with a double Gaussian function parametrised on their invariant-mass distributions from Monte Carlo simulations.

For the $M_{{\rm KK} \pi}$ distribution, an additional Gaussian was used to describe the $\Dplus\to \rm{K}^{+}\rm{K}^{-}\pi^+$ signal peak
present on the left side of the $\Ds$ signal. The extracted signal is denoted as $S$ and the background level under the signal peak is denoted as $B$.
The statistical significance of the observed
signals, here defined as ($S/\sqrt{S+B}$), varies from 3 to 62, depending on the meson species, the centrality and the $\pt$ interval.

The D-meson raw yields extracted in each $\pt$ interval were corrected to obtain the prompt D-meson cross sections according to

\begin{equation}
  \label{eq:topolcrosssectionPromptD}
  \frac{{\rm d^2}\sigma^{\rm prompt \, D}}{{\rm d}\pt {\rm d} y}=
  \frac{1}{\Delta\pt} \cdot \frac{f_{\rm prompt} (\pt) \cdot \frac{1}{2} \cdot N^{\rm D+\overline{D},raw}(\pt)}{c_{\Delta y}(\pt)} \cdot \frac{1}{({\rm Acc}\times\epsilon)_{\rm prompt}(\pt)} \cdot \frac{1}{{\rm BR} \cdot \Lint}\,.
\end{equation}
In the formula, $N^{\rm D+\overline{D},raw}$ is the raw yield (sum of particles and antiparticles) in the laboratory rapidity interval 
$|y_{\rm lab}|<y_{\rm fid}(\pt)$ in a $\pt$ interval of width $\Delta \pt$.
The raw yield includes contributions from prompt and non-prompt D mesons.
Non-prompt D mesons originating from beauty-hadron decays are labeled as `feed-down' in the following. The $f_{\rm prompt}$ term is the fraction of prompt D mesons in the raw yield.
The rapidity acceptance correction factor $c_{\Delta y}$ was computed using
the PYTHIA v6.4.21 event generator \cite{Sjostrand:2006za} with the Perugia-2011 tune as the ratio between the
generated D-meson yield in $\Delta y = 2\,y_{\rm fid}$,
and that in $|y_{\rm lab}|<0.5$.
The $c_{\Delta y}$ correction factor
has a uniform D-meson rapidity distribution in $|y_{\rm lab}|<y_{\rm fid}$
in the range $|y_{\rm lab}|<0.8$ as shown in \cite{Adam:2016ich}. The factor $1/2$ accounts for the fact that the measured yields include
particles and antiparticles while the cross sections are given for particles
only. The $({\rm Acc}\times\epsilon)_{\rm prompt}$ is the product of the acceptance of the detectors and the
efficiency of prompt D mesons, where $\epsilon$ accounts for primary vertex reconstruction,
D-meson decay track reconstruction and selection, as well as for D-meson candidate
selection efficiencies.
Finally, BR is the branching ratio of the considered decay channel.

The acceptance and the efficiency were obtained by means of Monte Carlo simulations,
that include a detailed description of the apparatus geometry, the detector response,
as well as the LHC beam conditions.
Proton--proton collisions requiring a $\rm{c\overline{c}}$ or $\rm{b\overline{b}}$ pair satisfying $|y| < 1$ were generated using a PYTHIA v6.4.21 event generator~\cite{Sjostrand:2006za} with the Perugia-2011 tune. 
An underlying p--Pb collision, generated with HIJING 1.36~\cite{Wang:1991hta},
was superimposed to each PYTHIA event
in order to describe the charged-particle multiplicity and detector occupancy observed in data.
To reproduce the primary vertex resolution found in data which improves with increasing multiplicity, generated events were weighted on the basis of their charged particle multiplicity.
The shape of the generated D-meson $\pt$ distribution
is consistent with that of FONLL pQCD calculations~\cite{Cacciari:1998it} at $\sqrt{s} = 5.02$~TeV. The results from FONLL calculations are found to be consistent with \pp data at $\sqrt{s} = 5.02$~TeV though at upper edge of uncertainties as described in ~\cite{pp:2019}.

\begin{figure}[!t]
     \begin{center}
        	\label{fig:D0AccEff}
	\includegraphics[width=0.4\columnwidth]{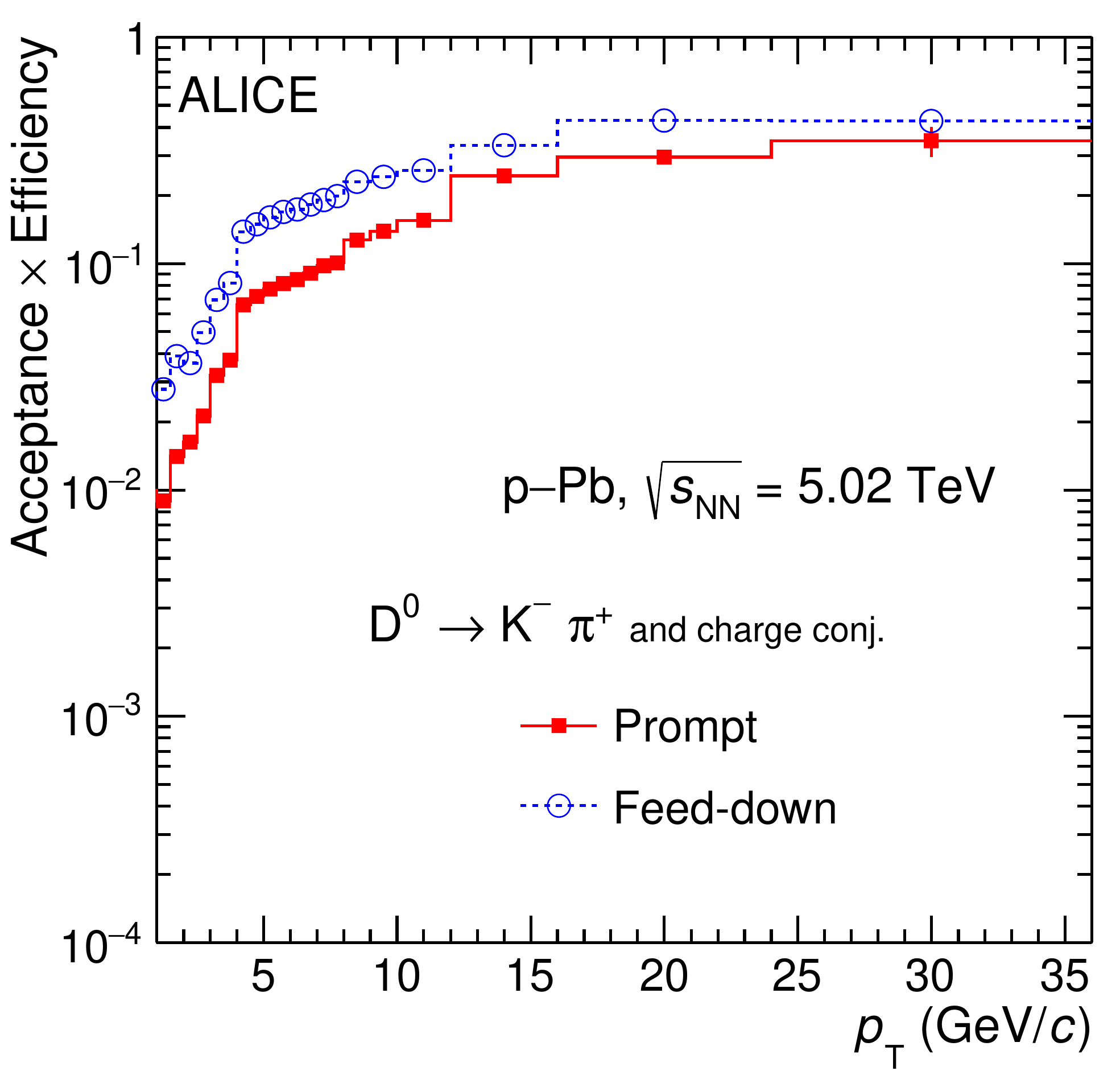}%Dzero_AccTimesEfficiencyFile.png}
	\label{fig:DplusAccEff}
	\includegraphics[width=0.4\columnwidth]{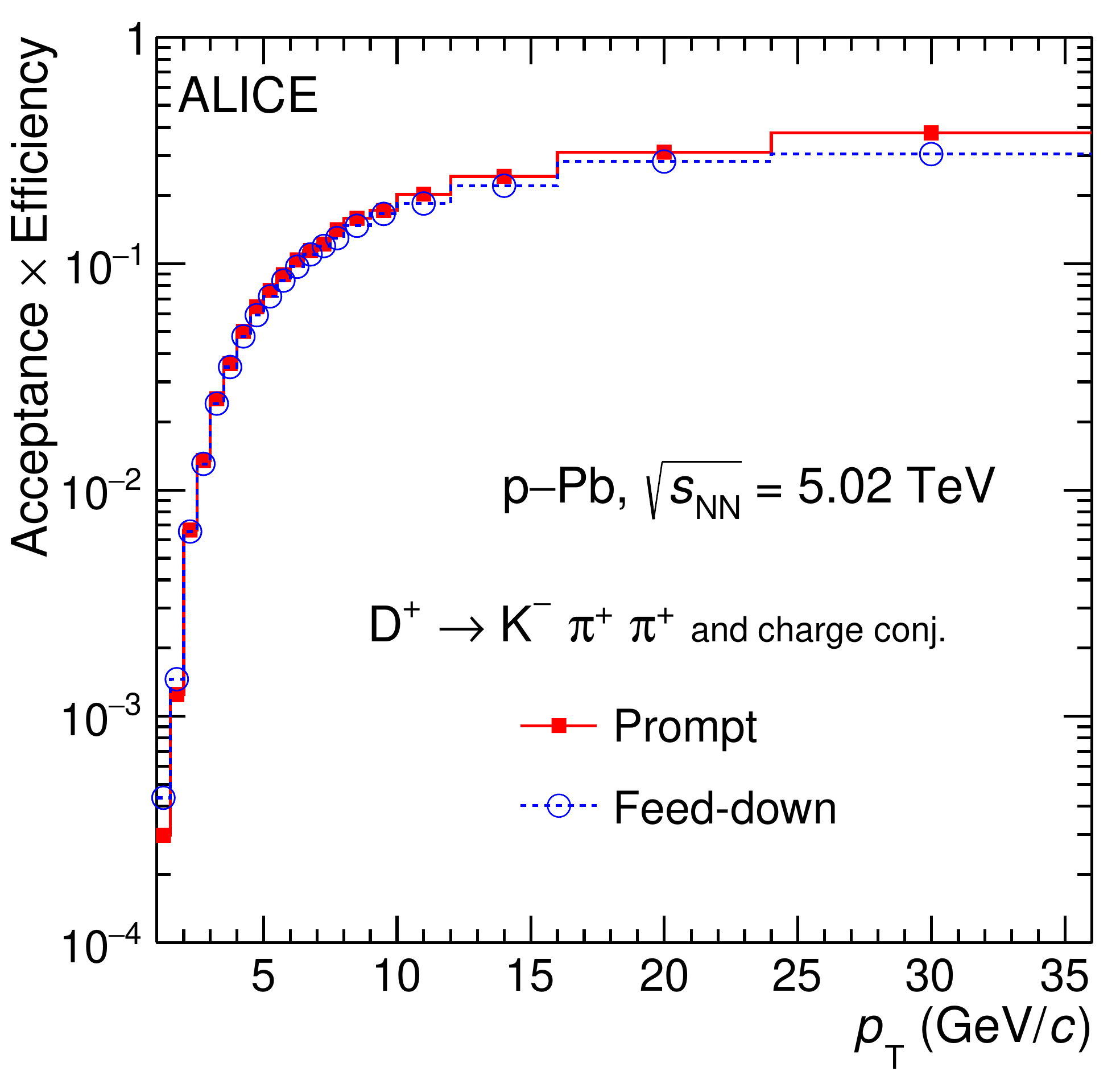}%Dplus_AccTimesEfficiencyFile.png}
	\\
	\label{fig:DstarAccEff}
	\includegraphics[width=0.4\columnwidth]{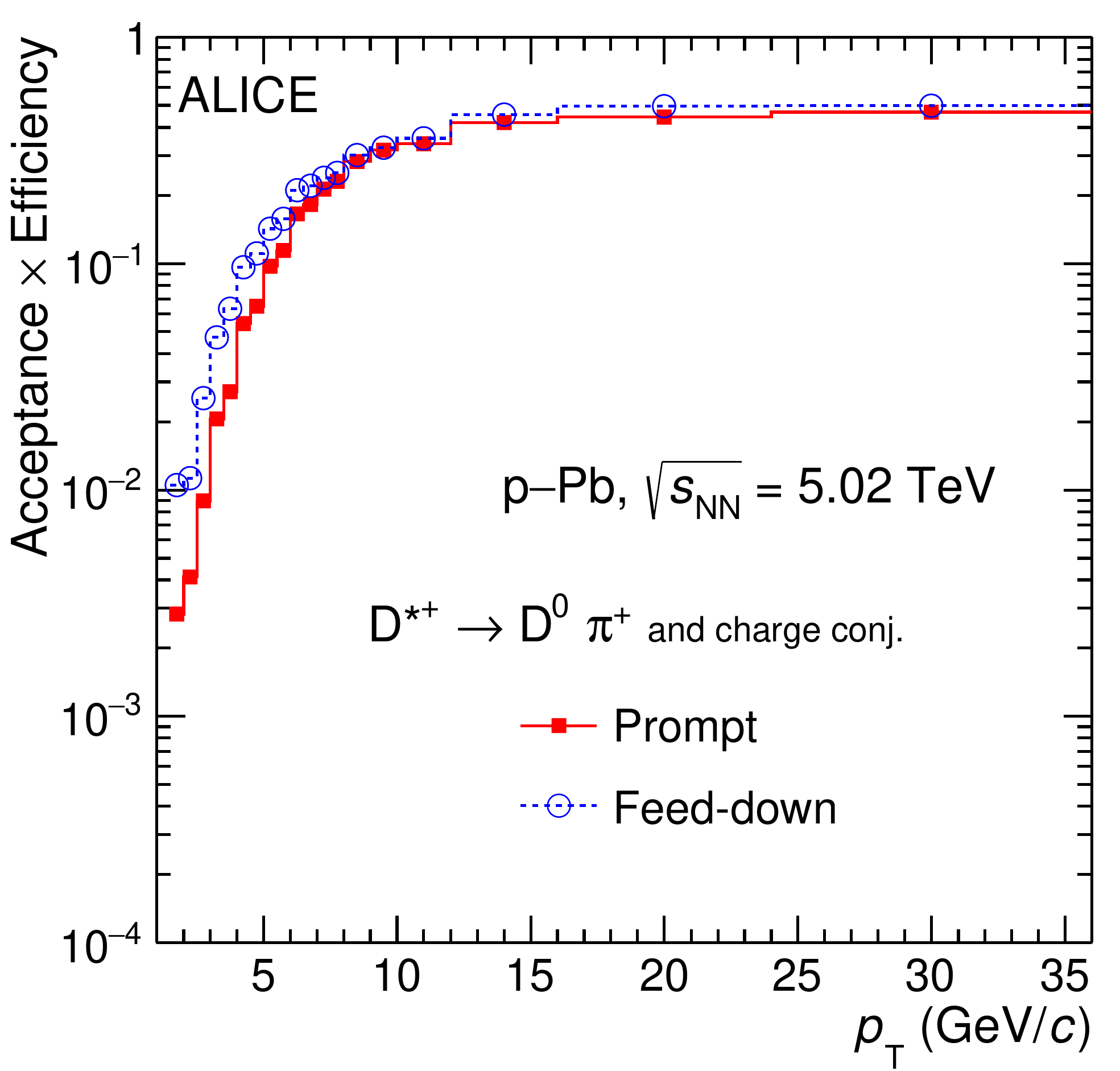}%Dstar_AccTimesEfficiencyFile.png}
	\label{fig:DsAccEff}
	\includegraphics[width=0.4\columnwidth]{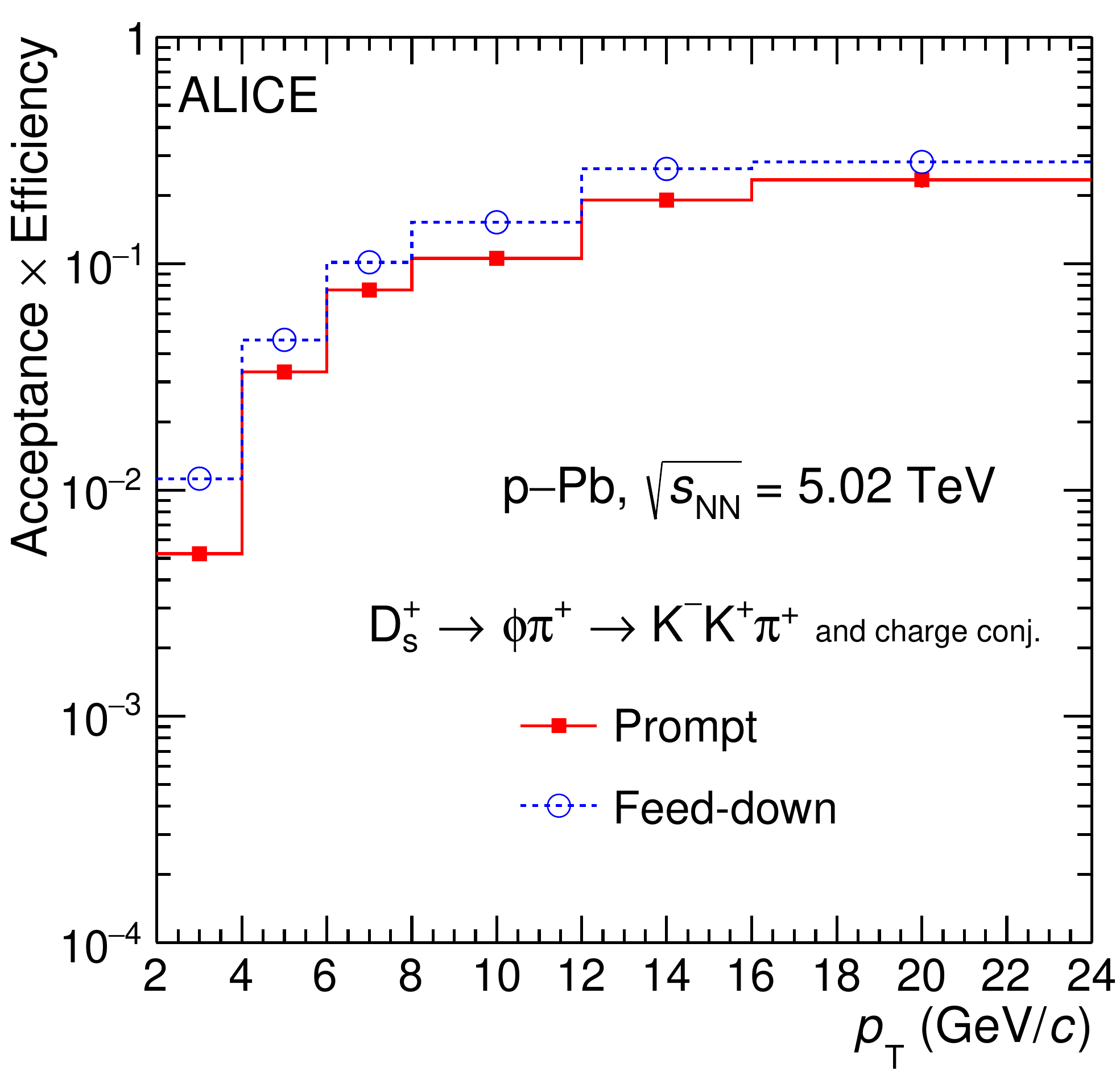}%Ds_AccTimesEfficiencyFile.png}
        \caption{The product of acceptance and efficiency for $\Dzero$, $\Dplus$, $\Dstar$, and $\Ds$ mesons as a function of transverse momentum in p--Pb collisions at $\sqrtsNN=5.02~\tev$.
        The values for prompt (solid line) and feed-down (dashed line) D mesons are shown.
        \label{fig:Deff}
        }
        \end{center}
    \end{figure}

Figure~\ref{fig:Deff} shows the
product of acceptance and efficiency $({\rm Acc}\times\varepsilon)$ for prompt and feed-down D mesons
with rapidity $|y_{\rm lab}| < y_{\rm fid}(\pt)$. The $\Dzero$, $\Dstar$, and $\Ds$ distributions are overall 
higher for the feed-down contribution compared to that of the prompt D mesons, while the opposite is true for the $\Dplus$ efficiency because of the topological selection.

The correction factor $f_{\mathrm{prompt}}$ was calculated per $p_{\rm T}$ interval using a FONLL-based method as described in~\cite{Acharya:2017jgo}.
The procedure uses the B-meson production cross section in pp collisions at $\sqrts = 5.02~\TeV$ estimated utilising FONLL calculations, the $\mathrm{B} \rightarrow \mathrm{D} + X$ decay kinematics from the EvtGen package~\cite{Ryd:2005zz}, the efficiencies for D mesons from beauty-hadron decays and a hypothesis on the nuclear modification factor $R_{\mathrm{pPb}}^\text{feed-down}$ of D mesons from B decays. 
The $\RpPb$ of prompt and feed-down D mesons were assumed to be equal on the basis of calculations including initial-state effects via the EPS09 nPDF parametrisations~\cite{Eskola:2009uj} or the Colour Glass Condensate formalism~\cite{Fujii:2013yja}, 
as well as the measurements of the $\rm B^0$-meson production in p--Pb 
collisions at $\sqrtsNN = 5.02~\TeV$ published by the CMS Collaboration~\cite{Khachatryan:2015uja}. 
Further details are given in Sec.~\ref{sec:Syst}. 
The resulting $f_{\rm prompt}$ values vary between 0.8 to 0.96 in the $|y_{\rm lab}|<y_{\mathrm{fid}}(\pt)$ interval depending on the $\pt$ range and the D-meson species.

%%----------------------------------------------------------
\subsection{Analysis without D-meson decay-vertex reconstruction}
\label{sec:lowpt}
\begin{figure}[!p]
\begin{center}
\includegraphics[width=\textwidth]{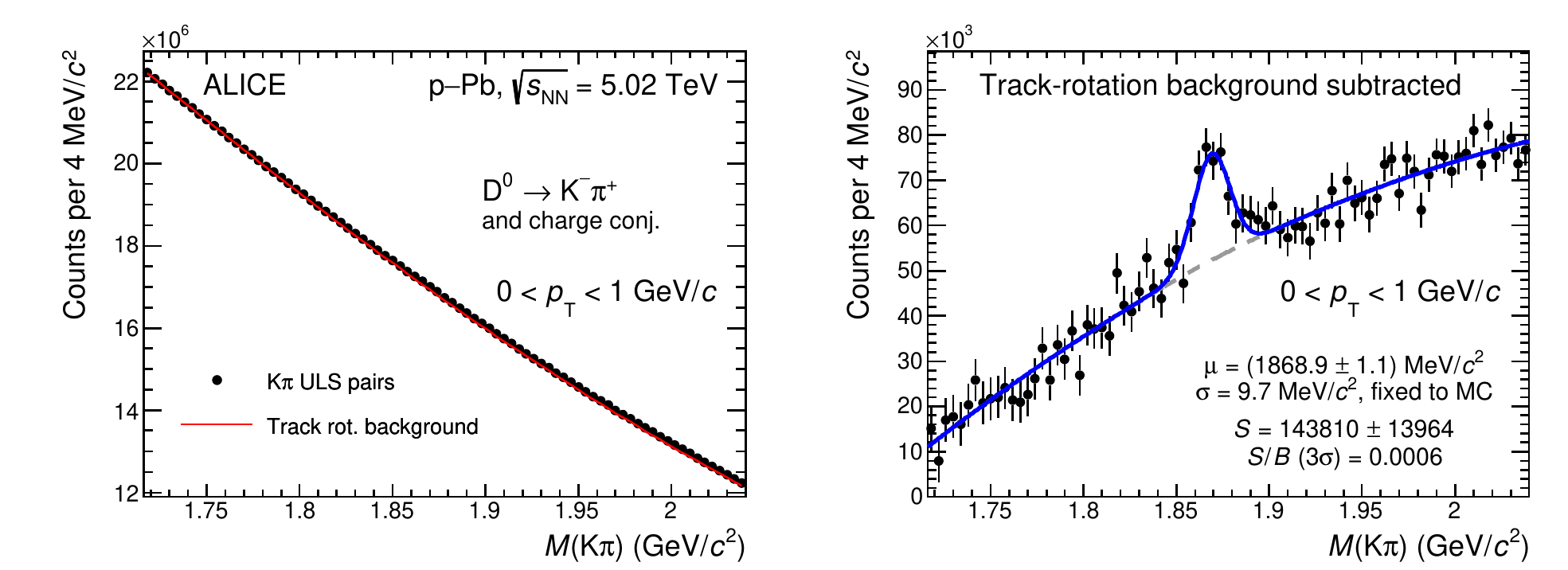}
\caption{Invariant-mass distributions of $\DtoKpi$ candidates (and
charge conjugates) for $0<\pt<1~\gev/c$. The left panel displays
the invariant-mass distribution of all ULS K$\pi$ pairs together with the
background distribution estimated with the 
track-rotation technique. The right panel shows the invariant-mass 
distribution after subtracting the background estimated with the track-rotation technique. The fit function is superimposed.}
\label{fig:invmassD0lowpt}
\end{center}
\end{figure}

In order to extend the cross section measurement down to $\pt=0$, a different analysis method, which does not employ geometrical selections on the displaced decay-vertex topology, was utilized for the two-body decay $\DtoKpi$ (and its charge conjugate)~\cite{Adam:2016ich}.
This analysis technique is based on particle identification and on the estimation and subtraction of the combinatorial background of K$\pi$ pairs.
Tracks with $|\eta|<0.8$ and $\pt>0.4~\gev/c$ were selected by applying the same track-quality cuts and pion and kaon identification criteria described above for the analysis with decay-vertex reconstruction. 
The $\Dzero$ and $\Dzerobar$ candidates were formed by combining kaon and pion tracks with opposite charge sign (UnLike Sign, ULS).
The resulting candidates were selected by applying the $\pt$-dependent fiducial acceptance selection, $|y_{\rm lab}|<y_{\rm fid}(\pt)$, adopted for the analyses with decay-vertex reconstruction.
No selections based on secondary-vertex displacement were applied
because at very low $\pt$ the D-meson decay topology cannot be efficiently 
resolved due to the insufficient resolution of the track impact parameter 
and the small Lorentz boost.
The combinatorial background was estimated with the track-rotation technique.
For each $\Dzero$ (and $\Dzerobar$) candidate, up to 19 combinatorial-background-like candidates were created by rotating the kaon track by different angles in the range between $\frac{\pi}{10}$ and $\frac{19\pi}{10}$ radians in azimuth.
The invariant-mass distribution of ULS K$\pi$ pairs in the transverse
momentum interval $0<\pt<1~\GeV/c$ is shown in the left panel of Fig.~\ref{fig:invmassD0lowpt} together with the one of the background estimated with the track-rotation technique, which was normalised to match the yield of ULS pairs at one edge of the invariant-mass interval considered for the extraction of the $\Dzero$ signal.

The invariant-mass distribution of background candidates was subtracted from the one of ULS K$\pi$ pairs and the resulting distribution, which contains the $\Dzero$ signal and the remaining background, is shown in the right panel of Fig.~\ref{fig:invmassD0lowpt}.
The $\Dzero$-meson raw signal (sum of particle and antiparticle contributions) was extracted via a fit to the background-subtracted invariant-mass distribution.
The fit function is composed of a Gaussian term to describe the signal, 
a second-order polynomial function to model the remaining background, and
a term describing the contribution of signal candidates passing the
selection criteria with swapped mass hypotheses of the final-state kaon and pion (reflections), 
whose invariant-mass distribution was taken from simulation.
The signal-to-background ratio increases from $6\cdot 10^{-4}$ to $3\cdot10^{-2}$ with increasing $\pt$ and the statistical significance is about 9 in $0<\pt<1~\gev/c$ and is greater than 15 for $\pt>2~\gev/c$.

\begin{figure}[!p]
\begin{center}
\includegraphics[width=0.48\textwidth]{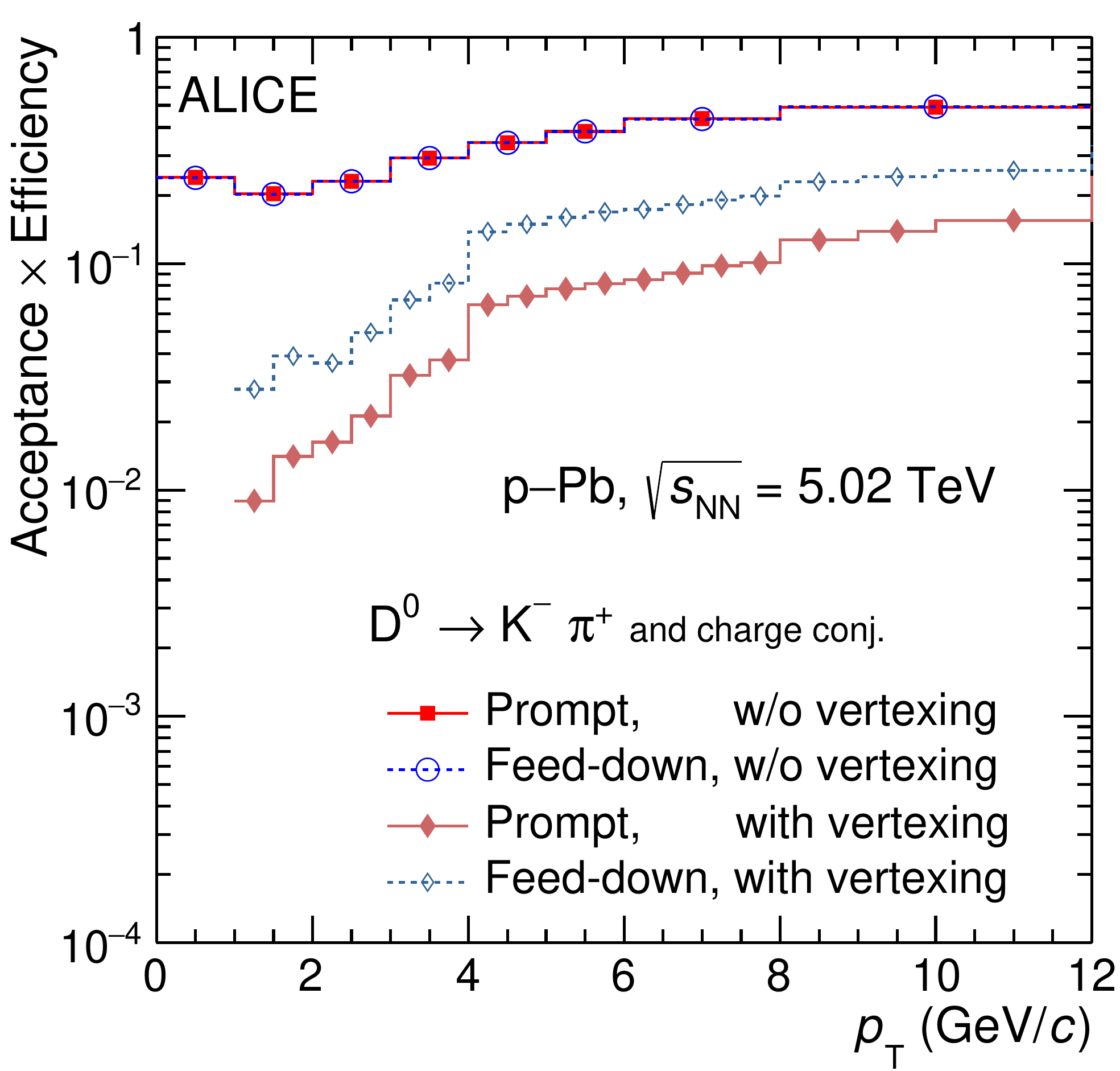}
\caption{
\label{fig:accefflowpt} 
Product of acceptance and efficiency of $\DtoKpi$ (and charge conjugates) in \pPb~collisions for the analyses with and without reconstruction of decay vertex. 
}
\end{center}
\end{figure}

The acceptance and efficiency were determined from the same
Monte Carlo simulations used for the analysis with decay-vertex reconstruction.
The resulting $({\rm Acc}\times\epsilon)$ of prompt 
$\Dzero$ mesons is shown as a function of $\pt$ 
in Fig.~\ref{fig:accefflowpt}.
Compared to the analysis with decay-vertex reconstruction,
the efficiency is higher by a factor of about 20 (3) at low (high) $\pt$
and it demonstrates a less steep $\pt$ dependence.
Note that for the analysis without decay-vertex reconstruction the efficiency
$\epsilon$ is almost independent of $\pt$ and the increase of the 
$({\rm Acc}\times\epsilon)$ with increasing $\pt$ is mainly determined by the 
geometrical acceptance of the apparatus.
Unlike in the analysis with decay-vertex reconstruction,
the efficiency is the same for prompt $\Dzero$ mesons 
and $\Dzero$ mesons from beauty-hadron decays.
% \clearpage

The prompt contribution to the $\Dzero$-meson raw yield, $f_{\rm prompt}$,
was estimated  with the same FONLL-based method used for the analysis with 
decay-vertex reconstruction.
The resulting $f_{\rm prompt}$ values decrease with increasing $\pt$
(from about 0.96 for $\pt<3~\GeV/c$ to about 0.9 in the interval 
$8<\pt<12~\gev/c$) and are larger than 
in the analysis with decay-vertex reconstruction, since the feed-down
component is not enhanced by the selection criteria.

%%----------------------------------------------------------
\subsection{Measurement of the prompt D-meson fraction based on a data-driven method}
\label{sec:fprompt}
\begin{figure}[!ht]
\begin{center}
	\includegraphics[width=0.85\textwidth]{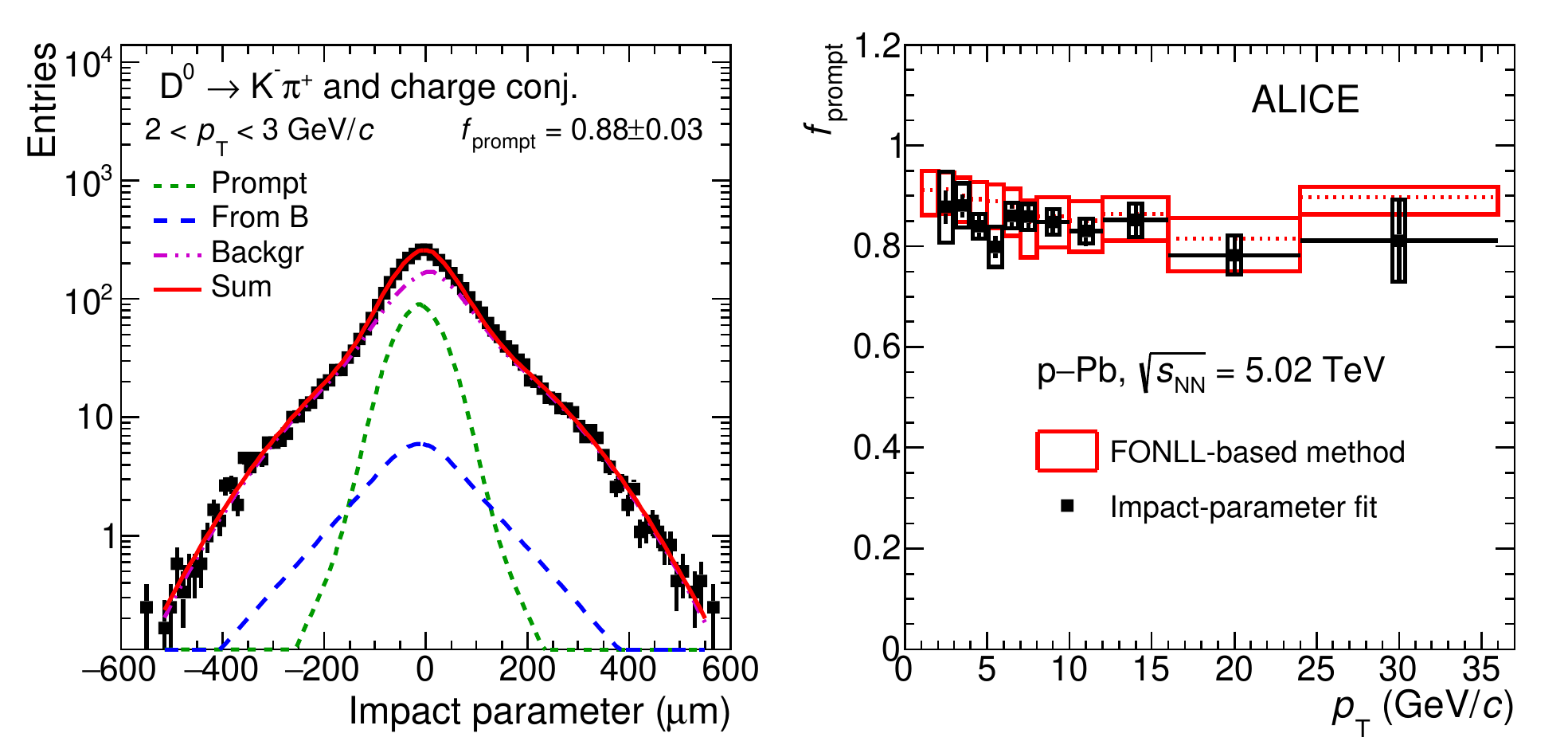}
	\includegraphics[width=0.85\textwidth]{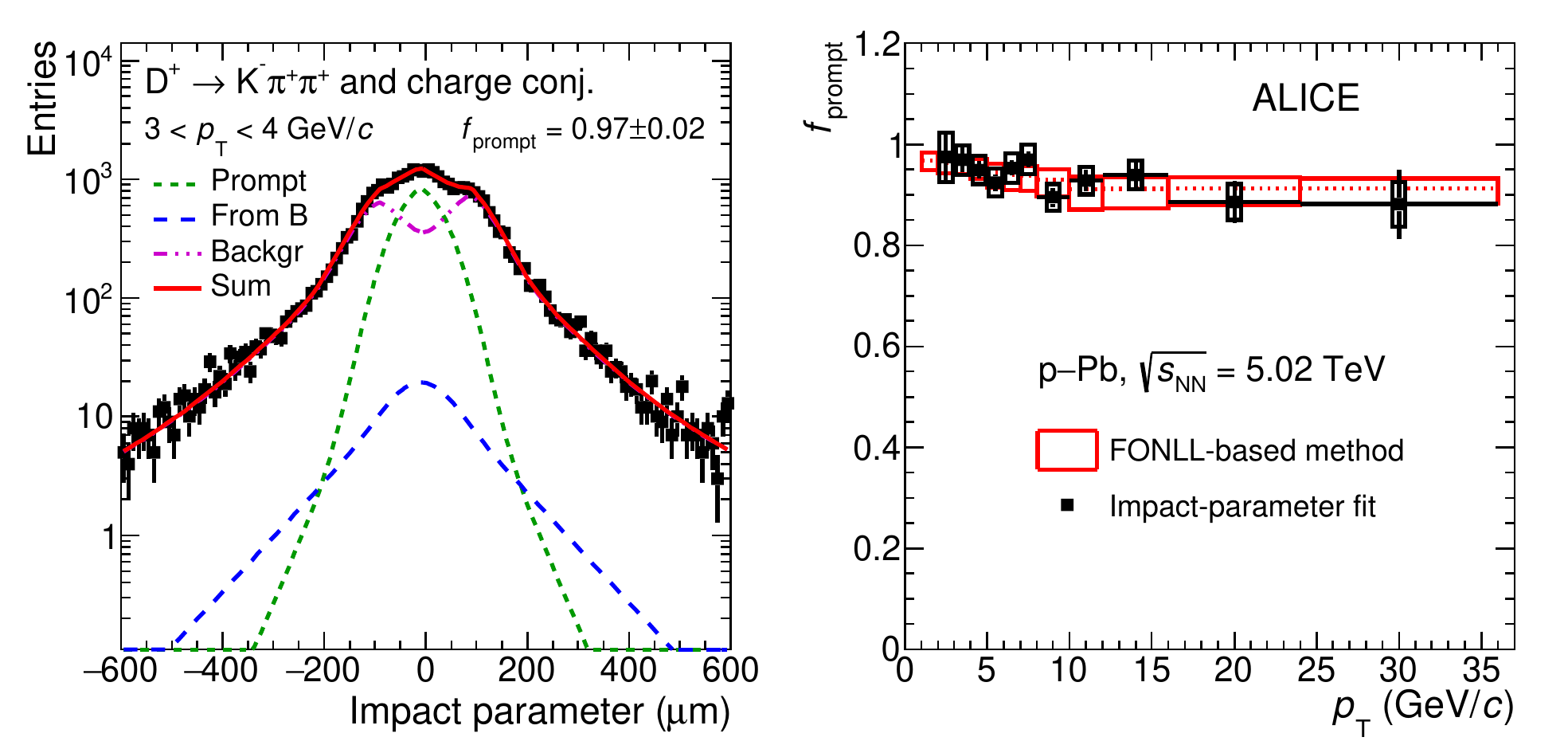}
	\includegraphics[width=0.85\textwidth]{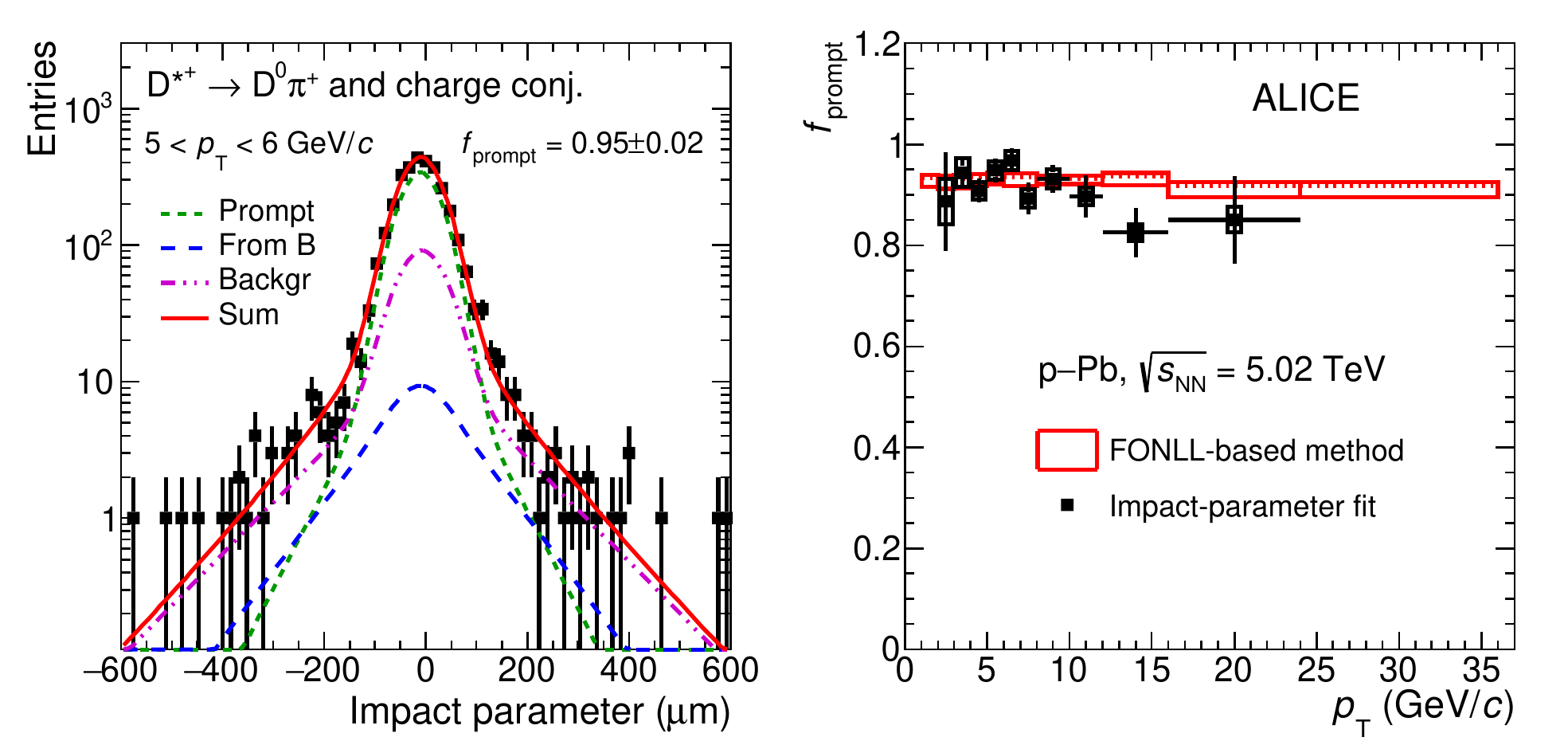}
\caption{Left: Exemplary fits to the impact-parameter distributions of $\Dzero$, $\Dplus$, and $\Dstar$ candidates.  
The curves show the fit functions describing the prompt, feed-down, and background contributions as well as their sum, as described in the text.
Right: fraction of prompt $\Dzero$, $\Dplus$, and $\Dstar$ raw yield as a function of transverse momentum $\pt$ compared with the values obtained with the FONLL-based approach. 
The results from the data-driven method are shown as square markers with the 
error bars (boxes) representing the statistical (systematic) uncertainty.  
The central values of $\fprompt$ from the FONLL-based approach are shown as the dashed lines and the uncertainty as red boxes.
\label{fig:fprompt} 
}
\end{center}
\end{figure}

The prompt fractions of $\Dzero$, $\Dplus$, and $\Dstar$ 
mesons, calculated via the FONLL-based method, were cross-checked for the analysis with decay vertex reconstruction utilizing a data-driven method that exploits 
the different shapes of the transverse-plane impact parameter to the primary vertex ($d_0$) of 
prompt and feed-down D mesons.  The D-meson candidates were selected using the same criteria described in Section~\ref{sec:topol}, with the exception that for $\Dplus$ the impact-parameter selection criteria were not applied. An additional selection was based on the candidate invariant-mass fits. The $\Dzero$ and $\Dplus$ mesons candidates were selected to have an invariant mass $|M-M_{\mathrm{D}}|<1.5 \sigma$, while for $\Dstar$-meson candidates a $|\Delta M-\Delta M_{\mathrm{D^{*+}}}|<2.5 \sigma$ selection was applied, where  $\sigma$ is the standard deviation of the Gaussian function describing the D-meson invariant-mass signal. The prompt fraction was estimated via an unbinned likelihood fit of the $d_0$ distribution of the D-meson candidates using the fit function 
\begin{equation}
 F(d_0) = S\cdot \left[ (1-f_{\rm prompt}) F^{\rm feed\mbox{-}down} (d_0) + f_{\rm prompt} F^{\rm prompt} (d_0) \right] + B\cdot F^{\rm backgr}(d_0)\,.
  \label{eq:ImpParFit}
\end{equation} 
In this function, 
$F^{\rm prompt} (d_0)$, $F^{\rm feed\mbox{-}down} (d_0)$  and $F^{\rm backgr}(d_0)$ 
are functions describing the impact-parameter distributions of prompt and feed-down D mesons and of background candidates. 
The function $F^{\rm prompt}$ consists of a detector resolution term modeled with a Gaussian function and a symmetric exponential term, 
$\frac{1}{2\lambda}\exp{\left( -\frac{\lvert d_0 \rvert}{\lambda} \right) }$ (with $\lambda$ as a free parameter), with the latter describing the tails of the impact-parameter distribution of prompt 
D mesons. The $F^{\rm feed\mbox{-}down}$ is the convolution of the detector resolution term with a symmetric double-exponential
function ($F^{\rm feed\mbox{-}down}_{\rm true}$) that describes the intrinsic impact-parameter distribution of D mesons from B-meson decays, which is determined by the decay length and decay kinematics of B mesons. 
The parameters of the $F^{\rm prompt}$ and $F^{\rm feed\mbox{-}down}_{\rm true}$ functions were fixed to the values obtained by fitting the distributions from Monte Carlo simulations,
with the exception of the Gaussian width of the detector-resolution term, which was kept free when applying the fit to the data in order to compensate for a possible imperfect description of the impact-parameter resolution in the simulation. 
The function $F^{\rm backgr}$ was parametrised on the impact-parameter distribution of background candidates, which were selected from side bands relative to the signal peak in the invariant-mass distributions, and in the case of $\Dstar$, the mass-difference distribution.  The function consists of a double Gaussian and a symmetric exponential term, which describes the tails, as reported in Ref.~\cite{Adam:2016ich}. 
In the case of the $\Dplus$, the function presents a double-peak structure with a depletion around zero that is induced by the selections applied.  

The left panels of Fig.~\ref{fig:fprompt} 
show examples of fits to the impact-parameter distributions of $\Dzero$, $\Dplus$, and $\Dstar$ 
mesons in the transverse-momentum intervals $2<\pt<3~\gevc$, $3<\pt<4~\gevc$, and $5<\pt<6~\gevc$, respectively.
The prompt fraction estimated using the data-driven approach has systematic uncertainties due to 
(i) the impact-parameter distribution assumed for prompt and feed-down D mesons and background candidates; 
(ii) the uncertainty on the signal and background yields extracted from the invariant-mass fits; and
(iii) the consistency of the procedure, evaluated via a Monte Carlo closure test.
These uncertainties were estimated using the procedures described in Ref.~\cite{Adam:2016ich}.
The total systematic uncertainty  on $f_{\rm prompt}$ based on the data-driven 
approach for the three D-meson species  is about 2--3\% in the interval $3<\pt<16~\gev/c$ and about 5\% in the interval $2<\pt<3~\gev/c$ and above 16~GeV/$c$.

The prompt fraction of $\Dzero$, $\Dplus$, and $\Dstar$ mesons 
measured utilizing the data-driven method is compared with the one calculated with the FONLL-based approach in the right panels of Fig.~\ref{fig:fprompt}. 
For $\Dzero$, $\Dplus$, and $\Dstar$ in $1<\pt<2~\gev/c$ and for the $\Dstar$ in $24<\pt<36~\gev/c$, given the poor precision of 
the impact-parameter fit, it was not possible to determine $f_{\rm prompt}$ with the data-driven approach.
The prompt fraction measured with the impact-parameter fits is
compatible with the FONLL-based estimation within 1$\sigma$ for almost all points.

%%---------------------------
\section{Systematic uncertainties}
\label{sec:Syst}

\begin{table}[ht!]
    \centering
     \caption{Summary of relative systematic uncertainties on $\Dzero$, $\Dplus$, $\Dstar$, and $\Ds$ production cross sections. The event centrality-dependent uncertainties are marked by the symbol $\diamond$.}
    \begin{tabular}{l|ccc|cc|cc|cc}
         & \multicolumn{3}{c|}          {$\Dzero$}   & \multicolumn{2}{c|}{$\Dplus$} & \multicolumn{2}{c|}{$\Dstar$} & \multicolumn{2}{c}{$\Ds$} \\
         $\pt$ ($\gevc$)             & 0--1 & 2--2.5  & 10--12                  & 2--2.5 & 10--12                             & 2--2.5 &  10--12               & 2--4 & 8--12 \\
         \hline
         Signal yield $\diamond$ &  5\% & 3\%  & 3\%                   & 2\% & 3\%                 & 7\% & 2\%                   &  3\% & 2\% \\
         Tracking efficiency  & 2.5\% & 2.5\% & 2.5\%                       & 3.7\% & 4\%                   & 3.2\% & 4.5\%                & 3.7\% & 4\% \\
         Selection efficiency         & negl. & 3\% & 3\%                                     & 7\% & 4\%                     & 2\% & 2\%              & 6\% & 4\% \\
         PID efficiency         &negl. & negl. &negl.                                             & negl. & negl.                               & negl. & negl.                         & 1\% & 1\% \\
         $\pt$ shape in MC           &negl. & negl. & negl.                                             & negl. & negl.                           & negl.& negl.                            & negl. & negl. \\
         Feed-down $\diamond$  & $^{+1.3}_{-1.7}$\% & $^{+4.2}_{-4.9}$\% & $^{+4.1}_{-5.6}$\% & $^{+1.8}_{-2.1}$\% & $^{+2.3}_{-3.2}$\% & $^{+3.0}_{-3.5}$\% & $^{+1.9}_{-2.6}$\% & $^{+3.6}_{-4.2}$\% & $^{+4.4}_{-5.6}$\% \\
         Branching ratio & \multicolumn{3}{c|}{1.0\%} &  \multicolumn{2}{c|}{3.1\%} & \multicolumn{2}{c|}{1.3\%} & \multicolumn{2}{c}{3.5\%} \\
         \hline
         Normalisation & \multicolumn{9}{c}{3.7\%} \\
         \hline 
    \end{tabular}

\label{sysunc_yieldtable}
\end{table}

Systematic uncertainties on the D-meson production cross sections were 
estimated considering the following sources:\\
(i) extraction of the raw yield from the invariant-mass distributions; (ii) track reconstruction efficiency; 
(iii) D-meson selection efficiency; 
(iv) PID efficiency; 
(v) the assumption on the shape of the D-meson $\pt$ spectrum generated  in the simulation; 
(vi) subtraction of the feed-down from beauty-hadron decays.\\
In addition, the $\pt$-differential cross sections have a systematic uncertainty 
on the overall normalisation induced by the uncertainties on the integrated luminosity of 3.7\%~\cite{Abelev:2014epa}
and on the branching ratios of the considered D-meson decays~\cite{Tanabashi:2018xmw}. 
The estimated values of the relative systematic uncertainties are summarised in Table \ref{sysunc_yieldtable}.
The contributions of the different sources were summed in quadrature to obtain
the total systematic uncertainty.

The systematic uncertainties on the raw yield extraction were evaluated for
each D-meson species by repeating the invariant-mass distribution fits, for each $\pt$ and centrality interval, varying the lower and upper limits 
of the fit range and the functional form of the background fit function.
In addition, the same approach was used with a bin-counting method, in which 
the signal yield was obtained by integrating the invariant-mass distribution 
after subtracting the background estimated from a fit to the side bands.
For $\Dzero$ mesons, an additional contribution due to the description of signal reflections 
in the invariant-mass distribution was estimated by varying the ratio of the 
integral of the reflections over the integral of the signal and the shape of 
the templates used in the invariant-mass fits. 
The systematic uncertainty was defined as the root mean square\,of the distribution of 
the signal yields obtained from the described variations. 
The uncertainty ranges between 1\% and 15\% depending on the D-meson species, $\pt$, event centrality and charged-particle 
multiplicity intervals of the measurement. An increase in the raw yield extraction uncertainties was observed in the most central collisions due to the lower $S$/$B$ ratio. 
For the $\Dzero$-meson analysis without decay-vertex reconstruction, different configurations of the rotation angle were used to estimate the background with the track-rotation technique. Furthermore, three alternative approaches were tested to estimate the background distribution: like-sign (LS) pairs, event mixing, and side-band fit~\cite{Adam:2016ich}. The raw yield values obtained subtracting these alternative background distributions were found to be consistent with those from the default configuration of the track-rotation method within the uncertainty estimated by varying the fit conditions and therefore no additional systematic uncertainty was assigned.
The systematic uncertainty on the track reconstruction efficiency was estimated 
by varying the track-quality selection criteria and by comparing the 
probability to match the tracks from the TPC to the hits in the ITS, in the data and simulation.
The comparison of the matching efficiency in the data and simulation was made 
after weighting the relative abundances of primary and secondary particles in 
the simulation to match those in the data, which were estimated via fits 
to the track impact-parameter distributions ~\cite{ALICE-PUBLIC-2017-005}.
The estimated uncertainty depends on the D-meson $\pt$ and it ranges from 
2.5\% to 4\% for the two-body decay of $\Dzero$ mesons and from 3.7\% to 4.5\% 
for the three-body decays of $\Dplus$, $\Dstar$, and $\Ds$ mesons.

The uncertainty on the selection efficiency originates 
from imperfections in the description of the D-meson kinematic and decay properties
and of the detector resolution and alignment in the simulation. 
For the analyses based on the decay-vertex reconstruction, the uncertainty was estimated by 
comparing the corrected yields obtained by repeating
the analysis with different sets of selection criteria, resulting in a  
significant modification of efficiencies, raw yield, and background estimates. 

The assigned uncertainty for non-strange D mesons is 2--3\% in most of the $\pt$ 
intervals and it increases to 7\% at low $\pt$, where the efficiencies are 
low and steeply fall with decreasing $\pt$, because of the tighter geometrical selections. 
A larger uncertainty (ranging from 7\% at high $\pt$ to 14\% at low $\pt$) was
estimated for the $\Ds$ mesons, for which more stringent selection criteria
were used in the analysis, as compared to non-strange D mesons.
In the case of the $\Dzero$-meson analysis without decay-vertex reconstruction,
the stability of the corrected yield was tested against variations of the 
single-track $\pt$ selection and no systematic effect was observed.

In addition, the efficiency values could also be sensitive to the generated
shapes of the D-meson transverse-momentum distributions and to the 
multiplicity of particles produced in the collision.
The systematic uncertainty due to the generated D-meson $\pt$ spectrum shape
was estimated by considering different input distributions (PYTHIA, FONLL) 
and was found to be negligible. 
The effect of possible differences between the charged-particle multiplicity 
distributions in data and multiplicity-weighted simulation, used to compute the efficiencies in the different centrality classes, as explained in section \ref{sec:topol}, varied between 0 and 2\% depending on the D-meson species, $\pt$, event centrality, and charged-particle multiplicity intervals.

To estimate the uncertainty on the PID-selection efficiency the analysis was repeated without PID selection, or with less stringent criteria in the cases where the signal extraction was not reliable without PID, as for example for the $\Ds$ and the $\Dzero$-meson analysis without decay-vertex reconstruction.
In addition, the pion and kaon PID selection efficiencies were compared in the data and in simulation using high purity samples of pions from the decay of $\rm K^{0}_{s}$ and kaons identified with the TOF combined with the D-meson decay kinematics. The PID uncertainty was found to vary between 0 and 1.5\% depending on the PID selection criteria used for each D-meson species.

The systematic uncertainty on the subtraction of feed-down from beauty-hadron decays 
(i.e.\,the calculation of the $f_{\rm prompt}$ fraction) was estimated 
 by varying the FONLL parameters (b-quark mass, factorisation and 
renormalisation scales) as described in~\cite{Cacciari:2012ny} and by varying 
the hypothesis on the nuclear modification factor of feed-down D mesons in 
the range $0.9 < R(Q)^{\text{feed-down}}_{\mathrm{pPb}}/R(Q)^{\mathrm{prompt}}_{\mathrm{pPb}} < 1.3$ for the integrated centrality interval and central collisions, and between  $0.9 < Q^{\text{feed-down}}_{\mathrm{pPb}}/Q^{\mathrm{prompt}}_{\mathrm{pPb}} < 1.1$ for the peripheral collisions, where the possible differences of the D-meson production mechanisms in p--Pb with respect to pp collisions are expected to be reduced as observed for both charmed  mesons and charged particles. 
The uncertainty ranges between $2\%$ and $5\%$ depending on the D-meson species, $\pt$, event centrality and charged-particle multiplicity intervals.

%%%----------------------------------------------------------
\section{Results}
\label{sec:results}

\subsection{$\pt$-differential cross sections}

\begin{figure}[!tbhp]
\begin{center}
\includegraphics[width=0.48\textwidth]{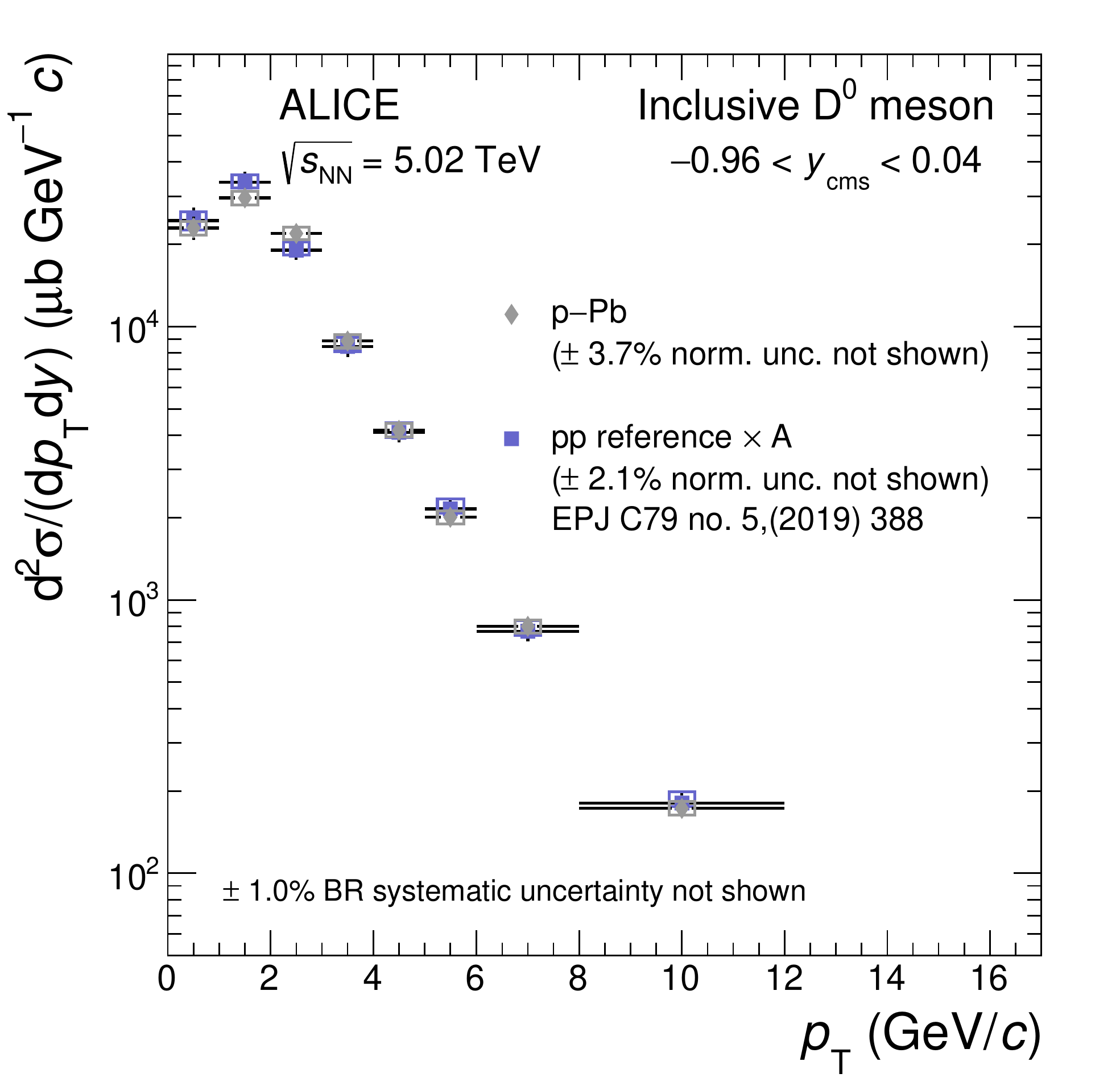}
\includegraphics[width=0.48\textwidth]{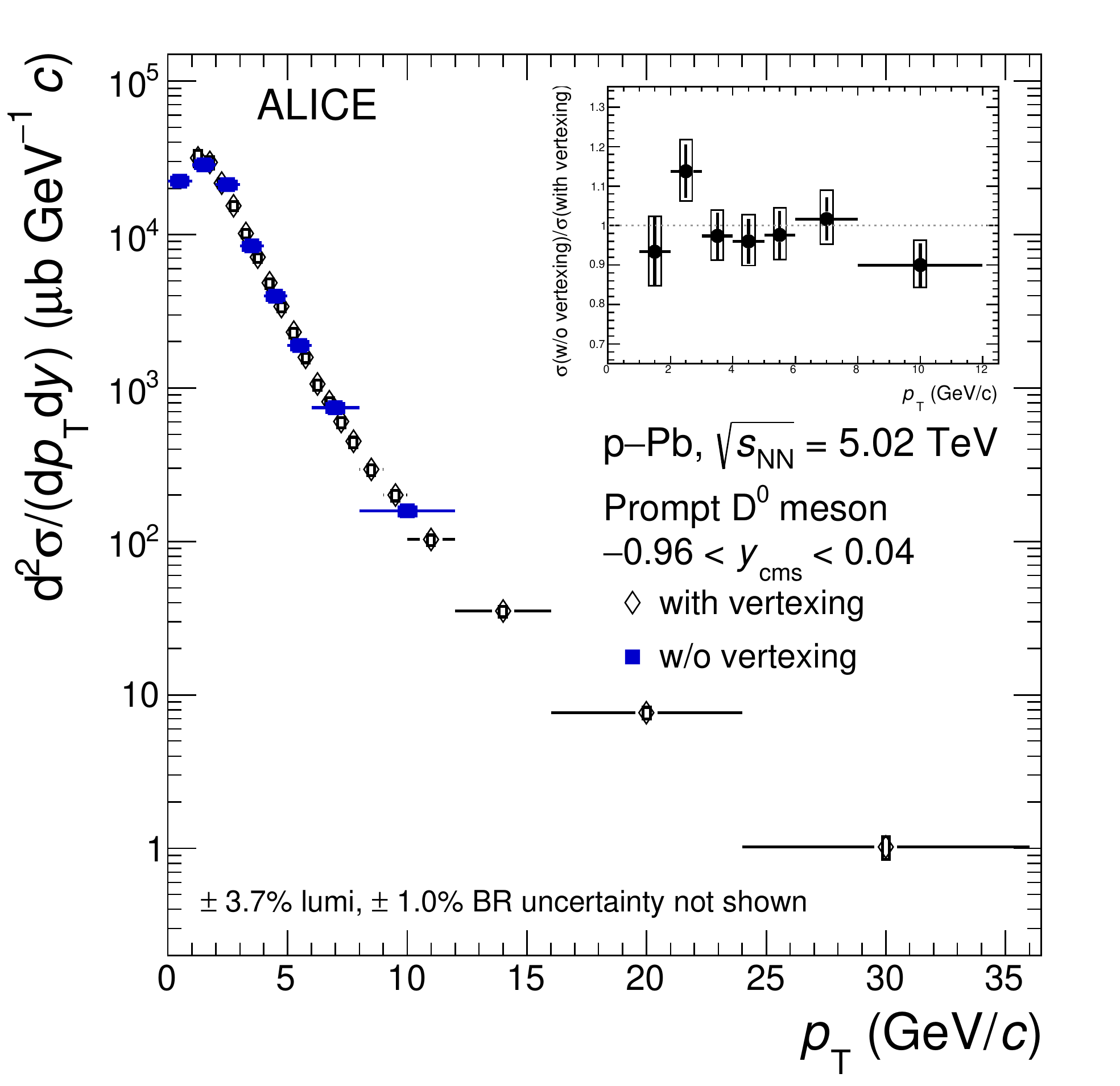}
\caption{ \label{fig:CrossSecPPB}
 Left: inclusive $\Dzero$-meson production cross sections from the analysis without decay-vertex reconstruction in p--Pb collisions and pp collisions, both at $\sqrtsNN =5.02~\tev$. The cross section measured in pp collision ~\cite{pp:2019} is  scaled by the Pb mass number ($A=208$) and corrected for the rapidity shift in p--Pb collisions using FONLL calculations. Right: $\pt$-differential production cross section of prompt $\Dzero$ mesons with $-0.96<y_{\rm cms}<0.04$ in p--Pb collisions at $\sqrtsNN =5.02~\tev$, measured with and without decay-vertex reconstruction. The vertical bars and the empty boxes represent the statistical and systematic uncertainties. The inset shows the ratio of the measurements in their common $\pt$ range.
 }

\end{center}
\end{figure}
The analysis without decay-vertex reconstruction allows for a direct measurement of the inclusive $\Dzero$-meson cross section because no selections that alter the fraction of prompt and feed-down D mesons are applied.
The inclusive $\Dzero$-meson cross section in p--Pb collisions is shown in the left panel of Fig.~\ref{fig:CrossSecPPB} and is compared with the measurement in pp collisions at the same centre-of-mass energy,  published in \cite{pp:2019}. The cross section in pp collisions was scaled by the Pb mass number $A=208$ and corrected for the rapidity shift in p--Pb collisions using FONLL calculations. The correction for the rapidity shift is a $\pt$-dependent factor of the order of 1--3\%. The uncertainty assigned on this correction is evaluated varying the quark mass and the perturbative scale parameters and including the PDFs uncertainty, and is 1\% at low $\pt$  and negligible at high $\pt$.\\ 
The total cross section for inclusive $\Dzero$-meson production in p--Pb collisions per unit of rapidity in $-0.96<y_{\rm cms}<0.04$ was obtained by integrating the $\pt$-differential cross section shown in the left panel of Fig.~\ref{fig:CrossSecPPB}.
The systematic uncertainty was defined by propagating the yield extraction uncertainty as uncorrelated among the $\pt$ intervals and all the other uncertainties as correlated. 
The cross section was then extrapolated to the whole $\pt$ range using FONLL calculations in order to take into account the fraction of cross section not measured for $\pt>12$ GeV/$c$. An uncertainty was estimated for the extrapolation varying the quark mass and the perturbative scale parameters and including the PDFs uncertainty. The resulting cross section is

\begin{equation}
{\rm d}\sigma^{\rm inclusive\,D^0}_{\rm p-Pb,\,5.02\,TeV}/{\rm d}y=91.2\pm 3.4\,({\rm stat.}) \pm 3.2\,({\rm syst.})\pm 3.4\,({\rm lumi.})\pm 0.9\,({\rm BR}) ^{+0.4}_{-0.2}\,({\rm extrap.})\,~{\rm mb}.
\end{equation}

The right panel of Fig.~\ref{fig:CrossSecPPB} shows the comparison of the $\pt$-differential production cross sections for prompt $\Dzero$ mesons with $-0.96<y_{\rm cms}<0.04$ in p--Pb collisions at $\sqrtsNN=5.02~\tev$ obtained from the analysis with and without decay-vertex reconstruction. The results are consistent within statistical uncertainties. 

    \begin{figure}[!tbhp]
        \begin{center}
            \includegraphics[width=0.48\textwidth]{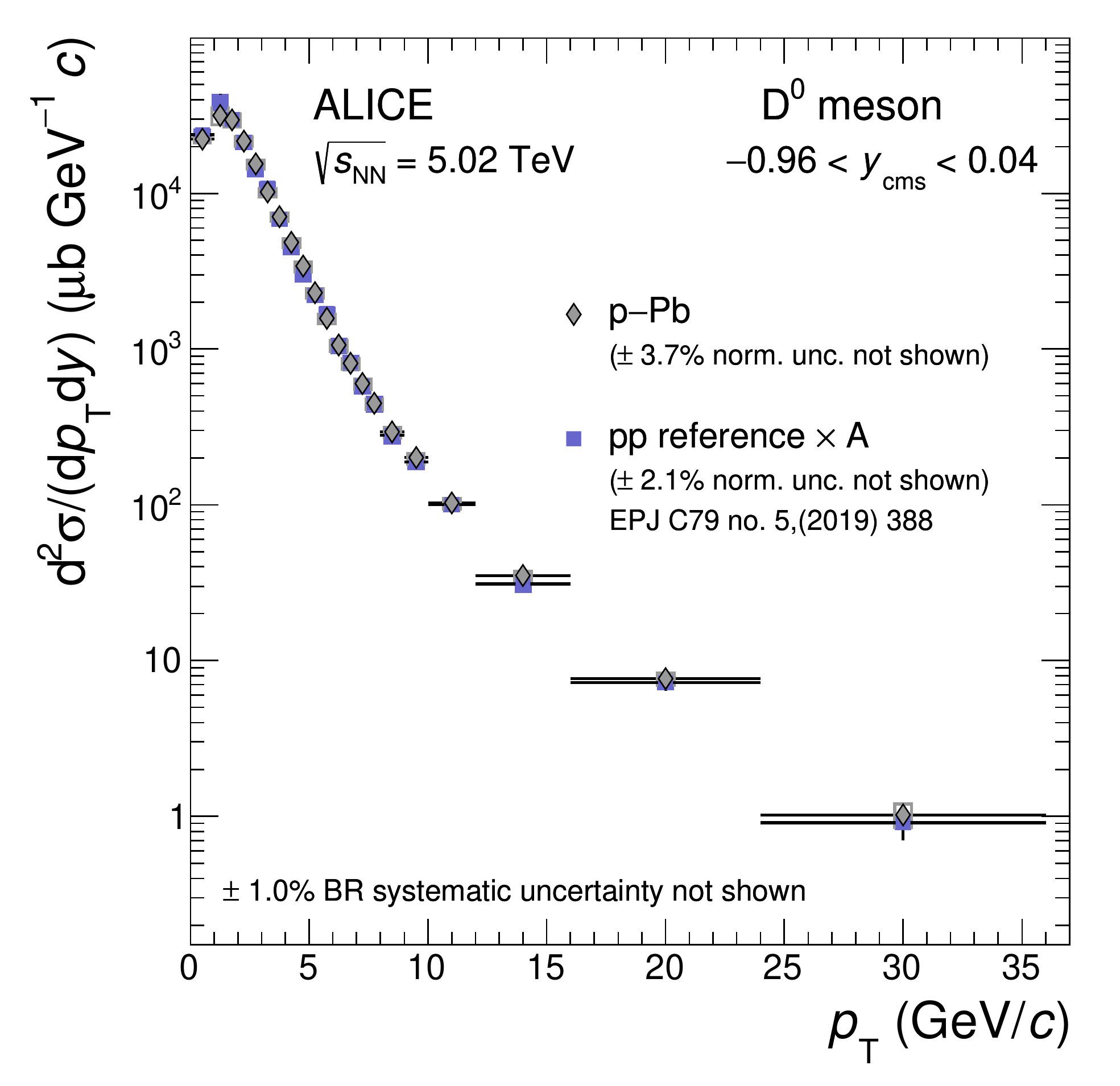}
            \includegraphics[width=0.48\textwidth]{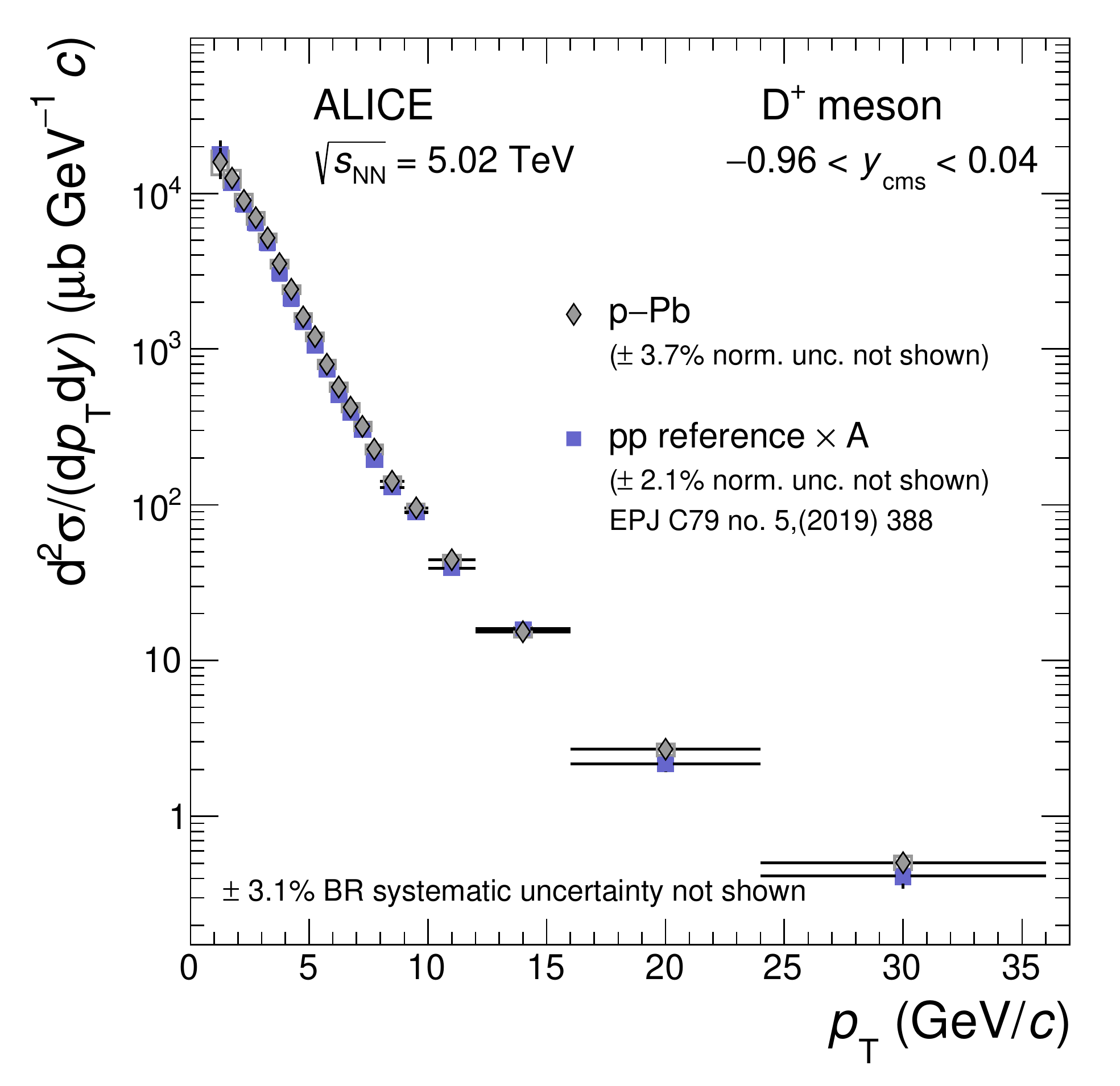}
            \includegraphics[width=0.48\textwidth]{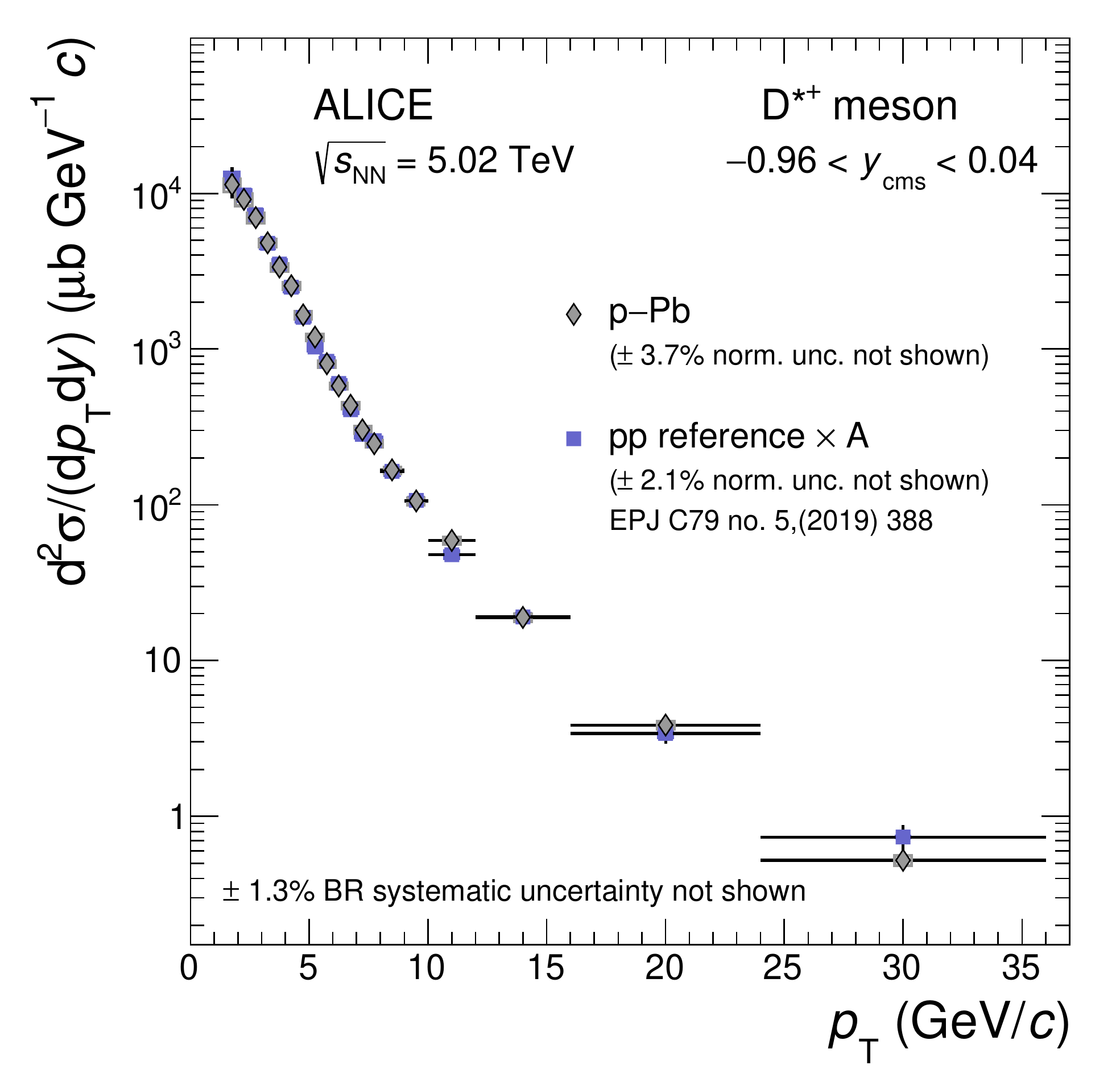}
            \includegraphics[width=0.48\textwidth]{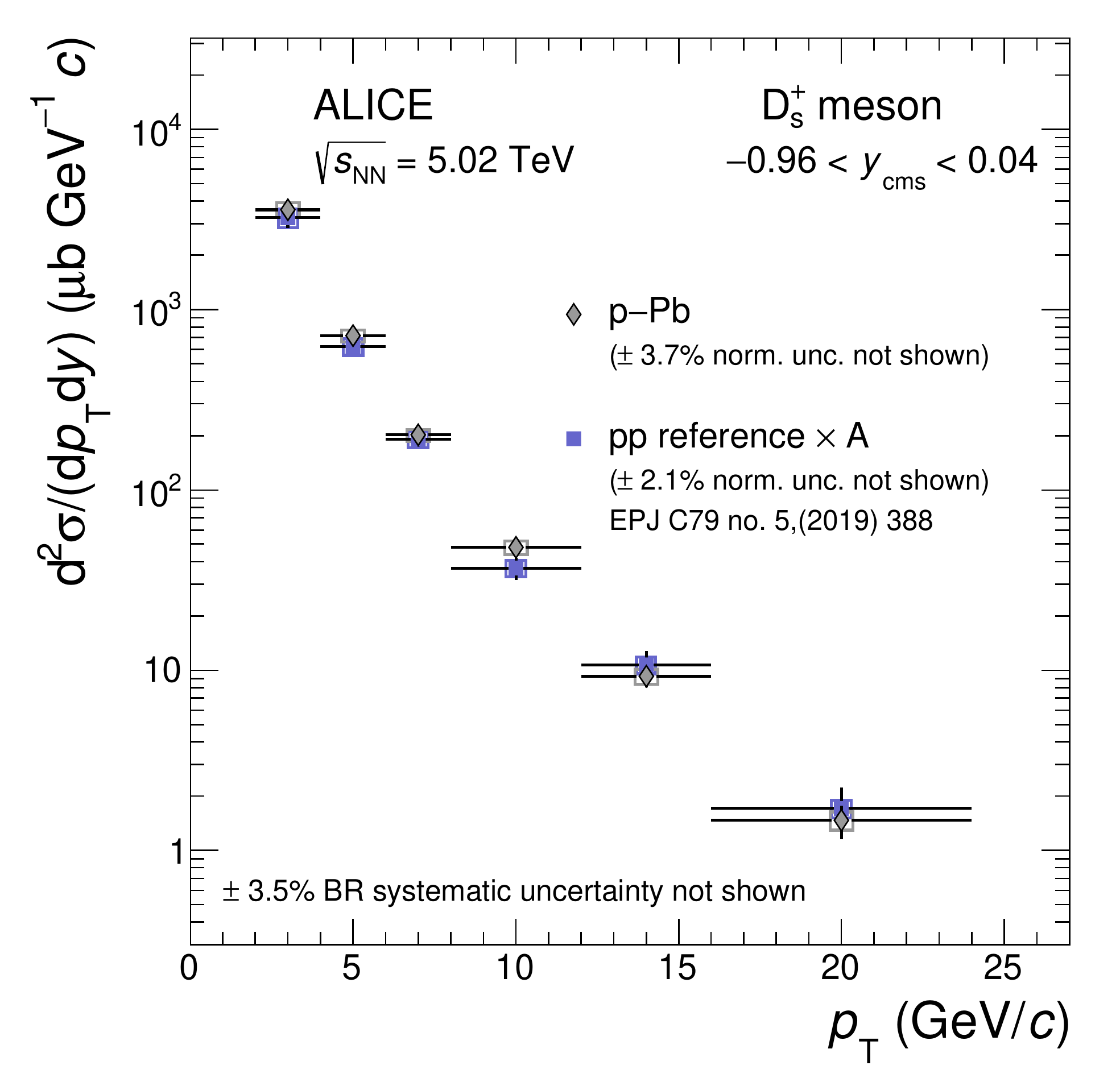}
        \caption{        \label{fig:dsdptppref}
        $\pt$-differential production cross sections of prompt $\Dzero$, $\Dplus$,
        $\Dstar$, and $\Ds$ mesons with $-0.96<y_{\rm cms}<0.04$ in p--Pb collisions at $\sqrtsNN=5.02~\tev$
        compared with the respective pp reference cross sections~\cite{pp:2019} scaled by the Pb mass number ($A=208$) and corrected for the rapidity shift. 
        For the $\Dzero$ meson, the results in the range $0<\pt<1~\gev/c$ are obtained from the analysis that was performed without decay-vertex reconstruction,
        while those in the range \mbox{$1<\pt<36~\gev/c$} are taken from the analysis with decay-vertex reconstruction.
        The vertical bars and the empty boxes represent the statistical and systematic uncertainties.
        }
        \end{center}
    \end{figure}

Considering the statistical and systematic uncertainties obtained in the two analyses, the most precise measurement of the prompt $\Dzero$ production cross section is obtained
using the results from the analysis without decay-vertex reconstruction in the interval  $0<\pt<1~\gev/c$ and
the analysis with decay-vertex reconstruction for $\pt>1~\gev/c$.
The resulting cross section is shown in the top-left panel of Fig.~\ref{fig:dsdptppref}.

The total cross section for prompt $\Dzero$-meson production per unit of rapidity in $-0.96<y_{\rm cms}<0.04$ was also calculated by integrating the $\pt$-differential measurement reported in the top-left panel of Fig.~\ref{fig:dsdptppref}, obtained combining the methods with and without decay-vertex reconstruction.
The systematic uncertainties were propagated as described above for the total cross section of inclusive $\Dzero$ mesons.
The resulting value is
\begin{equation}
{\rm d}\sigma^{\rm prompt\,D^0}_{\rm p-Pb,\,5.02\,TeV}/{\rm d}y=88.5\pm 2.7\,({\rm stat.}) ^{+5.3}_{-6.1}\,({\rm syst.})\pm 3.3\,({\rm lumi.})\pm 0.9\,({\rm BR})~{\rm mb}.
\label{eq:sigD0pPb}
\end{equation}

In Ref.~\cite{Adam:2016ich}, the ${\rm c\overline{c}}$ production cross section  in the rapidity interval $-0.96<y_{\rm cms}<0.04$ was reported. This calculation used the fraction of charm quarks hadronising into $\Dzero$ mesons to be  $f$(c $\rightarrow \Dzero$) = $0.542\pm 0.024$ which was derived in Ref.~\cite{Gladilin:2014tba} by averaging the measurements in $e^+e^-$ collisions at LEP. Recent measurements of the $\Lambda_c$-baryon production cross section in pp collisions at $\sqrt{s}= 7$ TeV and in p--Pb collisions at $\sqrt{s_{\rm NN}}= 5.02$ TeV~\cite{Acharya2018,Aaij:2018iyy} suggest that the fragmentation fractions of charm quarks into charmed baryons in pp collisions at LHC energies might differ significantly from the LEP results. Therefore, more precise measurements of charmed-baryon production cross sections are needed for an accurate calculation of the charm production cross section.

The average transverse momentum $\meanpt$ of prompt $\Dzero$ mesons was obtained by fitting the cross section reported in the top-left panel of Fig.~\ref{fig:dsdptppref} with a power-law function $f(\pt)=C\, \pt / (1+(\pt/p_0)^2)^n$, 
where $C$, $p_0$ and $n$ are the free parameters.
The result is:
\begin{equation}
\meanpt^{\rm prompt\,D^0}_{\rm p-Pb,\,5.02\,TeV} = 2.07 \pm 0.02\,({\rm stat.})\, \pm 0.04\,({\rm syst.})~\gev/c\,.
\end{equation}
where the systematic uncertainties were estimated with the procedure described in Ref.~\cite{Adam:2016ich}.
The result is compatible within statistical uncertainties with the one obtained in pp collisions at the same centre-of-mass energy: $\meanpt^{\rm prompt\,D^0}_{\rm pp,\,5.02\,TeV} = 2.06 \pm 0.03\,({\rm stat.})\, \pm 0.03\,({\rm syst.})~\gev/c$~\cite{pp:2019}. 

The $\pt$-differential cross sections for the other three D-meson species ($\Dplus$, $\Dstar$, and $\Ds$) are shown in the other panels of Fig.~\ref{fig:dsdptppref}. The cross sections measured in p--Pb collisions are compatible with the measurements published using the 2013 p--Pb data sample  
~\cite{Abelev:2014hha,Adam:2016mkz}, while having a factor 1.5--2 smaller statistical and systematic uncertainties and an extended $\pt$ reach.
The cross sections in p--Pb collisions are compared with the corresponding pp reference cross sections  at the same centre-of-mass energy~\cite{pp:2019} and rapidity interval.

\subsection{The $\pt$-differential nuclear modification factor}

The nuclear modification factor is computed as:
\begin{equation}
\RpPb = \frac{1}{A}\,\frac{{\rm d^2}\sigma_{\rm p-Pb}^{\rm prompt\,D}/{\rm d}\pt{\rm d}y}{{\rm d^2}\sigma^{\rm prompt\,D}_{\rm pp}/{\rm d}\pt{\rm d}y},
\label{eq:RpPb}
\end{equation}

where ${\rm d^2}\sigma_{\rm p-Pb}^{\rm prompt\,D}/{\rm d}\pt{\rm d}y$ is the  D-meson $\pt$-differential cross section in $-0.96<y_{\rm cms}<0.04$ in \pPb~collisions at $\sqrtsNN=5.02~\tev$, $A$ is the mass number of the Pb nucleus and ${\rm d^2}\sigma^{\rm prompt\,D}_{\rm pp}/{\rm d}\pt{\rm d}y$ is the cross section in pp collisions at the same centre-of-mass energy from ~\cite{pp:2019} corrected for the rapidity shift in p--Pb collisions. %using FONLL calculations.
The systematic uncertainties of the p--Pb and pp measurements were considered to be independent and were propagated quadratically, with the exception for the BR uncertainty, which cancels out in the ratio, and the uncertainty on the feed-down correction, which was recalculated for the ratio of cross sections by consistently varying the FONLL calculation parameters in the numerator and the denominator.

\begin{figure}[tbhp]
        \begin{center}
            \includegraphics[width=0.48\textwidth]{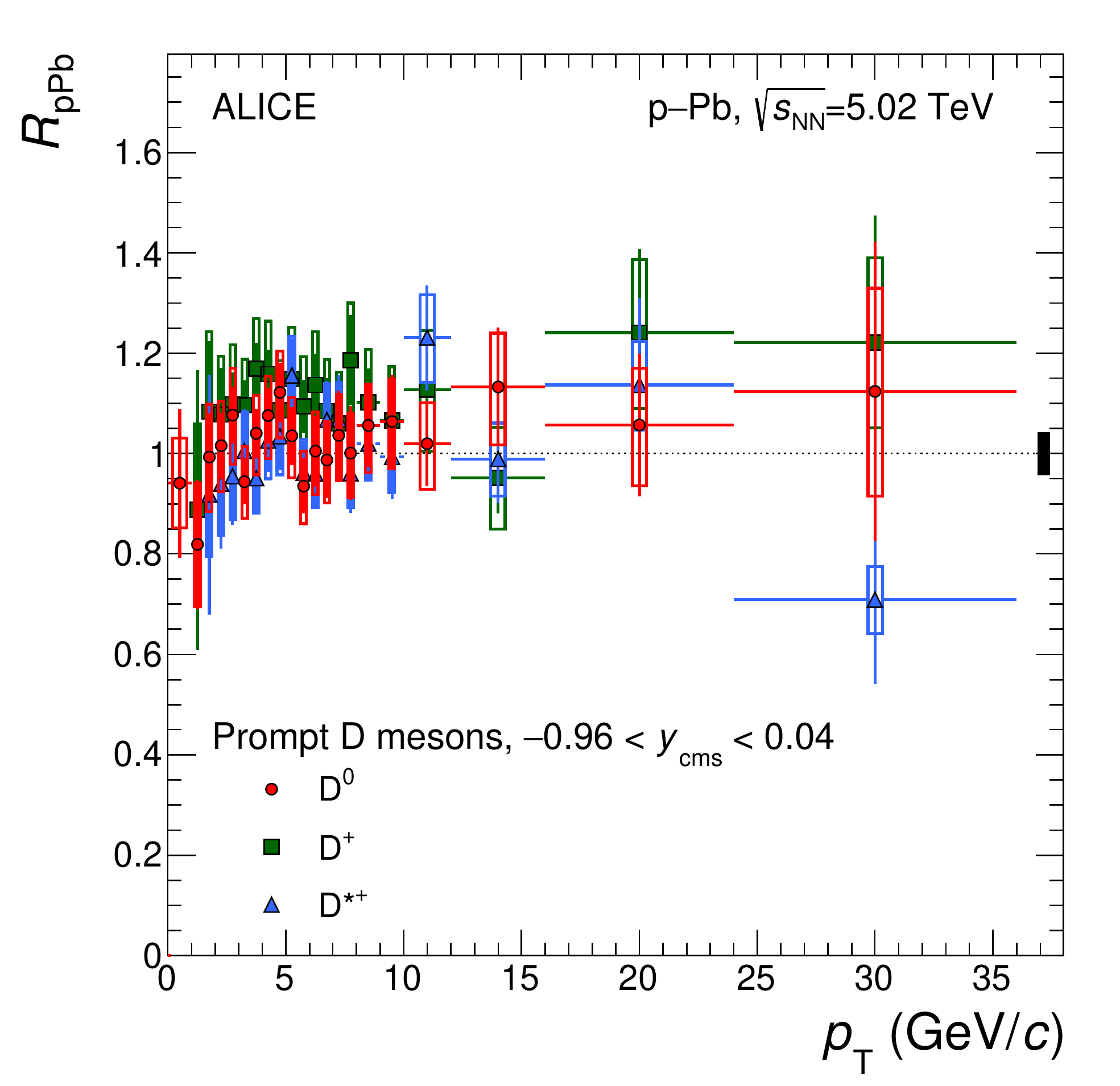}
            \includegraphics[width=0.48\textwidth]{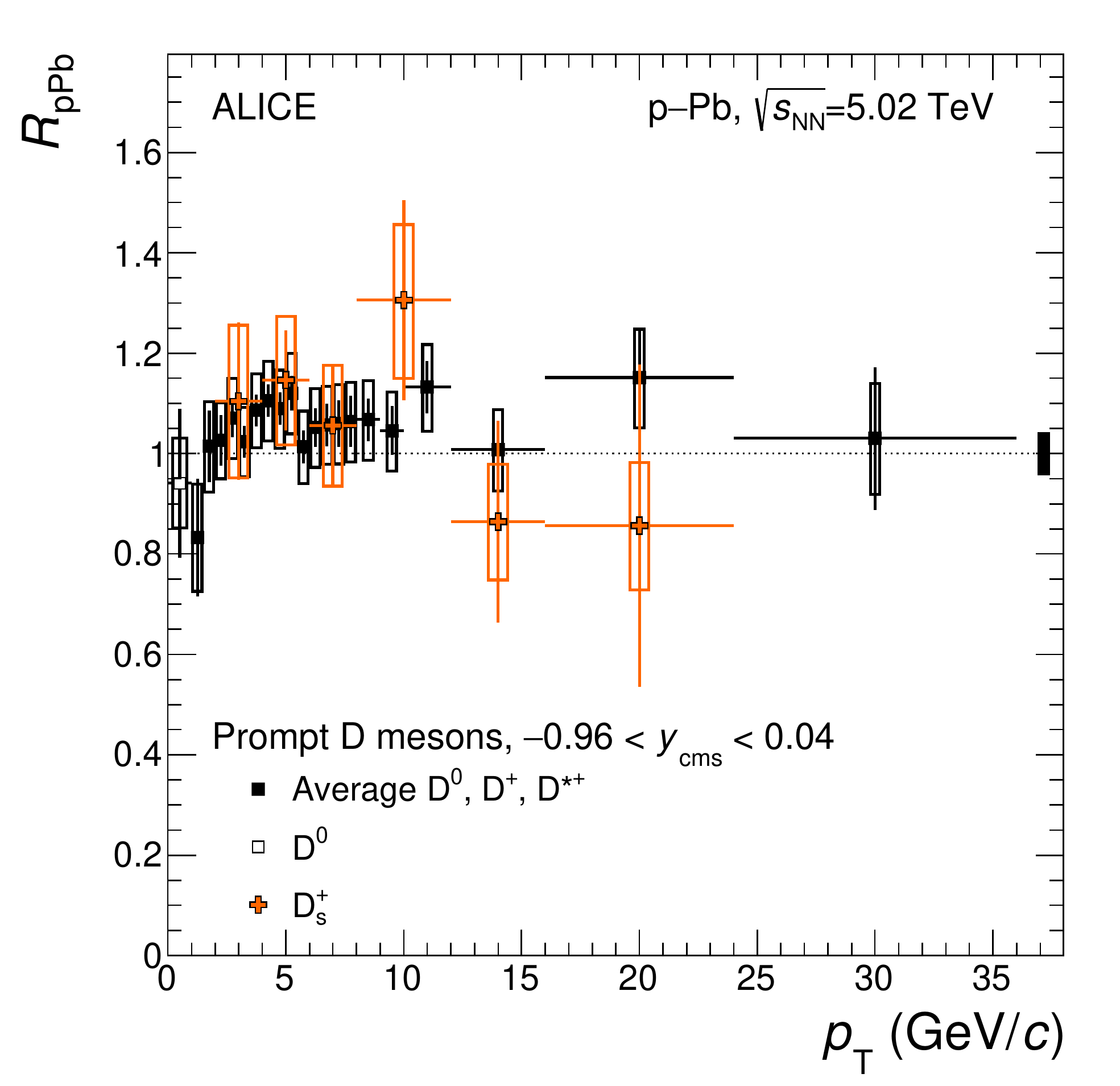}
        \caption{    \label{fig:RpAallmesons}
        Nuclear modification factors $\RpPb$ of prompt D mesons in p--Pb collisions at $\sqrtsNN=5.02~\tev$.
        Left: $\RpPb$ of $\Dzero$, $\Dplus$, and $\Dstar$ mesons. Right:  average $\RpPb$ of non-strange D-meson species in the interval $1<\pt<36~\gev/c$~\cite{Abelev:2014hha}, shown together with the $\Dzero$ $R_{\rm pPb}$ in $0<\pt<1~\GeV/c$ and the $\RpPb$ of $\Ds$ mesons in the interval $2<\pt<24~\gev/c$.
        The vertical bars and the empty boxes represent the statistical and systematic uncertainties.
        The black-filled box at $\RpPb=1$ represents the normalisation uncertainty. 
        }
       \end{center}
    \end{figure}

Figure~\ref{fig:RpAallmesons} shows the nuclear modification factors $\RpPb$ of prompt $\Dzero$, $\Dplus$, and $\Dstar$ mesons in the left panel and their average, along with the $\RpPb$ of $\Ds$ mesons, in the right panel.

With the current uncertainties it is not possible to disentangle a possible mass dependent effect originating from a collective expansion of the system that would modify the $\Dstar$ spectrum differently with respect to the $\Dzero$ and $\Dplus$ spectra. Therefore, the average of the nuclear modification factors of the three non-strange D-meson species is considered and it was calculated
using the inverse of the relative statistical uncertainties as weights. The systematic uncertainty of the average was calculated by propagating the uncertainties through the weighted average, while considering the contributions from tracking efficiency and beauty feed-down correction as fully correlated among the three species.
The D-meson $\RpPb$ is compatible with unity over the entire measured $\pt$ interval within 2 standard deviations. 
The $\RpPb$ of strange and non-strange D mesons are compatible among each other within statistical and systematic uncertainties.

\begin{figure}[!tbhp]
\begin{center}
\includegraphics[width=0.48\textwidth]{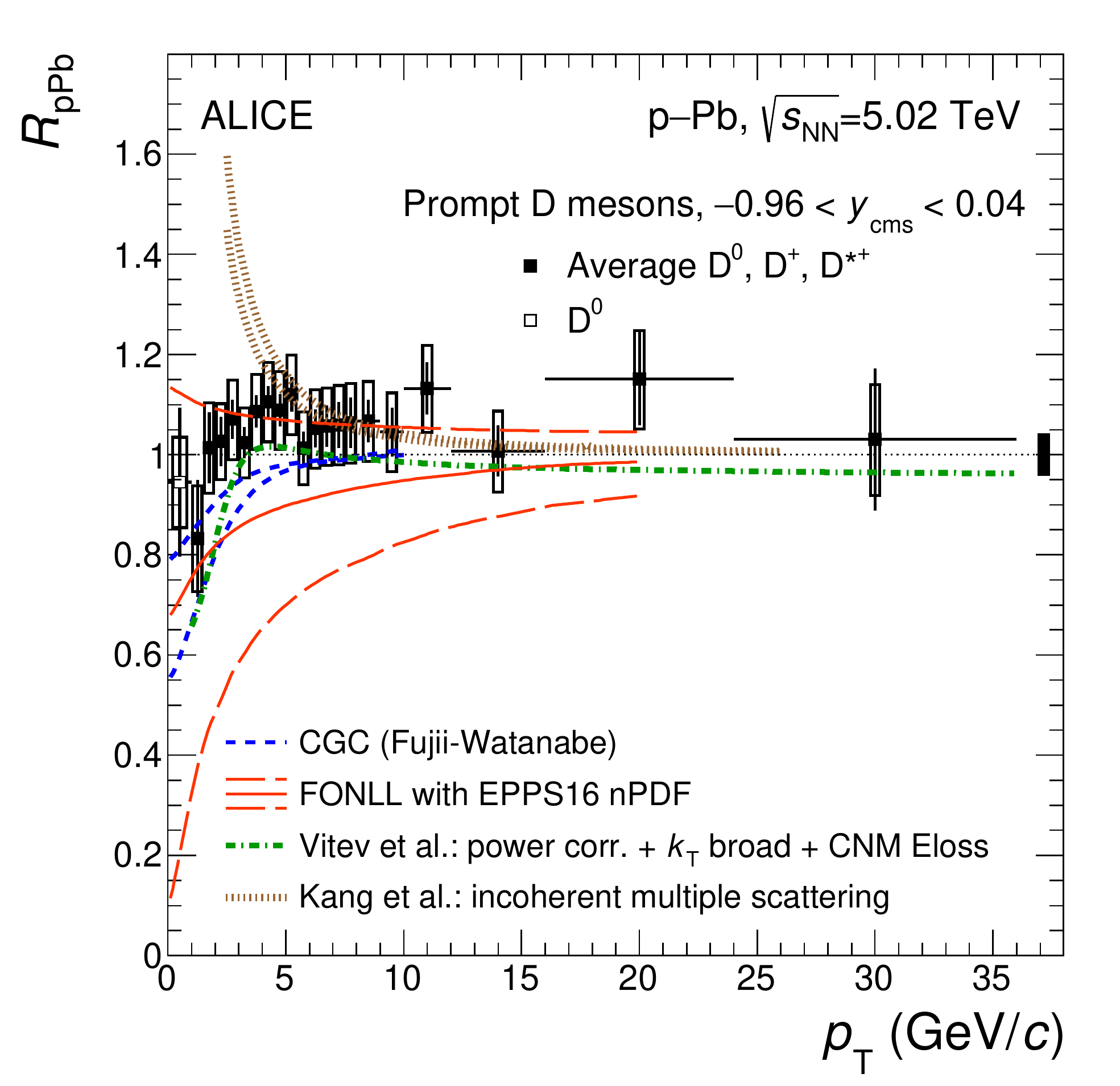}
\includegraphics[width=0.48\textwidth]{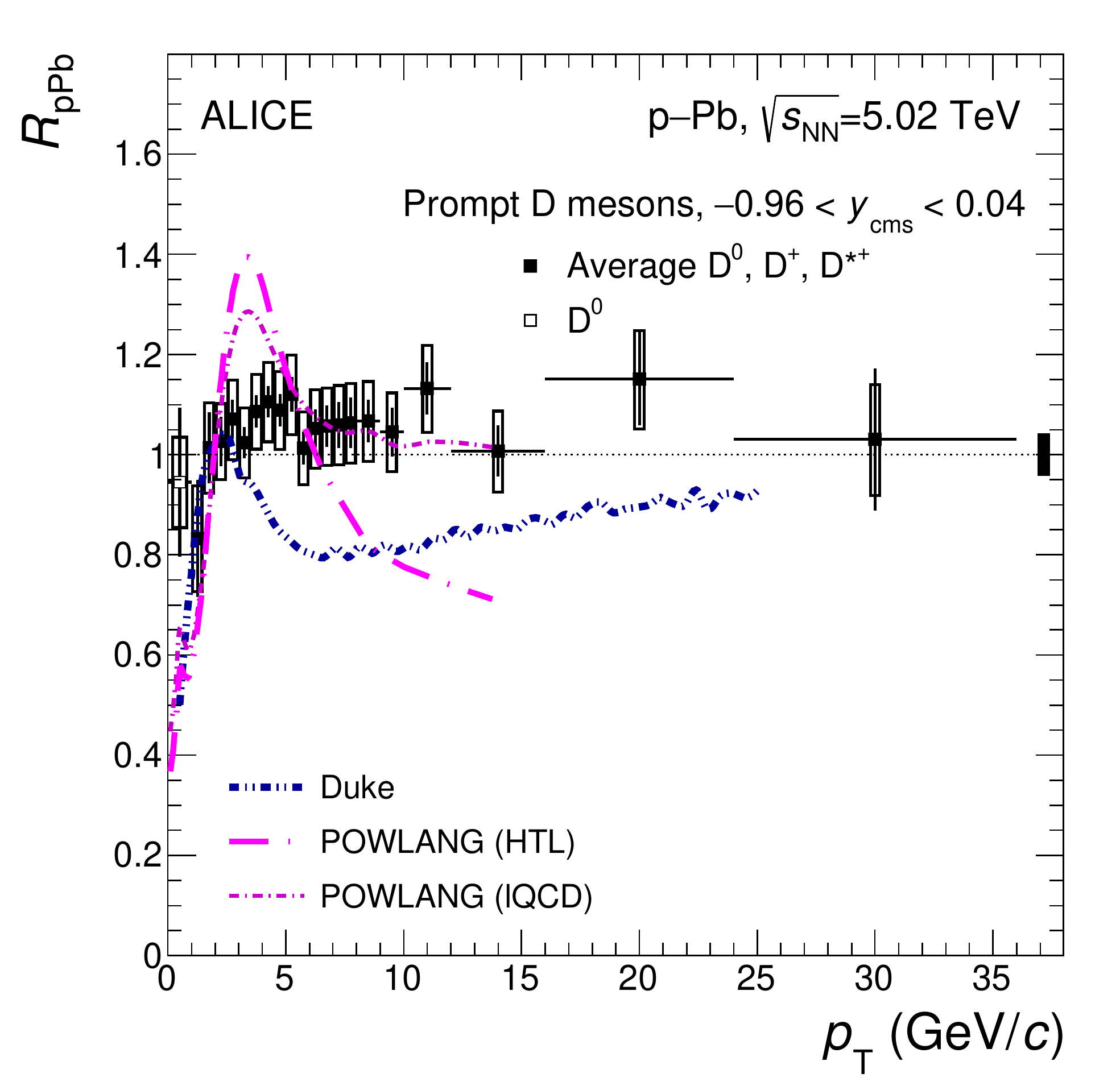}
\caption{\label{fig:RpPbVsPtModels}
Nuclear modification factor $\RpPb$ of prompt non-strange D mesons in p--Pb collisions at $\sqrtsNN=5.02~\tev$. In the left panel, the data are compared with calculations of theoretical models that include only CNM effects: CGC~\cite{Fujii:2017rqa}, FONLL~\cite{Cacciari:2012ny} with EPPS16 nPDFs~\cite{Eskola:2016oht}, a LO pQCD calculation (Vitev et al.)~\cite{Sharma:2009hn}, and a calculation based on incoherent multiple scatterings (Kang et al.)~\cite{Kang:2014hha}. In the right panel, the predictions of the Duke~\cite{Xu:2015iha} and POWLANG~\cite{Beraudo:2015wsd} transport models are compared with the measured D-meson $\RpPb$.
The vertical bars and the empty boxes represent the statistical and systematic uncertainties.
The black-filled box at $\RpPb=1$ represents the normalisation uncertainty. 
}
\end{center}
\end{figure}

The D-meson nuclear modification factor is also compared with theoretical calculations, shown in Fig.~\ref{fig:RpPbVsPtModels}.
In the left panel, four theoretical calculations that include only CNM effects are displayed.
A calculation based on the Colour Glass Condensate
formalism~\cite{Fujii:2013yja,Fujii:2017rqa} describes the data within a 2$\sigma$ uncertainty in the entire $\pt$ range, although the model underestimates systematically the measured points at low $\pt$ ($\pt<6$ GeV/$c$). A FONLL calculation  with CTEQ6M PDFs~\cite{Pumplin:2002vw} and EPPS16 NLO nuclear modification ~\cite{Eskola:2016oht} is compatible with the data within the uncertainties. The measurement lies on the upper limit of the EPPS16 nPDF uncertainty band, while this is not the case for the D-meson $\RpPb$ at forward rapidity measured by LHCb ~\cite{Aaij:2017gcy,Kusina:2017gkz}. The data are also described within the uncertainties by a LO pQCD calculation with intrinsic $k_{\rm T}$ broadening, nuclear shadowing, and energy loss of the charm quarks in cold nuclear matter (Vitev et al.)~\cite{Sharma:2009hn}. 
The calculation by Kang et al., that consists of a higher-twist calculation based on incoherent multiple scatterings, has a different trend with respect to the other models and it is excluded by the data for $\pt<4~\gev/c$.

In the right panel of Fig.~\ref{fig:RpPbVsPtModels}, the measurements are compared with the calculations of two transport models, Duke~\cite{Xu:2015iha} and POWLANG~\cite{Beraudo:2015wsd}, both of which assume that a QGP is formed in p--Pb collisions. These models are both based on the Langevin approach for the transport of heavy quarks through an expanding deconfined medium described by relativistic viscous hydrodynamics.  The Duke model includes both collisional and radiative energy loss.
The POWLANG model considers only collisional processes with two choices for the transport coefficients, based on hard-thermal-loop (HTL) and
lattice-QCD (lQCD) calculations. For both the Duke and the HTL based POWLANG estimates, the D-meson nuclear modification factor distribution has a peak structure, with the 
maximum at $\pt \approx 2.5~\GeV/c$ and $\pt \approx 3.5~\GeV/c$, respectively, possibly followed by a moderate ($<20$--30\%) suppression at higher $\pt$, resulting from the interplay of CNM effects and interactions of charm quarks with the radially expanding medium.
The trend suggested by these models is not supported by the data. The strong enhancement at $\pt\sim3-4$ GeV/$c$ observed in the model calculations is not consistent with the measured $\RpPb$, and a suppression larger than 10\% for $\pt > 8$ GeV/$c$ is excluded by the data with a 98\% confidence level.

The $\pt$-integrated nuclear modification factor of prompt
$\Dzero$ mesons in $-0.96<y_{\rm cms}<0.04$ was obtained from Eq. \ref{eq:RpPb} by integrating the $\pt$-differential cross sections in pp  and p--Pb collisions.
The result is
\begin{equation}
R_{\rm pPb}^{\rm prompt\,D^0}(\pt>0,\,-0.96<y_{\rm cms}<0.04) = 0.96 \pm 0.05\,({\rm stat.})\,^{+0.07}_{-0.07}({\rm syst.})\,
\end{equation}
and it is consistent with the atomic mass number scaling of the total charm cross section.

\subsection{The $\pt$ and centrality-dependent nuclear modification factor}
\label{subsec:QpPb}

The measurement of the nuclear modification factor was also computed in various centrality intervals, where the centrality is defined using the energy deposited by neutrons in the ZDC positioned in the Pb-going side (ZN energy), as described in Section \ref{sec:detector}. For each centrality class the nuclear modification factor, $\QpPb$, is defined as
\begin{equation}
\QpPb = \frac{ ({\rm d}^2 N^{\rm prompt \, D}/ {\rm d} \pt  {\rm d} y)^{\mathrm{i}}_{\mathrm{p-Pb} } }
{ \langle \TpPb \rangle_{\rm i} \, \times \, (\mathrm{d}^2 \sigma^{\rm prompt \, D}_{\mathrm{pp}} / \mathrm{d} \pt  {\rm d} y)} \,,
\label{eq:QpPb}
\end{equation}
where $({\rm d}^2 N^{\rm prompt \, D}/ {\rm d} \pt  {\rm d} y)^{\mathrm{i}}_{\mathrm{p-Pb}}$ is the yield of prompt D mesons in \pPb~collisions and $\langle \TpPb \rangle_{\rm i}$ is the average nuclear overlap function in a given centrality class.

The $\langle \TpPb \rangle_{\rm i}$ is estimated with the hybrid approach described in Ref.~\cite{Adam:2014qja} and is based on the assumption that the charged-particle multiplicity measured at mid-rapidity ($-1<\eta_{\rm cms}<0$) scales with the number of participant nucleons, $\Npart$. The average nuclear overlap function is defined as $\langle \TpPb \rangle_{\rm i} = \frac{\langle \Ncoll \rangle_{\rm i}}{\sigma_{\rm NN}}\,$ where $\sigma_{\rm NN}=(67.6\pm0.6)~\mb$ is the interpolated inelastic nucleon--nucleon cross section at $\sqrtsNN=5.02~\tev$~\cite{Loizides:2295119} and $\langle \Ncoll \rangle_{\rm i}$ is the average number of binary nucleon--nucleon collisions in a given centrality class. The latter is obtained as

\begin{equation}
\langle \Ncoll \rangle_{\rm i} =  \langle \Npart \rangle_{\rm i} - 1 =  \langle \Npart^{\rm MB} \rangle \cdot  \bigg (\frac{\langle \dNdEta \rangle_{\rm i}}{\langle \dNdEta \rangle^{\rm MB}}\bigg )_{-1<\eta<0} - 1\,,
\end{equation}

where $\langle \Npart^{\rm MB} \rangle=7.7$ ~\cite{TpACDS} is the average number of participants in minimum bias collisions.  
The values of $\langle \TpPb \rangle$ used for the analyses are reported in Table \ref{tab:TpA} \cite{TpACDS}. 

\begin{table}[t!]

\centering
\caption{ $\langle \TpPb \rangle$ and relative uncertainties for each centrality class considered in the analysis.}
\begin{tabular}{l|c|c|c|c|c}

Centrality classes & 0--10\%&10--20\%&20--40\%& 40--60\% & 60--100\%\\
\hline
$\langle \TpPb \rangle$ (1/mb) &0.172&0.158&0.137& 0.102& 0.046\\
\hline
Rel. unc. &6.9\%& 3.7\%& 1.7\% &4.8\% &5.2\%\\
\hline
\end{tabular}

\label{tab:TpA}
\end{table}

\begin{figure}[!tbhp]
\begin{center}
  \includegraphics[width=0.95\columnwidth]{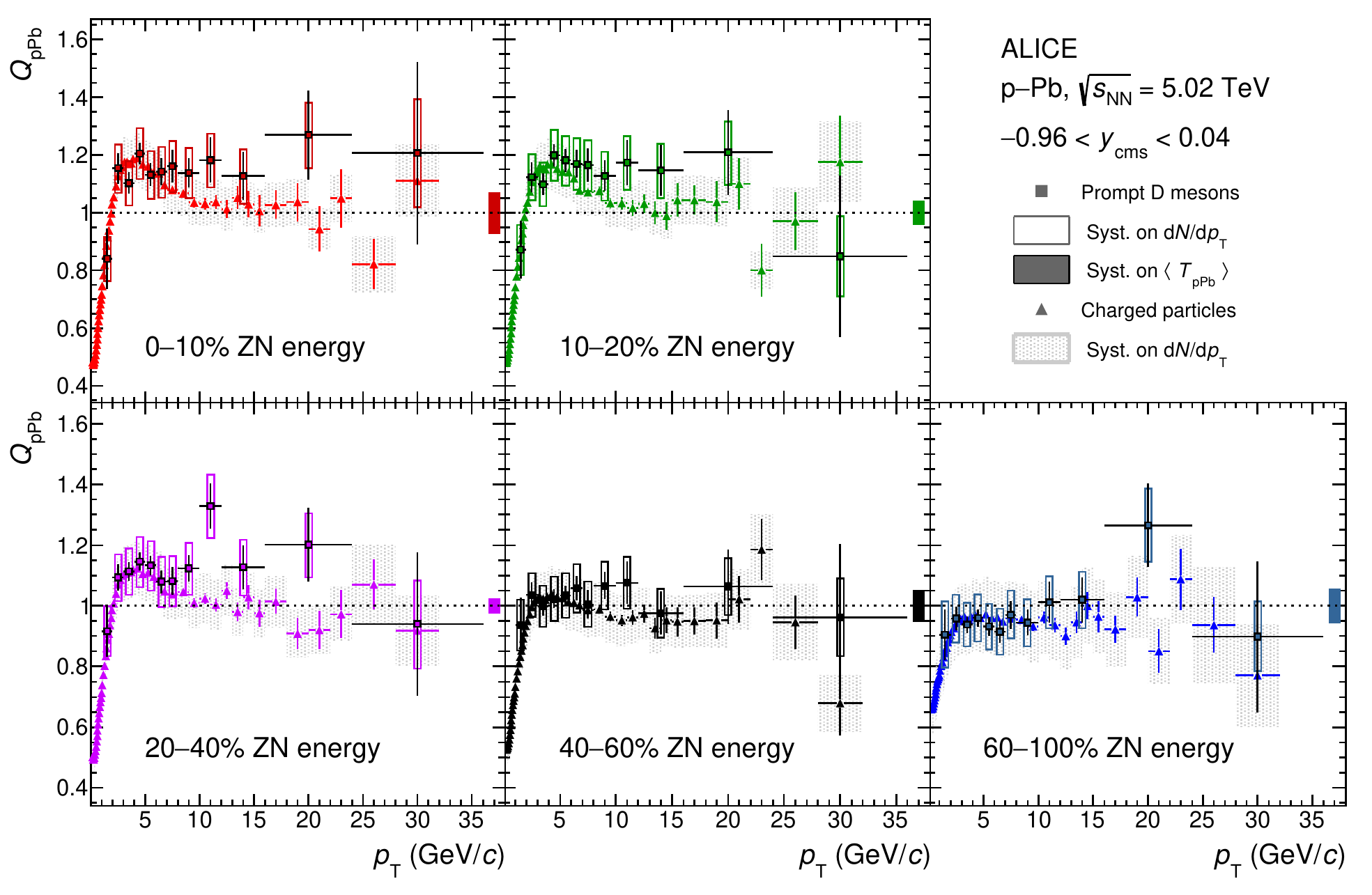}

\caption{
Nuclear modification factors of prompt D mesons as a function of $\pt$ in 0--10\%,  10--20\%,  20--40\%,  40--60\%, and 60--100\% centrality classes compared with those of charged particles~\cite{Adam:2014qja}.
The vertical bars represent the statistical uncertainties while the empty boxes and the shaded boxes represent the systematic uncertainties for the prompt D mesons and for the charged particles. The colour-filled boxes at $\QpPb$ $=1$ represent the normalisation uncertainties on the $\langle \TpPb \rangle$~\cite{TpACDS}. \label{fig:DAverageQpPb}
}
\end{center}
\end{figure}

The average of prompt $\Dzero$, $\Dplus$, and $\Dstar$ meson $\QpPb$ was calculated as a function of $\pt$ in the interval $1<\pt<36~\gev/c$ in  0--10\%,  10--20\%,  20--40\%,  40--60\%, and 60--100\% centrality classes, and is shown in Fig.~\ref{fig:DAverageQpPb}. The D-meson $\QpPb$ measurement shows a hint of suppression in the interval $1<\pt<2$ GeV/$c$. The observed suppression is strongest in the most central collisions. This is qualitatively expected from a stronger shadowing at low Bjorken-$x$ in central collisions. There is also a hint of enhancement at  $2<\pt<6$ GeV/$c$ in the most central classes (0--40\% centrality).
The results are also compared with the charged-particle $\QpPb$~\cite{Adam:2014qja} \footnote{The $\langle \TpPb \rangle$ values used to compute the charged-particles $\QpPb$ were updated with respect to ~\cite{Adam:2014qja} according to the values in \cite{TpACDS}.} in each centrality class. A similar trend is observed for prompt D mesons and charged particles in each centrality class. 

\begin{figure}[!h]
 \begin{center}
 \includegraphics[width=0.48\textwidth]{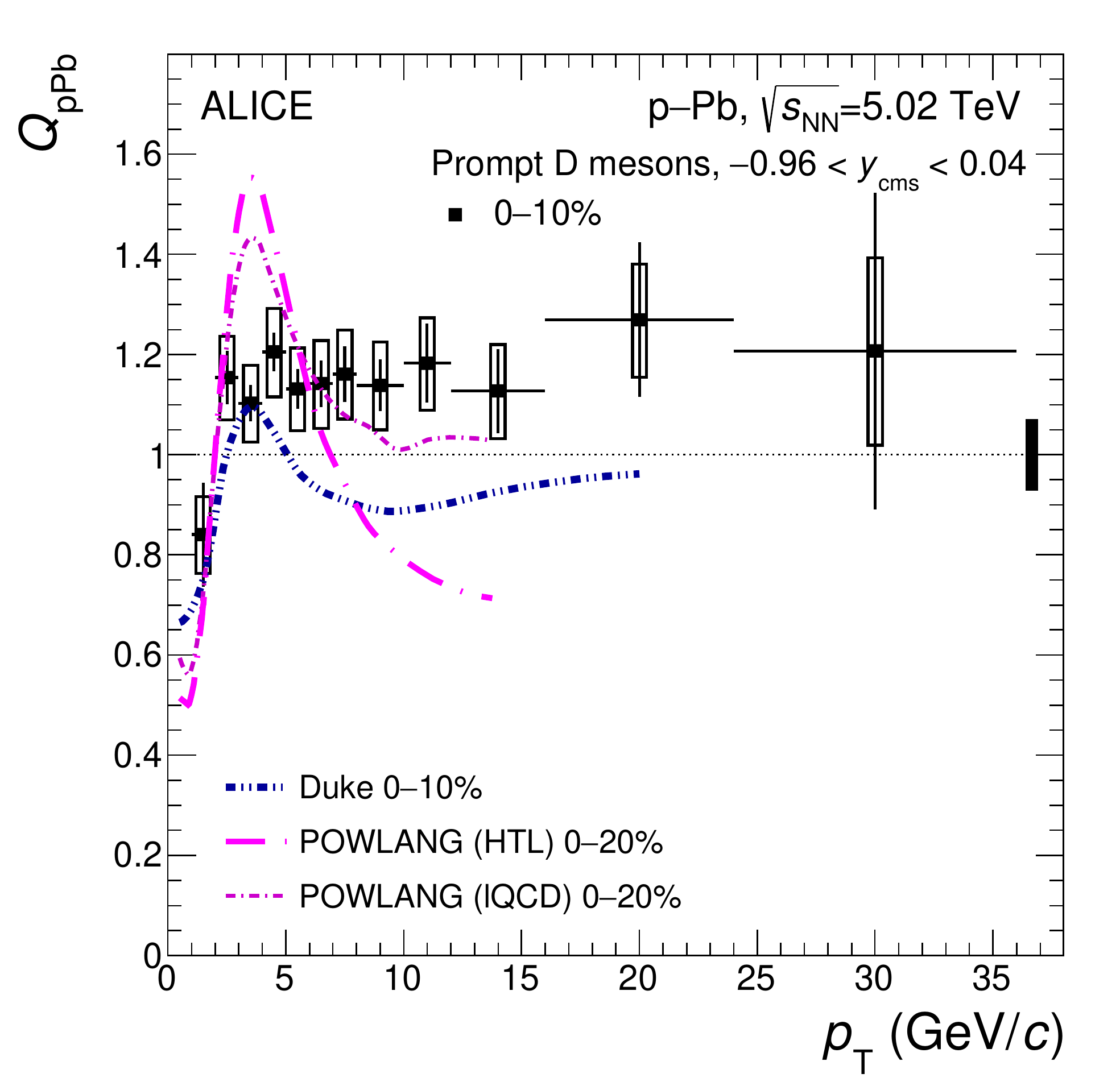}
 \caption{D-meson $\QpPb$ measured in the 0--10\% centrality class compared with the predictions of the Duke~\cite{Xu:2015iha} and POWLANG~\cite{Beraudo:2015wsd} transport models. The vertical bars and the empty boxes represent the statistical and systematic uncertainties. The colour-filled boxes at $\QpPb$ $=1$ represent the normalisation uncertainties.}
\label{fig:QpPbModel}
\end{center}
\end{figure}
The D-meson $\QpPb$ in the 0--10\% centrality class is  compared with the predictions of the Duke~\cite{Xu:2015iha} and POWLANG~\cite{Beraudo:2015wsd} transport models in Fig.~\ref{fig:QpPbModel}.
The POWLANG model predicts a pronounced bump in the nuclear modification factor at intermediate $\pt$ ($3<\pt<6$ GeV/$c$), which is not supported by the data, considering that the systematic uncertainties are mostly correlated among the $\pt$ intervals of the measurement.
At higher $\pt$ ($\pt>7$ GeV/$c$), POWLANG simulations with the HTL transport coefficients and the Duke model predict a suppression of the D-meson yield which is not observed in the data.

\begin{figure}[!h]
\begin{center}
  \includegraphics[width=0.9\columnwidth]{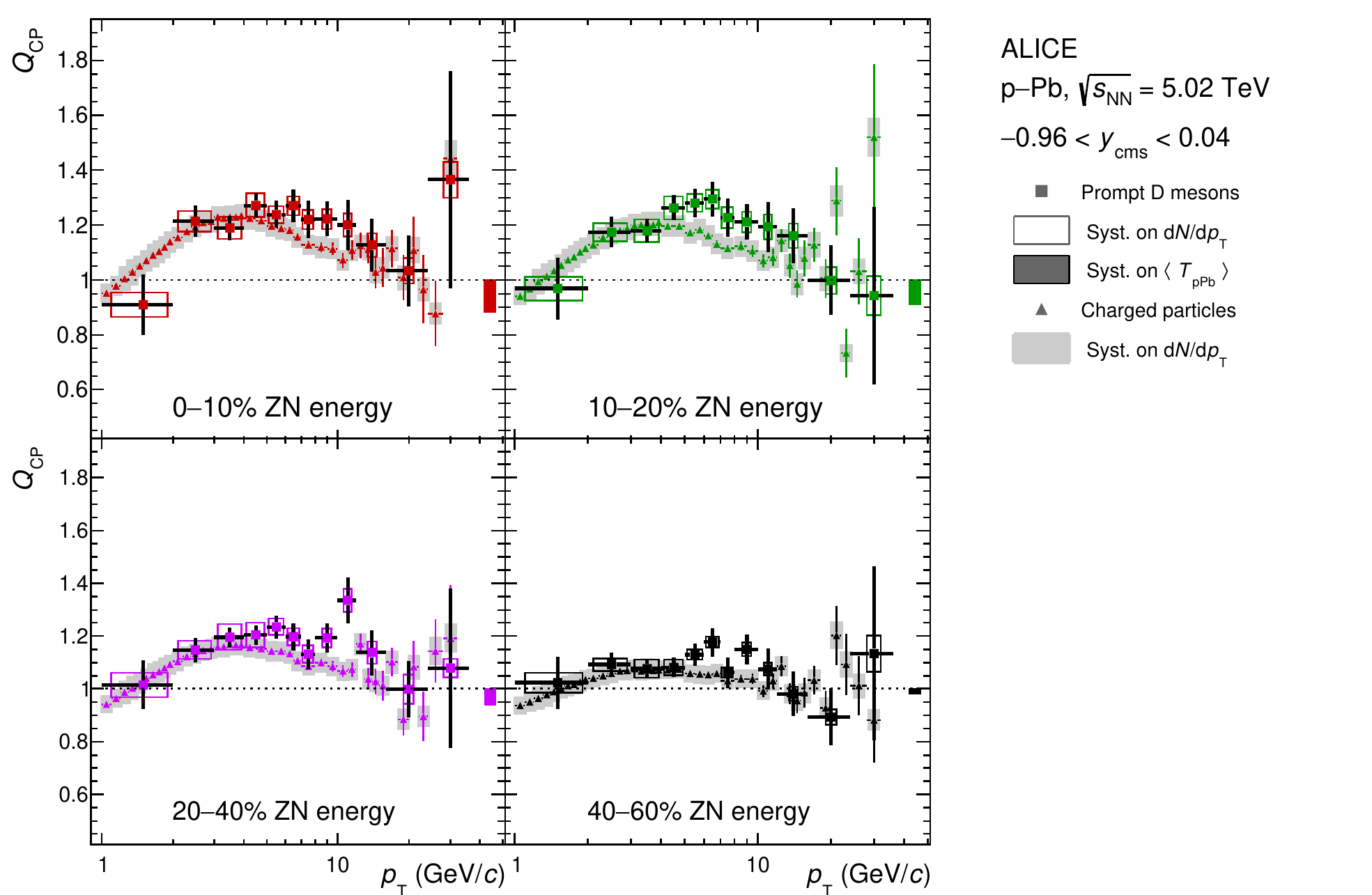}

\caption{
Average D-meson and charged-particles~\cite{Adam:2014qja} $\Qcp$ using the yields measured in 0--10\%,  10--20\%,  20--40\%, and 40--60\% as numerators and the yield in  60--100\% as the denominator. The vertical bars and the empty boxes represent the statistical and systematic uncertainties. The colour-filled boxes at $\Qcp$ $=1$ represent the normalisation uncertainties on the $ \langle \TpPb \rangle$.
\label{fig:DmesonQCP}
}
\end{center}
\end{figure}

The ratio of the D-meson yield in a given centrality class with respect to yield in the most peripheral centrality class (60--100\%), defined as

\begin{equation}
\Qcp = \frac{  ({\rm d}^2 N^{\rm prompt \, D} / {\rm d} \pt  {\rm d} y)^{\mathrm{i}}_{\mathrm{p-Pb} }  \big/ \langle \TpPb \rangle_{\mathrm{i}} }
{   ({\rm d}^2 N^{\rm prompt \, D}/ {\rm d} \pt  {\rm d} y)^{\mathrm{60-100\%}}_{\mathrm{p-Pb} } \big/ \langle \TpPb \rangle_{\mathrm{60-100\%}} },
\label{eq:QCP}
\end{equation}
was also calculated.
The $\Qcp$ observable is independent of the pp cross section and uses the yields in peripheral p--Pb collisions as a reference. Since the contributions from the track reconstruction, selection and PID efficiency cancel out in the ratio, the $\Qcp$ has reduced systematic uncertainties with respect to the $\QpPb$ ratio. 
The systematic uncertainties on the yield extraction were estimated by applying the fit variation procedure described in Section~\ref{sec:Syst}  directly on the signal yield ratio obtained from the invariant-mass distributions of the two centrality classes. 
To estimate the feed-down correction uncertainty, the contributions from the hypothesis on the nuclear modification factor of D mesons from B-hadron decays were considered as uncorrelated in each centrality class and were added in quadrature. \\
In Fig.~\ref{fig:DmesonQCP}, the average D-meson $\Qcp$ is shown for different centrality classes: 0--10\%,  10--20\%,  20--40\% and 40--60\%. The results are superimposed to those obtained for charged particles in the same centrality classes~\cite{Adam:2014qja}. A similar trend is observed for both measurements: when the results from the most central classes are used as the numerator, the $\Qcp$ increases in the interval 1--5~GeV/$c$, reaching values of about 1.3 and then shows a decreasing trend with increasing $\pt$. A $\Qcp>1$ with a significance of 3$\sigma$ is observed in the range $3<\pt<7$ GeV/$c$ when the 20--40\% centrality class is used as numerator. In this case, the normalisation uncertainty is smaller than the one of more central collisions due to the smaller separation between the centrality classes used in the calculation of $\Qcp$. When the  0--10\% and 10--20\% centrality classes are used as numerators,  a $\Qcp>1$ is observed in the same $\pt$ interval, with a significance of 1.5$\sigma$ and 2$\sigma$ due to the larger $ \langle \TpPb \rangle$ uncertainties. 
 A milder $\pt$ dependence is observed when the yields from more peripheral collisions are used as the numerator. 
A possible radial flow arising from a hydrodynamic evolution could modify the hadronisation dynamics of heavy quarks and potentially be the cause of the enhancement at intermediate $\pt$. 

\subsection{D-meson ratios}
\begin{figure}[!h]
 \begin{center}
 \includegraphics[width=\textwidth]{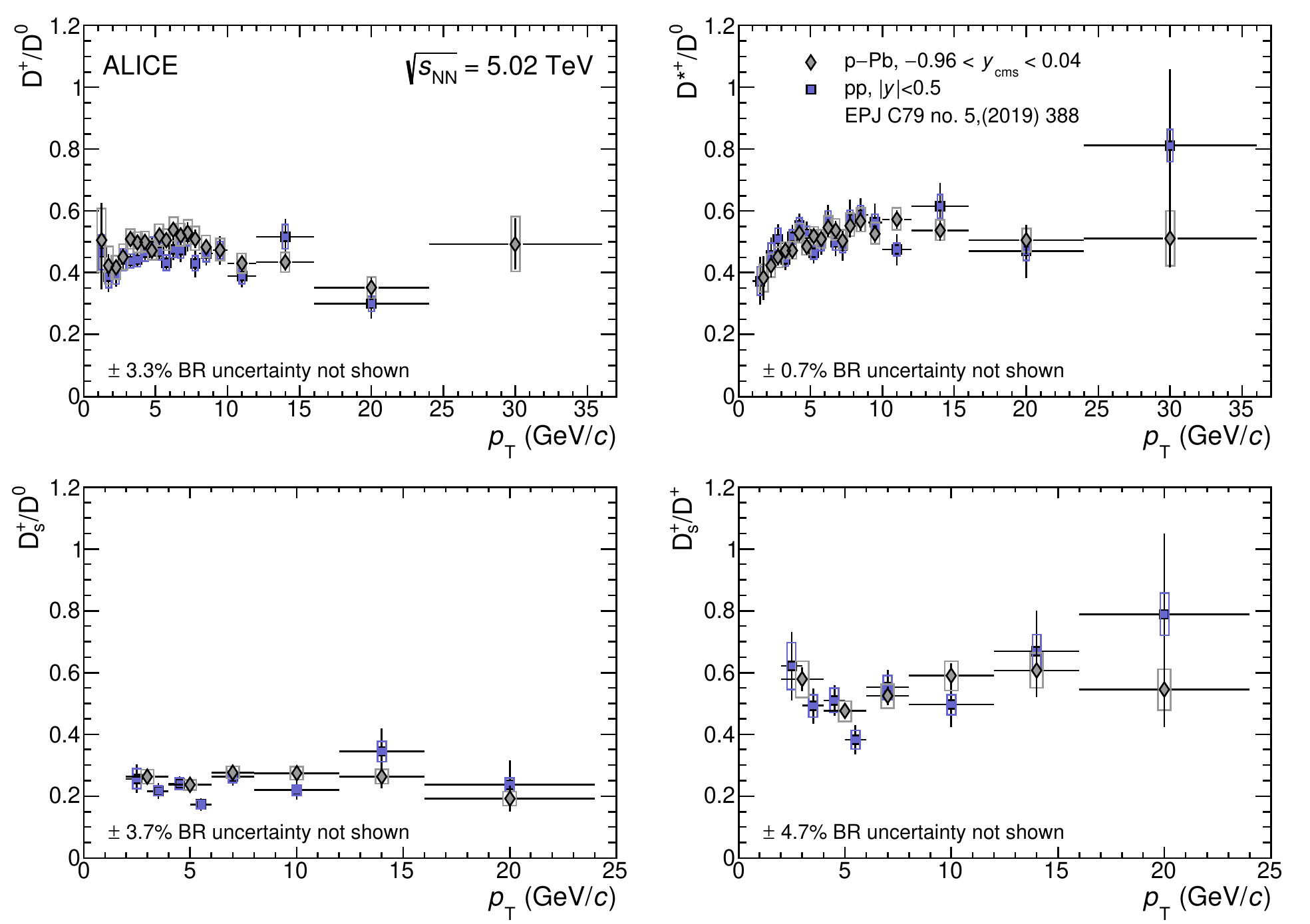} 
 \caption{Ratios of prompt D-meson production cross sections as a function of $\pt$ in p--Pb collisions at \mbox{$\sqrtsNN=5.02~\tev$}.
 The results are compared with those of pp collisions at the same centre-of-mass energy~\cite{pp:2019}. The vertical bars and the empty boxes represent the statistical and systematic uncertainties. }
\label{fig:Dratios}
\end{center}
\end{figure}

The ratios of the $\pt$-differential cross sections of $\Dzero$, $\Dplus$, $\Dstar$, and $\Ds$ mesons in the minimum bias sample are reported in Fig.~\ref{fig:Dratios}. In the evaluation of
the systematic uncertainties of the ratios, the contributions of the yield extraction and selection efficiency were considered as uncorrelated, while those of the feed-down from beauty-hadron decays and the tracking efficiency were treated as fully correlated among the different D-meson species. The measurements are compared to the ratios of D-meson cross sections
in pp collisions at $\sqrt s=5~\tev$~\cite{pp:2019}. The relative abundances of the four species are
unmodified in p--Pb with respect to pp collisions within statistical and systematic uncertainties.\\

The ratios of the $\Ds/\Dplus$-meson yields were also studied in different $\pt$ intervals as a function the multiplicity of charged particles produced in the collision. The charged-particle multiplicity, $\Nch$, was estimated at mid-rapidity by measuring the number of tracklets, $\Ntrk$ as in Refs.~\cite{Abelev:2012rz,Adam:2015ota}. The $\Ds/\Dplus$ ratios were extracted in three multiplicity classes defined as 1--40, 40--70, 70--200  tracklets.
A tracklet is defined as a track segment  that joins the reconstructed primary vertex with a pair of space points on the two SPD layers within the pseudorapidity range $|\eta|<1.0$. 
The measured $\Ntrk$ distribution is affected by the position of the interaction vertex along the beam axis and by the evolution of the detector conditions. The former is due to the collision system asymmetry and the limited SPD rapidity coverage, while the latter is a consequence of a variation in active SPD channels over time. 
To account for these effects, the $\Ntrk$ distributions were corrected offline on an event-by-event basis. 
The correlation between the measured $\Ntrk$ and $\Nch$, equivalent to the number of generated ``physical primaries'', 
was obtained from a Monte Carlo simulation and parametrised with a linear function. Here, physical primaries are defined as prompt particles produced in the collision, along with their decay products, but excluding those from weak decays of strange particles~\cite{ALICE-PUBLIC-2017-005}.
The systematic uncertainty on the conversion from $\Ntrk$ to $\Nch$ was calculated using different Monte Carlo generators and using different parameterisations of the correlation.
The total systematic uncertainty varies from 2\% in the highest multiplicity class to 7\% in the lowest multiplicity class.

\begin{figure}[!h]
 \begin{center}
 \includegraphics[width=\textwidth]{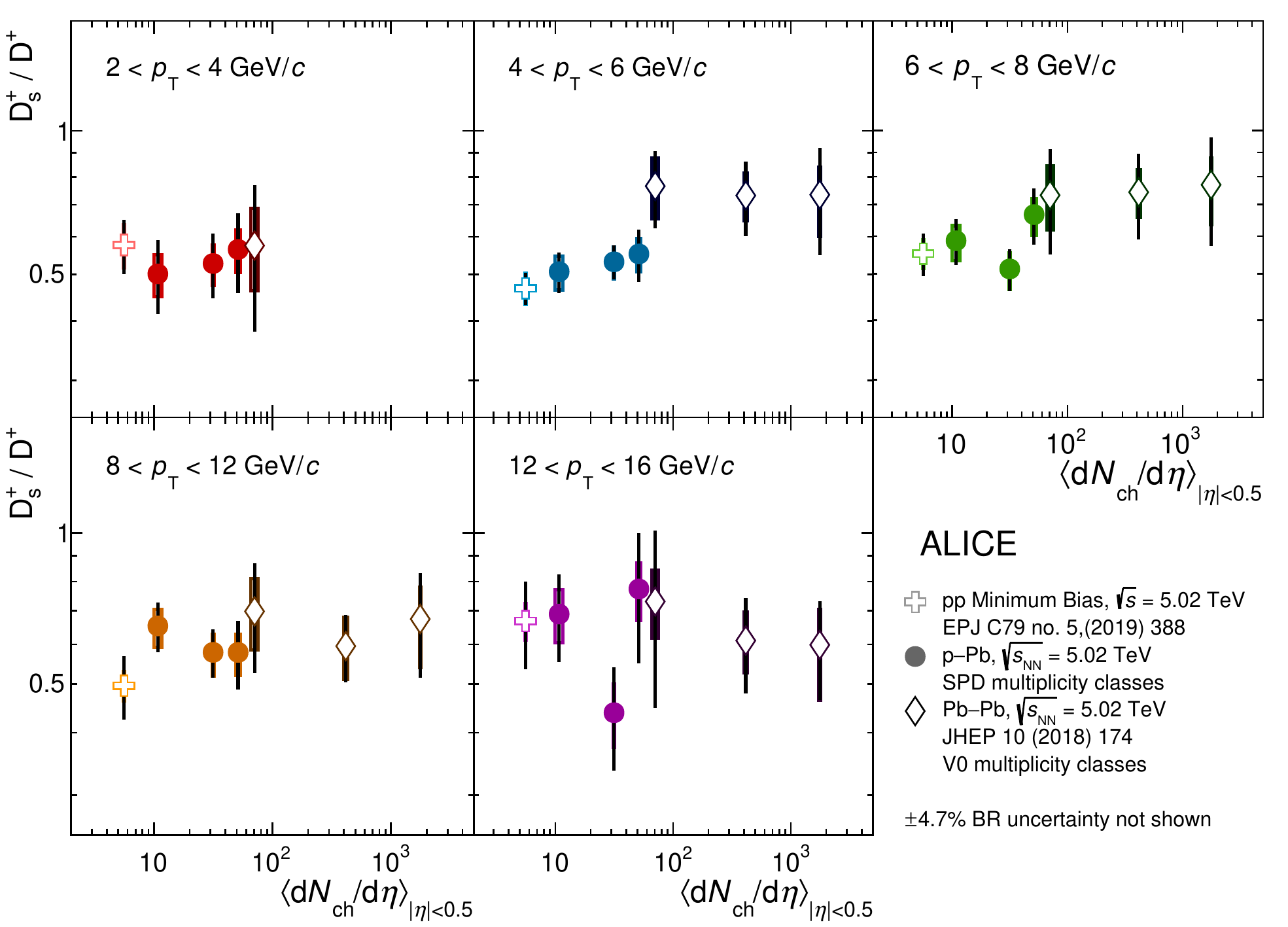}
 \caption{$\Ds/\Dplus$-meson yield ratios, as a function of primary charged particles per unity of pseudorapidity in pp~\cite{pp:2019}, p--Pb, and Pb--Pb~\cite{PbPb:2018} collisions at $\sqrtsNN=5.02 ~\tev$ in five different $\pt$ intervals from 2 to 16 $\gev/c$.}
\label{fig:DsDplusRatios}
\end{center}
\end{figure}

The ratios of the $\Ds/\Dplus$-meson yields are shown in Fig.~\ref{fig:DsDplusRatios}
 as a function of the number of primary charged particles per unity of pseudorapidity (d$\Nch$/d$\eta$$|_{|\eta|<0.5}$)
 in five $\pt$ intervals ranging from 2 to 16 $\gev/c$. As a comparison, the measured ratios in pp collisions~\cite{pp:2019} and in
Pb--Pb collisions~\cite{PbPb:2018} at $\sqrt{s_{\rm NN}}=$ 5.02 TeV are also shown in the figure.
Within uncertainties, there is no indication of a modification of the $\Ds/\Dplus$-yield ratios in pp and p--Pb collisions, up to the highest multiplicities that could be studied with the current p--Pb sample, which are similar to those of peripheral Pb--Pb collisions (60--80\% centrality class). A hint of an enhancement of the $\Ds/\Dplus$-yield ratios in Pb--Pb collisions in $4<\pt <8$ GeV/$c$ is observed, as already shown in~\cite{PbPb:2018}.  The larger data sample of Pb--Pb collisions collected by ALICE in 2018 will provide a more precise measurement.

%%%----------------------------------------------------------
\section{Summary}
\label{sec:summary}

The production cross sections of the prompt charmed mesons ($\Dzero$, $\Dplus$, $\Dstar$, and $\Ds$) in \pPb~collisions at a centre-of-mass energy per nucleon pair of $\sqrtsNN=5.02~\tev$ were measured as a function of $\pt$ in the rapidity interval $-0.96<y_{\rm cms}<0.04$ 
with luminosity of $\Lint= 292 \pm 11~\mu$b$^{-1}$. 
The $\pt$-differential production cross sections were reported in the transverse momentum range $0<\pt<36~\gev/c$ for $\Dzero$ mesons, $1<\pt<36~\gev/c$ for $\Dplus$ mesons, $1.5<\pt<36~\gev/c$ for $\Dstar$ mesons, and $2<\pt<24~\gev/c$ for $\Ds$ mesons. The larger sample used for this analysis, with respect to that collected in 2013, allowed for a significant reduction, by a factor 1.5--2, of the statistical and systematic uncertainties, along with an extension of the $\pt$ reach.

The $\pt$-differential nuclear modification factor $\RpPb$ of D mesons, calculated by using the pp reference measured at the same centre-of-mass energy, was found to be compatible with unity for $0<\pt<36~\gev/c$. 
The $\RpPb$ results are described within uncertainties by theoretical calculations that include initial-state effects. 
The $\RpPb$ is also compared with parton-transport model based calculations that assume 
the formation of a deconfined QCD medium in \pPb~collisions. The trend predicted by these models is not supported by the data. The strong enhancement at $\pt\sim$3--4 GeV/$c$ observed in the calculations is not consistent with the measured $\RpPb$, and a suppression larger than 10\% for $\pt > 8$ GeV/$c$ is excluded by the data at 98\% confidence level.

The centrality dependence of the D-meson yields was also studied in different centrality classes, from most central to peripheral collisions, in the interval  $1<\pt<36~\gev/c$. 
 The average $\QpPb$ of prompt $\Dzero$, $\Dplus$, and $\Dstar$ mesons is consistent with unity within the uncertainties for $\pt>2$ GeV/$c$. The measurements show a hint of suppression in $1<\pt<2$ GeV/$c$ stronger in the most central collisions with respect to the peripheral ones, as qualitatively expected from a stronger shadowing at low Bjorken-$x$ in central collisions~\cite{Eskola:1991ec,Helenius:2012wd}. There is also a hint of enhancement in the intermediate $\pt$ region in the most central collision classes (0--40\% centrality). The same trend is observed for the charged-particles $\QpPb$.
The average D-meson $\Qcp$  has been computed. For the most central collision classes, the $\Qcp$ increases in the $\pt$ interval 1--5~GeV/$c$, reaching values of about 1.3. Above a $\pt$ of 5 GeV/$c$ the distribution tends to decrease with increasing $\pt$. A milder $\pt$ dependence is observed for more peripheral collisions. A similar trend is observed for both charmed mesons and charged particles in all the centrality classes considered. A possible radial flow arising from hydrodynamic evolution could modify the hadronisation dynamics of heavy quarks and give rise to the enhancement measured in the intermediate $\pt$ interval~\cite{Xu:2015iha,Beraudo:2015wsd}.

The ratios of the $\pt$-differential cross sections of $\Dzero$, $\Dplus$, $\Dstar$, and $\Ds$ 
mesons were evaluated and compared to those measured in pp collisions at $\sqrt s=5.02~\tev$.
The relative abundances of the four species are 
unmodified in p--Pb collisions with respect to pp collisions, within the uncertainties. 
The ratios of $\Ds/\Dplus$-meson yields, as a function of the number of primary charged particles per unit of pseudorapidity, show no evidence of modifications in pp and p--Pb collisions, within the uncertainties.

%%%----------------------------------------------------------
%\newpage
\section{Appendix}
\label{sec:appendix}
\begin{figure}[!tbhp]
\begin{center}
\includegraphics[width=0.475\columnwidth]{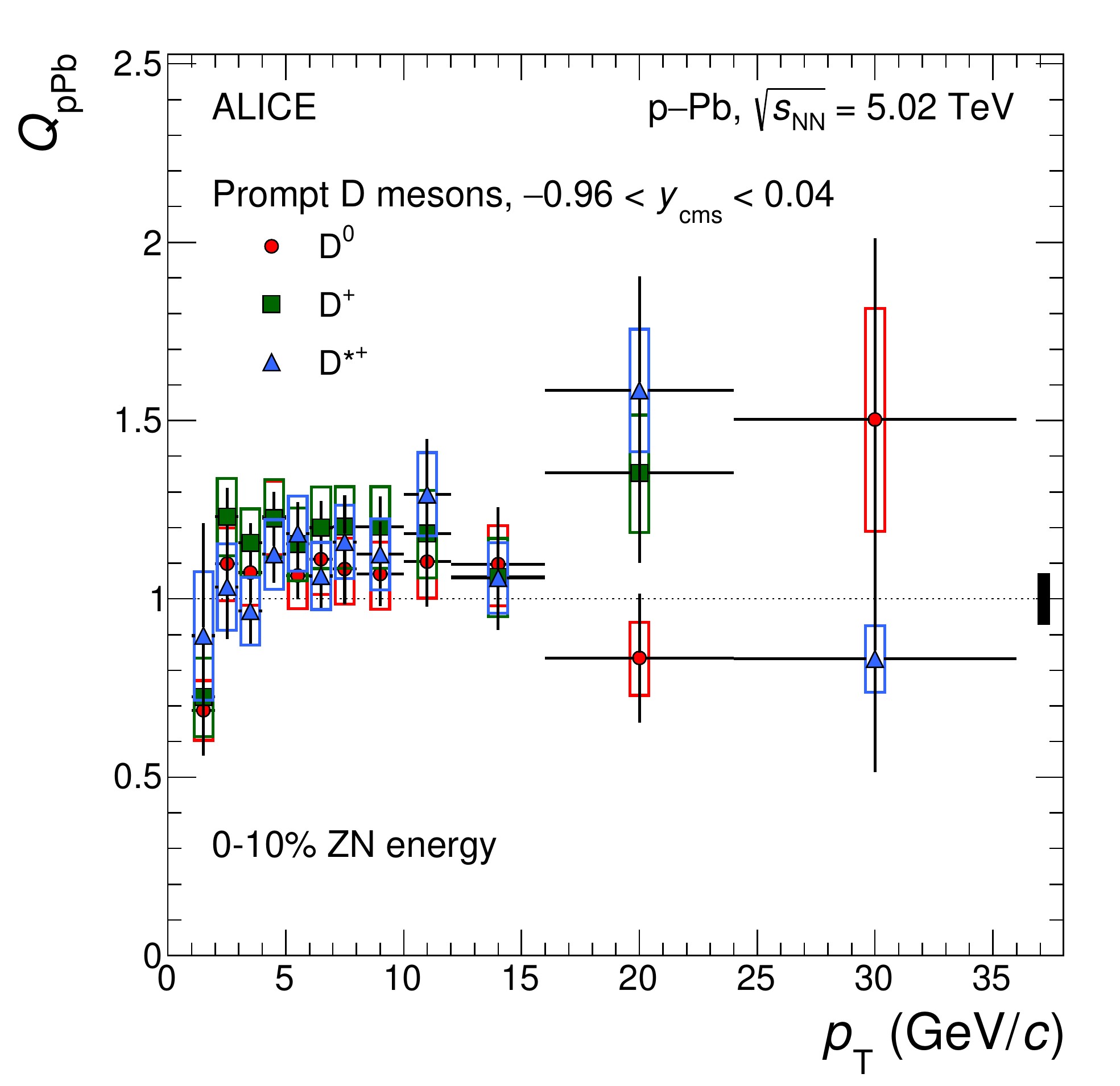}
  \includegraphics[width=0.475\columnwidth]{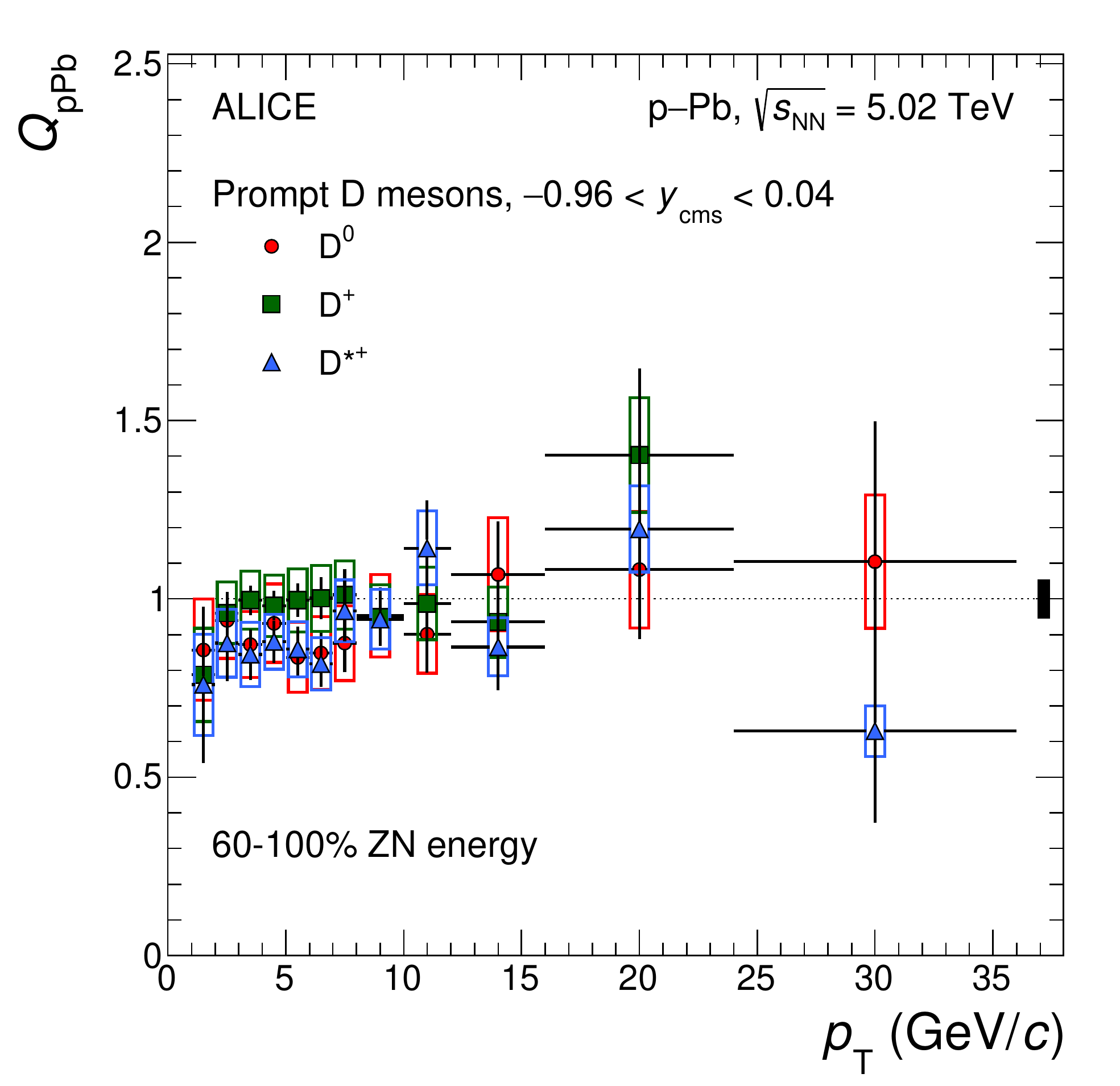}

\caption{
$\Dzero$, $\Dplus$, and $\Dstar$ meson nuclear modification factors as a function of $\pt$ in the 0--10\%~(left) and 60--100\%~(right) centrality classes.
The vertical bars and the empty boxes represent the statistical and systematic uncertainties. The colour-filled boxes at $\QpPb$ $=1$ represent the normalisation uncertainties.
\label{fig:DmesonQpPb2}
}
\end{center}
\end{figure}

\begin{figure}[!tbhp]
\begin{center}
\includegraphics[width=0.475\columnwidth]{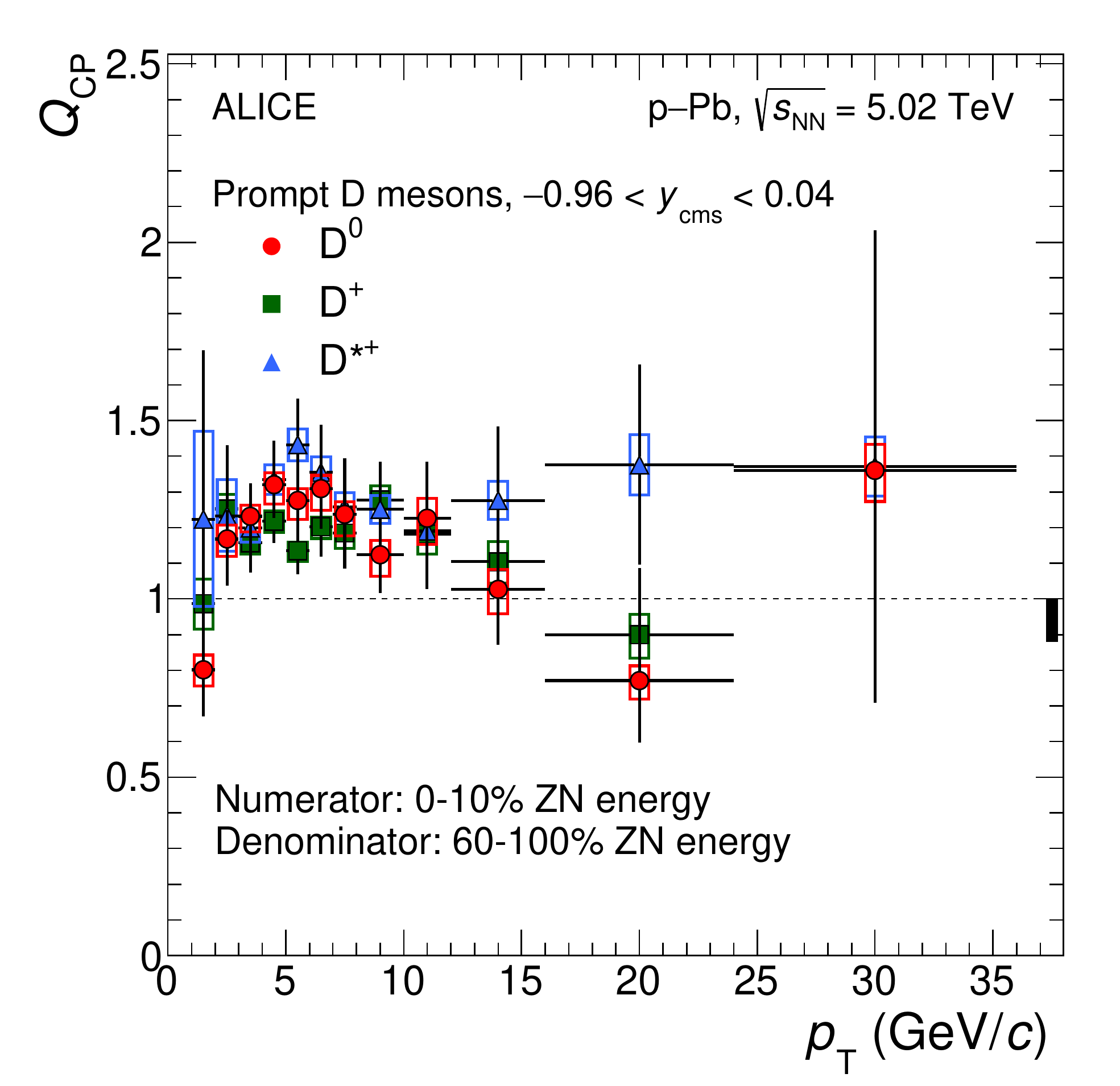}

\caption{
 $\Dzero$, $\Dplus$, and $\Dstar$ meson $\Qcp$, obtained using 0--10\% as the central class and 60--100\% as the peripheral class.
The vertical bars and the empty boxes represent the statistical and systematic uncertainties, respectively. The colour-filled boxes at $\Qcp$$=1$ represent the normalisation uncertainties.
\label{fig:DmesonQCP2}
}
\end{center}
\end{figure}

Figure~\ref{fig:DmesonQpPb2} presents the $\QpPb$ results for $\Dzero$, $\Dplus$, and $\Dstar$ as a function of $\pt$ for the 0--10\% and 60--100\% centrality classes.
Figure~\ref{fig:DmesonQCP2} shows the $\Qcp$ for the three non-strange D mesons, obtained using 0--10\% as central class and 60--100\% as peripheral class. The results are compatible within uncertainties between the three D--meson species. 

\FloatBarrier
%%
%\vspace{1cm}%
------------------------------------------------
%%%%%%%%% acknowledgements
\newenvironment{acknowledgement}{\relax}{\relax}
\begin{acknowledgement}
\section*{Acknowledgements}
% Version: 2019-05-23

The ALICE Collaboration would like to thank all its engineers and technicians for their invaluable contributions to the construction of the experiment and the CERN accelerator teams for the outstanding performance of the LHC complex.
The ALICE Collaboration gratefully acknowledges the resources and support provided by all Grid centres and the Worldwide LHC Computing Grid (WLCG) collaboration.
The ALICE Collaboration acknowledges the following funding agencies for their support in building and running the ALICE detector:
A. I. Alikhanyan National Science Laboratory (Yerevan Physics Institute) Foundation (ANSL), State Committee of Science and World Federation of Scientists (WFS), Armenia;
Austrian Academy of Sciences, Austrian Science Fund (FWF): [M 2467-N36] and Nationalstiftung f\"{u}r Forschung, Technologie und Entwicklung, Austria;
Ministry of Communications and High Technologies, National Nuclear Research Center, Azerbaijan;
Conselho Nacional de Desenvolvimento Cient\'{\i}fico e Tecnol\'{o}gico (CNPq), Universidade Federal do Rio Grande do Sul (UFRGS), Financiadora de Estudos e Projetos (Finep) and Funda\c{c}\~{a}o de Amparo \`{a} Pesquisa do Estado de S\~{a}o Paulo (FAPESP), Brazil;
Ministry of Science \& Technology of China (MSTC), National Natural Science Foundation of China (NSFC) and Ministry of Education of China (MOEC) , China;
Croatian Science Foundation and Ministry of Science and Education, Croatia;
Centro de Aplicaciones Tecnol\'{o}gicas y Desarrollo Nuclear (CEADEN), Cubaenerg\'{\i}a, Cuba;
Ministry of Education, Youth and Sports of the Czech Republic, Czech Republic;
The Danish Council for Independent Research | Natural Sciences, the Carlsberg Foundation and Danish National Research Foundation (DNRF), Denmark;
Helsinki Institute of Physics (HIP), Finland;
Commissariat \`{a} l'Energie Atomique (CEA), Institut National de Physique Nucl\'{e}aire et de Physique des Particules (IN2P3) and Centre National de la Recherche Scientifique (CNRS) and R\'{e}gion des  Pays de la Loire, France;
Bundesministerium f\"{u}r Bildung und Forschung (BMBF) and GSI Helmholtzzentrum f\"{u}r Schwerionenforschung GmbH, Germany;
General Secretariat for Research and Technology, Ministry of Education, Research and Religions, Greece;
National Research, Development and Innovation Office, Hungary;
Department of Atomic Energy Government of India (DAE), Department of Science and Technology, Government of India (DST), University Grants Commission, Government of India (UGC) and Council of Scientific and Industrial Research (CSIR), India;
Indonesian Institute of Science, Indonesia;
Centro Fermi - Museo Storico della Fisica e Centro Studi e Ricerche Enrico Fermi and Istituto Nazionale di Fisica Nucleare (INFN), Italy;
Institute for Innovative Science and Technology , Nagasaki Institute of Applied Science (IIST), Japan Society for the Promotion of Science (JSPS) KAKENHI and Japanese Ministry of Education, Culture, Sports, Science and Technology (MEXT), Japan;
Consejo Nacional de Ciencia (CONACYT) y Tecnolog\'{i}a, through Fondo de Cooperaci\'{o}n Internacional en Ciencia y Tecnolog\'{i}a (FONCICYT) and Direcci\'{o}n General de Asuntos del Personal Academico (DGAPA), Mexico;
Nederlandse Organisatie voor Wetenschappelijk Onderzoek (NWO), Netherlands;
The Research Council of Norway, Norway;
Commission on Science and Technology for Sustainable Development in the South (COMSATS), Pakistan;
Pontificia Universidad Cat\'{o}lica del Per\'{u}, Peru;
Ministry of Science and Higher Education and National Science Centre, Poland;
Korea Institute of Science and Technology Information and National Research Foundation of Korea (NRF), Republic of Korea;
Ministry of Education and Scientific Research, Institute of Atomic Physics and Ministry of Research and Innovation and Institute of Atomic Physics, Romania;
Joint Institute for Nuclear Research (JINR), Ministry of Education and Science of the Russian Federation, National Research Centre Kurchatov Institute, Russian Science Foundation and Russian Foundation for Basic Research, Russia;
Ministry of Education, Science, Research and Sport of the Slovak Republic, Slovakia;
National Research Foundation of South Africa, South Africa;
Swedish Research Council (VR) and Knut \& Alice Wallenberg Foundation (KAW), Sweden;
European Organization for Nuclear Research, Switzerland;
National Science and Technology Development Agency (NSDTA), Suranaree University of Technology (SUT) and Office of the Higher Education Commission under NRU project of Thailand, Thailand;
Turkish Atomic Energy Agency (TAEK), Turkey;
National Academy of  Sciences of Ukraine, Ukraine;
Science and Technology Facilities Council (STFC), United Kingdom;
National Science Foundation of the United States of America (NSF) and United States Department of Energy, Office of Nuclear Physics (DOE NP), United States of America.   %%%%%%% get the lates version before submitting
\end{acknowledgement}

\bibliographystyle{utphys}
\bibliography{DpPb2016PAS}

\providecommand{\href}[2]{#2}\begingroup\raggedright\begin{thebibliography}{10}

\bibitem{Cacciari:1998it}
M.~Cacciari, M.~Greco, and P.~Nason, ``{The $p_{\rm T}$ spectrum in heavy
  flavor hadroproduction},''
  \href{http://dx.doi.org/10.1088/1126-6708/1998/05/007}{{\em JHEP} {\bfseries
  9805} (1998) 007},
\href{http://arxiv.org/abs/hep-ph/9803400}{{\ttfamily arXiv:hep-ph/9803400
  [hep-ph]}}.
%%CITATION = HEP-PH/9803400;%%.

\bibitem{Cacciari:2012ny}
M.~Cacciari, S.~Frixione, N.~Houdeau, M.~L. Mangano, P.~Nason, and G.~Ridolfi,
  ``{Theoretical predictions for charm and bottom production at the {LHC}},''
  \href{http://dx.doi.org/10.1007/JHEP10(2012)137}{{\em JHEP} {\bfseries 1210}
  (2012) 137},
\href{http://arxiv.org/abs/1205.6344}{{\ttfamily arXiv:1205.6344 [hep-ph]}}.
%%CITATION = ARXIV:1205.6344;%%.

\bibitem{Kniehl:2004fy}
B.~A. Kniehl, G.~Kramer, I.~Schienbein, and H.~Spiesberger, ``{Inclusive $\rm
  D^{*+}$- production in p anti-p collisions with massive charm quarks},''
  \href{http://dx.doi.org/10.1103/PhysRevD.71.014018}{{\em Phys. Rev.}
  {\bfseries D71} (2005) 014018},
\href{http://arxiv.org/abs/hep-ph/0410289}{{\ttfamily arXiv:hep-ph/0410289
  [hep-ph]}}.
%%CITATION = HEP-PH/0410289;%%.

\bibitem{Kniehl:2005mk}
B.~A. Kniehl, G.~Kramer, I.~Schienbein, and H.~Spiesberger, ``{Collinear
  subtractions in hadroproduction of heavy quarks},''
  \href{http://dx.doi.org/10.1140/epjc/s2005-02200-7}{{\em Eur. Phys. J.}
  {\bfseries C41} (2005) 199--212},
\href{http://arxiv.org/abs/hep-ph/0502194}{{\ttfamily arXiv:hep-ph/0502194
  [hep-ph]}}.
%%CITATION = HEP-PH/0502194;%%.

\bibitem{Kniehl:2012ti}
B.~A. Kniehl, G.~Kramer, I.~Schienbein, and H.~Spiesberger, ``{Inclusive
  Charmed-Meson Production at the CERN LHC},''
  \href{http://dx.doi.org/10.1140/epjc/s10052-012-2082-2}{{\em Eur. Phys. J.}
  {\bfseries C72} (2012) 2082},
\href{http://arxiv.org/abs/1202.0439}{{\ttfamily arXiv:1202.0439 [hep-ph]}}.
%%CITATION = ARXIV:1202.0439;%%.

\bibitem{Helenius:2018uul}
I.~Helenius and H.~Paukkunen, ``{Revisiting the D-meson hadroproduction in
  general-mass variable flavour number scheme},''
  \href{http://dx.doi.org/10.1007/JHEP05(2018)196}{{\em JHEP} {\bfseries 05}
  (2018) 196},
\href{http://arxiv.org/abs/1804.03557}{{\ttfamily arXiv:1804.03557 [hep-ph]}}.
%%CITATION = ARXIV:1804.03557;%%.

\bibitem{Andronic:2015wma}
A.~Andronic {\em et~al.}, ``{Heavy-flavour and quarkonium production in the LHC
  era: from proton-proton to heavy-ion collisions},''
  \href{http://dx.doi.org/10.1140/epjc/s10052-015-3819-5}{{\em Eur. Phys. J.}
  {\bfseries C76} no.~3, (2016) 107},
\href{http://arxiv.org/abs/1506.03981}{{\ttfamily arXiv:1506.03981 [nucl-ex]}}.
%%CITATION = ARXIV:1506.03981;%%.

\bibitem{Prino:2016cni}
F.~Prino and R.~Rapp, ``{Open Heavy Flavor in QCD Matter and in Nuclear
  Collisions},'' \href{http://dx.doi.org/10.1088/0954-3899/43/9/093002}{{\em J.
  Phys.} {\bfseries G43} no.~9, (2016) 093002},
\href{http://arxiv.org/abs/1603.00529}{{\ttfamily arXiv:1603.00529 [nucl-ex]}}.
%%CITATION = ARXIV:1603.00529;%%.

\bibitem{Arneodo:1992wf}
M.~Arneodo, ``{Nuclear effects in structure functions},''
\href{http://dx.doi.org/10.1016/0370-1573(94)90048-5}{{\em Phys. Rept.}
  {\bfseries 240} (1994) 301--393}.
%%CITATION = PRPLC,240,301;%%.

\bibitem{Malace:2014uea}
S.~Malace, D.~Gaskell, D.~W. Higinbotham, and I.~Cloet, ``{The Challenge of the
  EMC Effect: existing data and future directions},''
  \href{http://dx.doi.org/10.1142/S0218301314300136}{{\em Int. J. Mod. Phys.}
  {\bfseries E23} (2014) 1430013},
\href{http://arxiv.org/abs/1405.1270}{{\ttfamily arXiv:1405.1270 [nucl-ex]}}.
%%CITATION = ARXIV:1405.1270;%%.

\bibitem{Eskola:2009uj}
K.~Eskola, H.~Paukkunen, and C.~Salgado, ``{EPS09: A New Generation of NLO and
  LO Nuclear Parton Distribution Functions},''
  \href{http://dx.doi.org/10.1088/1126-6708/2009/04/065}{{\em JHEP} {\bfseries
  0904} (2009) 065},
\href{http://arxiv.org/abs/0902.4154}{{\ttfamily arXiv:0902.4154 [hep-ph]}}.
%%CITATION = ARXIV:0902.4154;%%.

\bibitem{Hirai:2007sx}
M.~Hirai, S.~Kumano, and T.~H. Nagai, ``{Determination of nuclear parton
  distribution functions and their uncertainties in next-to-leading order},''
  \href{http://dx.doi.org/10.1103/PhysRevC.76.065207}{{\em Phys. Rev.}
  {\bfseries C76} (2007) 065207},
\href{http://arxiv.org/abs/0709.3038}{{\ttfamily arXiv:0709.3038 [hep-ph]}}.
%%CITATION = ARXIV:0709.3038;%%.

\bibitem{deFlorian:2003qf}
D.~de~Florian and R.~Sassot, ``{Nuclear parton distributions at next-to-leading
  order},'' \href{http://dx.doi.org/10.1103/PhysRevD.69.074028}{{\em Phys.
  Rev.} {\bfseries D69} (2004) 074028},
\href{http://arxiv.org/abs/hep-ph/0311227}{{\ttfamily arXiv:hep-ph/0311227
  [hep-ph]}}.
%%CITATION = HEP-PH/0311227;%%.

\bibitem{Eskola:2016oht}
K.~J. Eskola, P.~Paakkinen, H.~Paukkunen, and C.~A. Salgado, ``{EPPS16: Nuclear
  parton distributions with LHC data},''
  \href{http://dx.doi.org/10.1140/epjc/s10052-017-4725-9}{{\em Eur. Phys. J.}
  {\bfseries C77} no.~3, (2017) 163},
\href{http://arxiv.org/abs/1612.05741}{{\ttfamily arXiv:1612.05741 [hep-ph]}}.
%%CITATION = ARXIV:1612.05741;%%.

\bibitem{Kusina:2017gkz}
A.~Kusina, J.-P. Lansberg, I.~Schienbein, and H.-S. Shao, ``{Gluon Shadowing in
  Heavy-Flavor Production at the LHC},''
  \href{http://dx.doi.org/10.1103/PhysRevLett.121.052004}{{\em Phys. Rev.
  Lett.} {\bfseries 121} no.~5, (2018) 052004},
\href{http://arxiv.org/abs/1712.07024}{{\ttfamily arXiv:1712.07024 [hep-ph]}}.
%%CITATION = ARXIV:1712.07024;%%.

\bibitem{Eskola:2019bgf}
K.~J. Eskola, I.~Helenius, P.~Paakkinen, and H.~Paukkunen, ``{A QCD analysis of
  LHCb D-meson data in p+Pb collisions},''
\href{http://arxiv.org/abs/1906.02512}{{\ttfamily arXiv:1906.02512 [hep-ph]}}.
%%CITATION = ARXIV:1906.02512;%%.

\bibitem{Gelis:2010nm}
F.~Gelis, E.~Iancu, J.~Jalilian-Marian, and R.~Venugopalan, ``{The Color Glass
  Condensate},''
  \href{http://dx.doi.org/10.1146/annurev.nucl.010909.083629}{{\em Ann. Rev.
  Nucl. Part. Sci.} {\bfseries 60} (2010) 463--489},
\href{http://arxiv.org/abs/1002.0333}{{\ttfamily arXiv:1002.0333 [hep-ph]}}.
%%CITATION = ARXIV:1002.0333;%%.

\bibitem{Tribedy:2011aa}
P.~Tribedy and R.~Venugopalan, ``{QCD saturation at the LHC: Comparisons of
  models to pp and A A data and predictions for p--Pb collisions},''
  \href{http://dx.doi.org/10.1016/j.physletb.2012.02.047,
  10.1016/j.physletb.2012.12.004}{{\em Phys. Lett.} {\bfseries B710} (2012)
  125--133}, \href{http://arxiv.org/abs/1112.2445}{{\ttfamily arXiv:1112.2445
  [hep-ph]}}.
[Erratum: Phys. Lett.B718,1154(2013)].
%%CITATION = ARXIV:1112.2445;%%.

\bibitem{Albacete:2012xq}
J.~L. Albacete, A.~Dumitru, H.~Fujii, and Y.~Nara, ``{CGC predictions for p--Pb
  collisions at the LHC},''
  \href{http://dx.doi.org/10.1016/j.nuclphysa.2012.09.012}{{\em Nucl. Phys.}
  {\bfseries A897} (2013) 1--27},
\href{http://arxiv.org/abs/1209.2001}{{\ttfamily arXiv:1209.2001 [hep-ph]}}.
%%CITATION = ARXIV:1209.2001;%%.

\bibitem{Rezaeian:2012ye}
A.~H. Rezaeian, ``{CGC predictions for p--A collisions at the LHC and signature
  of QCD saturation},''
  \href{http://dx.doi.org/10.1016/j.physletb.2012.11.066}{{\em Phys. Lett.}
  {\bfseries B718} (2013) 1058--1069},
\href{http://arxiv.org/abs/1210.2385}{{\ttfamily arXiv:1210.2385 [hep-ph]}}.
%%CITATION = ARXIV:1210.2385;%%.

\bibitem{Fujii:2013yja}
H.~Fujii and K.~Watanabe, ``{Heavy quark pair production in high energy pA
  collisions: Open heavy flavors},''
  \href{http://dx.doi.org/10.1016/j.nuclphysa.2013.10.006}{{\em Nucl. Phys.}
  {\bfseries A920} (2013) 78--93},
\href{http://arxiv.org/abs/1308.1258}{{\ttfamily arXiv:1308.1258 [hep-ph]}}.
%%CITATION = ARXIV:1308.1258;%%.

\bibitem{Vitev:2007ve}
I.~Vitev, ``{Non-Abelian energy loss in cold nuclear matter},''
  \href{http://dx.doi.org/10.1103/PhysRevC.75.064906}{{\em Phys. Rev.}
  {\bfseries C75} (2007) 064906},
\href{http://arxiv.org/abs/hep-ph/0703002}{{\ttfamily arXiv:hep-ph/0703002
  [hep-ph]}}.
%%CITATION = HEP-PH/0703002;%%.

\bibitem{Lev:1983hh}
M.~Lev and B.~Petersson, ``{Nuclear Effects at Large Transverse Momentum in a
  {QCD} Parton Model},''
\href{http://dx.doi.org/10.1007/BF01648792}{{\em Z. Phys.} {\bfseries C21}
  (1983) 155}.
%%CITATION = ZEPYA,C21,155;%%.

\bibitem{Wang:1998ww}
X.-N. Wang, ``{Systematic study of high $p_{T}$ hadron spectra in $p p$, $p$A
  and AA collisions from SPS to RHIC energies},''
  \href{http://dx.doi.org/10.1103/PhysRevC.61.064910}{{\em Phys.Rev.}
  {\bfseries C61} (2000) 064910},
\href{http://arxiv.org/abs/nucl-th/9812021}{{\ttfamily arXiv:nucl-th/9812021
  [nucl-th]}}.
%%CITATION = NUCL-TH/9812021;%%.

\bibitem{Kopeliovich:2002yh}
B.~Z. Kopeliovich, J.~Nemchik, A.~Schafer, and A.~V. Tarasov, ``{Cronin effect
  in hadron production off nuclei},''
  \href{http://dx.doi.org/10.1103/PhysRevLett.88.232303}{{\em Phys. Rev. Lett.}
  {\bfseries 88} (2002) 232303},
\href{http://arxiv.org/abs/hep-ph/0201010}{{\ttfamily arXiv:hep-ph/0201010
  [hep-ph]}}.
%%CITATION = HEP-PH/0201010;%%.

\bibitem{CMS:2012qk}
{\bfseries CMS} Collaboration, S.~Chatrchyan {\em et~al.}, ``{Observation of
  long-range near-side angular correlations in proton-lead collisions at the
  LHC},'' \href{http://dx.doi.org/10.1016/j.physletb.2012.11.025}{{\em Phys.
  Lett.} {\bfseries B718} (2013) 795--814},
\href{http://arxiv.org/abs/1210.5482}{{\ttfamily arXiv:1210.5482 [nucl-ex]}}.
%%CITATION = ARXIV:1210.5482;%%.

\bibitem{Abelev:2012ola}
{\bfseries ALICE} Collaboration, B.~Abelev {\em et~al.}, ``{Long-range angular
  correlations on the near and away side in $p$-Pb collisions at $\sqrt{s_{\rm
  NN}}=5.02$ TeV},''
  \href{http://dx.doi.org/10.1016/j.physletb.2013.01.012}{{\em Phys. Lett.}
  {\bfseries B719} (2013) 29--41},
\href{http://arxiv.org/abs/1212.2001}{{\ttfamily arXiv:1212.2001 [nucl-ex]}}.
%%CITATION = ARXIV:1212.2001;%%.

\bibitem{ABELEV:2013wsa}
{\bfseries ALICE} Collaboration, B.~Abelev {\em et~al.}, ``{Long-range angular
  correlations of $\rm \pi$, K and p in p--Pb collisions at $\sqrt{s_{\rm NN}}$
  = 5.02 TeV},'' \href{http://dx.doi.org/10.1016/j.physletb.2013.08.024}{{\em
  Phys. Lett.} {\bfseries B726} (2013) 164--177},
\href{http://arxiv.org/abs/1307.3237}{{\ttfamily arXiv:1307.3237 [nucl-ex]}}.
%%CITATION = ARXIV:1307.3237;%%.

\bibitem{Aad:2012gla}
{\bfseries ATLAS} Collaboration, G.~Aad {\em et~al.}, ``{Observation of
  Associated Near-Side and Away-Side Long-Range Correlations in $\sqrt{s_{\rm
  NN}}$=5.02 TeV Proton-Lead Collisions with the ATLAS Detector},''
  \href{http://dx.doi.org/10.1103/PhysRevLett.110.182302}{{\em Phys. Rev.
  Lett.} {\bfseries 110} no.~18, (2013) 182302},
\href{http://arxiv.org/abs/1212.5198}{{\ttfamily arXiv:1212.5198 [hep-ex]}}.
%%CITATION = ARXIV:1212.5198;%%.

\bibitem{Adam:2015bka}
{\bfseries ALICE} Collaboration, J.~Adam {\em et~al.}, ``{Forward-central
  two-particle correlations in p--Pb collisions at $\sqrt{s_{\rm NN}}$ = 5.02
  TeV},'' \href{http://dx.doi.org/10.1016/j.physletb.2015.12.010}{{\em Phys.
  Lett.} {\bfseries B753} (2016) 126--139},
\href{http://arxiv.org/abs/1506.08032}{{\ttfamily arXiv:1506.08032 [nucl-ex]}}.
%%CITATION = ARXIV:1506.08032;%%.

\bibitem{Aaij:2015qcq}
{\bfseries LHCb} Collaboration, R.~Aaij {\em et~al.}, ``{Measurements of
  long-range near-side angular correlations in $\sqrt{s_{\text{NN}}}=5$ TeV
  proton-lead collisions in the forward region},''
  \href{http://dx.doi.org/10.1016/j.physletb.2016.09.064}{{\em Phys. Lett.}
  {\bfseries B762} (2016) 473--483},
\href{http://arxiv.org/abs/1512.00439}{{\ttfamily arXiv:1512.00439 [nucl-ex]}}.
%%CITATION = ARXIV:1512.00439;%%.

\bibitem{Khachatryan:2015waa}
{\bfseries CMS} Collaboration, V.~Khachatryan {\em et~al.}, ``{Evidence for
  Collective Multiparticle Correlations in p--Pb Collisions},''
  \href{http://dx.doi.org/10.1103/PhysRevLett.115.012301}{{\em Phys. Rev.
  Lett.} {\bfseries 115} no.~1, (2015) 012301},
  \href{http://arxiv.org/abs/1502.05382}{{\ttfamily arXiv:1502.05382
  [nucl-ex]}}.

\bibitem{Aaboud:2017blb}
{\bfseries ATLAS} Collaboration, M.~Aaboud {\em et~al.}, ``{Measurement of
  long-range multiparticle azimuthal correlations with the subevent cumulant
  method in $pp$ and p--Pb collisions with the ATLAS detector at the CERN Large
  Hadron Collider},'' \href{http://dx.doi.org/10.1103/PhysRevC.97.024904}{{\em
  Phys. Rev.} {\bfseries C97} no.~2, (2018) 024904},
\href{http://arxiv.org/abs/1708.03559}{{\ttfamily arXiv:1708.03559 [hep-ex]}}.
%%CITATION = ARXIV:1708.03559;%%.

\bibitem{Abelev:2013haa}
{\bfseries ALICE} Collaboration, B.~Abelev {\em et~al.}, ``{Multiplicity
  Dependence of Pion, Kaon, Proton and Lambda Production in p--Pb Collisions at
  $\sqrt{s_{\rm NN}}$ = 5.02 TeV},''
  \href{http://dx.doi.org/10.1016/j.physletb.2013.11.020}{{\em Phys. Lett.}
  {\bfseries B728} (2014) 25--38},
\href{http://arxiv.org/abs/1307.6796}{{\ttfamily arXiv:1307.6796 [nucl-ex]}}.
%%CITATION = ARXIV:1307.6796;%%.

\bibitem{Chatrchyan:2013eya}
{\bfseries CMS} Collaboration, S.~Chatrchyan {\em et~al.}, ``{Study of the
  production of charged pions, kaons, and protons in p--Pb collisions at
  $\sqrt{s_{NN}} =$ 5.02 $\,\text {TeV}$},''
  \href{http://dx.doi.org/10.1140/epjc/s10052-014-2847-x}{{\em Eur. Phys. J.}
  {\bfseries C74} no.~6, (2014) 2847},
\href{http://arxiv.org/abs/1307.3442}{{\ttfamily arXiv:1307.3442 [hep-ex]}}.
%%CITATION = ARXIV:1307.3442;%%.

\bibitem{Abelev:2014zpa}
{\bfseries ALICE} Collaboration, B.~Abelev {\em et~al.}, ``{Suppression of
  $\psi$(2S) production in p--Pb collisions at $\sqrt{s_{\rm NN}}$ = 5.02
  TeV},'' \href{http://dx.doi.org/10.1007/JHEP12(2014)073}{{\em JHEP}
  {\bfseries 12} (2014) 073},
\href{http://arxiv.org/abs/1405.3796}{{\ttfamily arXiv:1405.3796 [nucl-ex]}}.
%%CITATION = ARXIV:1405.3796;%%.

\bibitem{Aaij:2016eyl}
{\bfseries LHCb} Collaboration, R.~Aaij {\em et~al.}, ``{Study of $\psi(2S)$
  production and cold nuclear matter effects in p--Pb collisions at
  $\sqrt{s_{NN}}=5~\mathrm{TeV}$},''
  \href{http://dx.doi.org/10.1007/JHEP03(2016)133}{{\em JHEP} {\bfseries 03}
  (2016) 133},
\href{http://arxiv.org/abs/1601.07878}{{\ttfamily arXiv:1601.07878 [nucl-ex]}}.
%%CITATION = ARXIV:1601.07878;%%.

\bibitem{Adam:2016ohd}
{\bfseries ALICE} Collaboration, J.~Adam {\em et~al.}, ``{Centrality dependence
  of $\mathbf{\psi}$(2S) suppression in p--Pb collisions at
  $\mathbf{\sqrt{{\textit s}_{\rm NN}}}$ = 5.02 TeV},''
  \href{http://dx.doi.org/10.1007/JHEP06(2016)050}{{\em JHEP} {\bfseries 06}
  (2016) 050},
\href{http://arxiv.org/abs/1603.02816}{{\ttfamily arXiv:1603.02816 [nucl-ex]}}.
%%CITATION = ARXIV:1603.02816;%%.

\bibitem{Borsanyi:2010bp}
{\bfseries Wuppertal-Budapest} Collaboration, S.~Borsanyi {\em et~al.}, ``{Is
  there still any $T_c$ mystery in lattice QCD? Results with physical masses in
  the continuum limit III},''
  \href{http://dx.doi.org/10.1007/JHEP09(2010)073}{{\em JHEP} {\bfseries 1009}
  (2010) 073},
\href{http://arxiv.org/abs/1005.3508}{{\ttfamily arXiv:1005.3508 [hep-lat]}}.
%%CITATION = ARXIV:1005.3508;%%.

\bibitem{Bazavov:2011nk}
A.~Bazavov, T.~Bhattacharya, M.~Cheng, C.~DeTar, H.~Ding, {\em et~al.}, ``{The
  chiral and deconfinement aspects of the QCD transition},''
  \href{http://dx.doi.org/10.1103/PhysRevD.85.054503}{{\em Phys.Rev.}
  {\bfseries D85} (2012) 054503},
\href{http://arxiv.org/abs/1111.1710}{{\ttfamily arXiv:1111.1710 [hep-lat]}}.
%%CITATION = ARXIV:1111.1710;%%.

\bibitem{Jaiswal:2016hex}
A.~Jaiswal and V.~Roy, ``{Relativistic hydrodynamics in heavy-ion collisions:
  general aspects and recent developments},''
  \href{http://dx.doi.org/10.1155/2016/9623034}{{\em Adv. High Energy Phys.}
  {\bfseries 2016} (2016) 9623034},
\href{http://arxiv.org/abs/1605.08694}{{\ttfamily arXiv:1605.08694 [nucl-th]}}.
%%CITATION = ARXIV:1605.08694;%%.

\bibitem{Busza:2018rrf}
W.~Busza, K.~Rajagopal, and W.~van~der Schee, ``{Heavy Ion Collisions: The Big
  Picture, and the Big Questions},''
  \href{http://dx.doi.org/10.1146/annurev-nucl-101917-020852}{{\em Ann. Rev.
  Nucl. Part. Sci.} {\bfseries 68} (2018) 339--376},
\href{http://arxiv.org/abs/1802.04801}{{\ttfamily arXiv:1802.04801 [hep-ph]}}.
%%CITATION = ARXIV:1802.04801;%%.

\bibitem{Nagle:2018nvi}
J.~L. Nagle and W.~A. Zajc, ``{Small System Collectivity in Relativistic
  Hadronic and Nuclear Collisions},''
  \href{http://dx.doi.org/10.1146/annurev-nucl-101916-123209}{{\em Ann. Rev.
  Nucl. Part. Sci.} {\bfseries 68} (2018) 211--235},
\href{http://arxiv.org/abs/1801.03477}{{\ttfamily arXiv:1801.03477 [nucl-ex]}}.
%%CITATION = ARXIV:1801.03477;%%.

\bibitem{Bozek:2012gr}
P.~Bozek and W.~Broniowski, ``{Correlations from hydrodynamic flow in p--Pb
  collisions},'' \href{http://dx.doi.org/10.1016/j.physletb.2012.12.051}{{\em
  Phys. Lett.} {\bfseries B718} (2013) 1557--1561},
\href{http://arxiv.org/abs/1211.0845}{{\ttfamily arXiv:1211.0845 [nucl-th]}}.
%%CITATION = ARXIV:1211.0845;%%.

\bibitem{Bozek:2013uha}
P.~Bozek and W.~Broniowski, ``{Collective dynamics in high-energy
  proton-nucleus collisions},''
  \href{http://dx.doi.org/10.1103/PhysRevC.88.014903}{{\em Phys. Rev.}
  {\bfseries C88} no.~1, (2013) 014903},
\href{http://arxiv.org/abs/1304.3044}{{\ttfamily arXiv:1304.3044 [nucl-th]}}.
%%CITATION = ARXIV:1304.3044;%%.

\bibitem{Weller:2017tsr}
R.~D. Weller and P.~Romatschke, ``{One fluid to rule them all: viscous
  hydrodynamic description of event-by-event central pp, p--Pb and Pb--Pb
  collisions at $\sqrt{s}=5.02$ TeV},''
  \href{http://dx.doi.org/10.1016/j.physletb.2017.09.077}{{\em Phys. Lett.}
  {\bfseries B774} (2017) 351--356},
\href{http://arxiv.org/abs/1701.07145}{{\ttfamily arXiv:1701.07145 [nucl-th]}}.
%%CITATION = ARXIV:1701.07145;%%.

\bibitem{Dusling:2012cg}
K.~Dusling and R.~Venugopalan, ``{Evidence for BFKL and saturation dynamics
  from dihadron spectra at the LHC},''
  \href{http://dx.doi.org/10.1103/PhysRevD.87.051502}{{\em Phys. Rev.}
  {\bfseries D87} no.~5, (2013) 051502},
\href{http://arxiv.org/abs/1210.3890}{{\ttfamily arXiv:1210.3890 [hep-ph]}}.
%%CITATION = ARXIV:1210.3890;%%.

\bibitem{Dusling:2015gta}
K.~Dusling, W.~Li, and B.~Schenke, ``{Novel collective phenomena in high-energy
  proton-proton and proton-nucleus collisions},''
  \href{http://dx.doi.org/10.1142/S0218301316300022}{{\em Int. J. Mod. Phys.}
  {\bfseries E25} no.~01, (2016) 1630002},
\href{http://arxiv.org/abs/1509.07939}{{\ttfamily arXiv:1509.07939 [nucl-ex]}}.
%%CITATION = ARXIV:1509.07939;%%.

\bibitem{He:2015hfa}
L.~He, T.~Edmonds, Z.-W. Lin, F.~Liu, D.~Molnar, and F.~Wang, ``{Anisotropic
  parton escape is the dominant source of azimuthal anisotropy in transport
  models},'' \href{http://dx.doi.org/10.1016/j.physletb.2015.12.051}{{\em Phys.
  Lett.} {\bfseries B753} (2016) 506--510},
\href{http://arxiv.org/abs/1502.05572}{{\ttfamily arXiv:1502.05572 [nucl-th]}}.
%%CITATION = ARXIV:1502.05572;%%.

\bibitem{Bierlich:2014xba}
C.~Bierlich, G.~Gustafson, L.~L\"{o}nnblad, and A.~Tarasov, ``{Effects of
  Overlapping Strings in pp Collisions},''
  \href{http://dx.doi.org/10.1007/JHEP03(2015)148}{{\em JHEP} {\bfseries 03}
  (2015) 148},
\href{http://arxiv.org/abs/1412.6259}{{\ttfamily arXiv:1412.6259 [hep-ph]}}.
%%CITATION = ARXIV:1412.6259;%%.

\bibitem{Bierlich:2017vhg}
C.~Bierlich, G.~Gustafson, and L.~L\"{o}nnblad, ``{Collectivity without plasma
  in hadronic collisions},''
  \href{http://dx.doi.org/10.1016/j.physletb.2018.01.069}{{\em Phys. Lett.}
  {\bfseries B779} (2018) 58--63},
\href{http://arxiv.org/abs/1710.09725}{{\ttfamily arXiv:1710.09725 [hep-ph]}}.
%%CITATION = ARXIV:1710.09725;%%.

\bibitem{Xu:2015iha}
Y.~Xu, S.~Cao, G.-Y. Qin, W.~Ke, M.~Nahrgang, J.~Auvinen, and S.~A. Bass,
  ``{Heavy-flavor dynamics in relativistic p--Pb collisions at
  $\sqrt{S_{NN}}=5.02$ TeV},''
  \href{http://dx.doi.org/10.1016/j.nuclphysbps.2016.05.050}{{\em Nucl. Part.
  Phys. Proc.} {\bfseries 276-278} (2016) 225--228},
\href{http://arxiv.org/abs/1510.07520}{{\ttfamily arXiv:1510.07520 [nucl-th]}}.
%%CITATION = ARXIV:1510.07520;%%.

\bibitem{Beraudo:2015wsd}
A.~Beraudo, A.~De~Pace, M.~Monteno, M.~Nardi, and F.~Prino, ``{Heavy-flavour
  production in high-energy d-Au and p--Pb collisions},''
  \href{http://dx.doi.org/10.1007/JHEP03(2016)123}{{\em JHEP} {\bfseries 03}
  (2016) 123},
\href{http://arxiv.org/abs/1512.05186}{{\ttfamily arXiv:1512.05186 [hep-ph]}}.
%%CITATION = ARXIV:1512.05186;%%.

\bibitem{Acharya:2017tfn}
{\bfseries ALICE} Collaboration, S.~Acharya {\em et~al.}, ``{Search for
  collectivity with azimuthal J/$\psi$-hadron correlations in high multiplicity
  p--Pb collisions at $\sqrt{s_{\rm NN}}$ = 5.02 and 8.16 TeV},''
  \href{http://dx.doi.org/10.1016/j.physletb.2018.02.039}{{\em Phys. Lett.}
  {\bfseries B780} (2018) 7--20},
\href{http://arxiv.org/abs/1709.06807}{{\ttfamily arXiv:1709.06807 [nucl-ex]}}.
%%CITATION = ARXIV:1709.06807;%%.

\bibitem{Sirunyan:2018toe}
{\bfseries CMS} Collaboration, A.~M. Sirunyan {\em et~al.}, ``{Elliptic flow of
  charm and strange hadrons in high-multiplicity p--Pb collisions at
  $\sqrt{s_{_\mathrm{NN}}} =$ 8.16 TeV},''
  \href{http://dx.doi.org/10.1103/PhysRevLett.121.082301}{{\em Phys. Rev.
  Lett.} {\bfseries 121} no.~8, (2018) 082301},
\href{http://arxiv.org/abs/1804.09767}{{\ttfamily arXiv:1804.09767 [hep-ex]}}.
%%CITATION = ARXIV:1804.09767;%%.

\bibitem{Acharya:2018dxy}
{\bfseries ALICE} Collaboration, S.~Acharya {\em et~al.}, ``{Azimuthal
  anisotropy of heavy-flavour decay electrons in p--Pb collisions at $
  \sqrt{s_{\rm NN}}$ = 5.02 TeV},''
  \href{http://dx.doi.org/10.1103/PhysRevLett.122.072301}{{\em Phys. Rev.
  Lett.} {\bfseries 122} no.~7, (2019) 072301},
\href{http://arxiv.org/abs/1805.04367}{{\ttfamily arXiv:1805.04367 [nucl-ex]}}.
%%CITATION = ARXIV:1805.04367;%%.

\bibitem{Greco:2003mm}
V.~Greco, C.~M. Ko, and P.~Levai, ``{Parton coalescence at RHIC},''
  \href{http://dx.doi.org/10.1103/PhysRevC.68.034904}{{\em Phys. Rev.}
  {\bfseries C68} (2003) 034904},
\href{http://arxiv.org/abs/nucl-th/0305024}{{\ttfamily arXiv:nucl-th/0305024
  [nucl-th]}}.
%%CITATION = NUCL-TH/0305024;%%.

\bibitem{Greco:2003vf}
V.~Greco, C.~M. Ko, and R.~Rapp, ``{Quark coalescence for charmed mesons in
  ultrarelativistic heavy ion collisions},''
  \href{http://dx.doi.org/10.1016/j.physletb.2004.06.064}{{\em Phys. Lett.}
  {\bfseries B595} (2004) 202--208},
\href{http://arxiv.org/abs/nucl-th/0312100}{{\ttfamily arXiv:nucl-th/0312100
  [nucl-th]}}.
%%CITATION = NUCL-TH/0312100;%%.

\bibitem{Andronic:2007zu}
A.~Andronic, P.~Braun-Munzinger, K.~Redlich, and J.~Stachel, ``{Charmonium and
  open charm production in nuclear collisions at SPS/FAIR energies and the
  possible influence of a hot hadronic medium},''
  \href{http://dx.doi.org/10.1016/j.physletb.2007.10.064}{{\em Phys. Lett.}
  {\bfseries B659} (2008) 149--155},
\href{http://arxiv.org/abs/0708.1488}{{\ttfamily arXiv:0708.1488 [nucl-th]}}.
%%CITATION = ARXIV:0708.1488;%%.

\bibitem{He:2012df}
M.~He, R.~J. Fries, and R.~Rapp, ``{$\mathbf{D_s}$-Meson as Quantitative Probe
  of Diffusion and Hadronization in Nuclear Collisions},''
  \href{http://dx.doi.org/10.1103/PhysRevLett.110.112301}{{\em Phys. Rev.
  Lett.} {\bfseries 110} no.~11, (2013) 112301},
\href{http://arxiv.org/abs/1204.4442}{{\ttfamily arXiv:1204.4442 [nucl-th]}}.
%%CITATION = ARXIV:1204.4442;%%.

\bibitem{Adam:2015vsf}
{\bfseries ALICE} Collaboration, J.~Adam {\em et~al.}, ``{Multi-strange baryon
  production in p--Pb collisions at $\sqrt{s_\mathbf{NN}}=5.02$ TeV},''
  \href{http://dx.doi.org/10.1016/j.physletb.2016.05.027}{{\em Phys. Lett.}
  {\bfseries B758} (2016) 389--401},
\href{http://arxiv.org/abs/1512.07227}{{\ttfamily arXiv:1512.07227 [nucl-ex]}}.
%%CITATION = ARXIV:1512.07227;%%.

\bibitem{ALICE:2017jyt}
{\bfseries ALICE} Collaboration, J.~Adam {\em et~al.}, ``{Enhanced production
  of multi-strange hadrons in high-multiplicity proton-proton collisions},''
  \href{http://dx.doi.org/10.1038/nphys4111}{{\em Nature Phys.} {\bfseries 13}
  (2017) 535--539},
\href{http://arxiv.org/abs/1606.07424}{{\ttfamily arXiv:1606.07424 [nucl-ex]}}.
%%CITATION = ARXIV:1606.07424;%%.

\bibitem{Abelev:2014hha}
{\bfseries ALICE} Collaboration, B.~Abelev {\em et~al.}, ``{Measurement of
  prompt D-meson production in p--Pb collisions at $\sqrt{s_{\rm NN}}$ = 5.02
  TeV},'' \href{http://dx.doi.org/10.1103/PhysRevLett.113.232301}{{\em
  Phys.Rev.Lett.} {\bfseries 113} no.~23, (2014) 232301},
\href{http://arxiv.org/abs/1405.3452}{{\ttfamily arXiv:1405.3452 [nucl-ex]}}.
%%CITATION = ARXIV:1405.3452;%%.

\bibitem{Adam:2016mkz}
{\bfseries ALICE} Collaboration, J.~Adam {\em et~al.}, ``{Measurement of
  D-meson production versus multiplicity in p--Pb collisions at $
  \sqrt{s_{\mathrm{NN}}}=5.02 $ TeV},''
  \href{http://dx.doi.org/10.1007/JHEP08(2016)078}{{\em JHEP} {\bfseries 08}
  (2016) 078},
\href{http://arxiv.org/abs/1602.07240}{{\ttfamily arXiv:1602.07240 [nucl-ex]}}.
%%CITATION = ARXIV:1602.07240;%%.

\bibitem{Adam:2016ich}
{\bfseries ALICE} Collaboration, J.~Adam {\em et~al.}, ``{D-meson production in
  p--Pb collisions at $\sqrt{s_{\rm NN}}=$5.02 TeV and in pp collisions at
  $\sqrt{s}=$7 TeV},'' \href{http://dx.doi.org/10.1103/PhysRevC.94.054908}{{\em
  Phys. Rev.} {\bfseries C94} no.~5, (2016) 054908},
\href{http://arxiv.org/abs/1605.07569}{{\ttfamily arXiv:1605.07569 [nucl-ex]}}.
%%CITATION = ARXIV:1605.07569;%%.

\bibitem{Aamodt:2008zz}
{\bfseries ALICE} Collaboration, K.~Aamodt {\em et~al.}, ``{The ALICE
  experiment at the CERN LHC},''
\href{http://dx.doi.org/10.1088/1748-0221/3/08/S08002}{{\em JINST} {\bfseries
  3} (2008) S08002}.
%%CITATION = JINST,3,S08002;%%.

\bibitem{Abelev:2014ffa}
{\bfseries ALICE} Collaboration, B.~Abelev {\em et~al.}, ``{Performance of the
  ALICE Experiment at the CERN LHC},''
  \href{http://dx.doi.org/10.1142/S0217751X14300440}{{\em Int.J.Mod.Phys.}
  {\bfseries A29} (2014) 1430044},
\href{http://arxiv.org/abs/1402.4476}{{\ttfamily arXiv:1402.4476 [nucl-ex]}}.
%%CITATION = ARXIV:1402.4476;%%.

\bibitem{Abelev:2014epa}
{\bfseries ALICE} Collaboration, B.~Abelev {\em et~al.}, ``{Measurement of
  visible cross sections in proton-lead collisions at $\sqrt{s_{\rm NN}}$ =
  5.02 TeV in van der Meer scans with the ALICE detector},''
  \href{http://dx.doi.org/10.1088/1748-0221/9/11/P11003}{{\em JINST} {\bfseries
  9} no.~11, (2014) P11003},
\href{http://arxiv.org/abs/1405.1849}{{\ttfamily arXiv:1405.1849 [nucl-ex]}}.
%%CITATION = ARXIV:1405.1849;%%.

\bibitem{Adam:2014qja}
{\bfseries ALICE} Collaboration, J.~Adam {\em et~al.}, ``{Centrality dependence
  of particle production in p--Pb collisions at $\sqrt{s_{\rm NN} }$= 5.02
  TeV},'' \href{http://dx.doi.org/10.1103/PhysRevC.91.064905}{{\em Phys.Rev.}
  {\bfseries C91} (2015) 064905},
\href{http://arxiv.org/abs/1412.6828}{{\ttfamily arXiv:1412.6828 [nucl-ex]}}.
%%CITATION = ARXIV:1412.6828;%%.

\bibitem{pp:2019}
{\bfseries ALICE} Collaboration, S.~Acharya {\em et~al.}, ``{Measurement of
  ${{\mathrm{D}}^0}$ , ${{\mathrm{D}}^+}$ , ${{\mathrm{D}}^{*+}}$ and
  ${{\mathrm{D}}^+_{\mathrm{s}}}$ production in pp collisions at
  ${\sqrt{{\textit{s}}}~=~5.02~{\text {TeV}}}$ with ALICE},''
  \href{http://dx.doi.org/10.1140/epjc/s10052-019-6873-6}{{\em Eur. Phys. J.}
  {\bfseries C79} no.~5, (2019) 388},
\href{http://arxiv.org/abs/1901.07979}{{\ttfamily arXiv:1901.07979 [nucl-ex]}}.
%%CITATION = ARXIV:1901.07979;%%.

\bibitem{Tanabashi:2018xmw}
{\bfseries Particle Data Group} Collaboration, M.~Tanabashi {\em et~al.},
  ``{Review of Particle Physics},''
  \href{http://dx.doi.org/10.1103/PhysRevD.98.030001}{{\em Phys. Rev. D}
  {\bfseries 98} (2018) 030001}.
  \url{https://link.aps.org/doi/10.1103/PhysRevD.98.030001}.

\bibitem{Adam:2015ota}
{\bfseries ALICE} Collaboration, J.~Adam {\em et~al.}, ``{Measurement of charm
  and beauty production at central rapidity versus charged-particle
  multiplicity in proton-proton collisions at $\mathbf{\sqrt{{\textit s}}}=7$
  TeV},'' \href{http://dx.doi.org/10.1007/JHEP09(2015)148}{{\em JHEP}
  {\bfseries 09} (2015) 148},
\href{http://arxiv.org/abs/1505.00664}{{\ttfamily arXiv:1505.00664 [nucl-ex]}}.
%%CITATION = ARXIV:1505.00664;%%.

\bibitem{Sjostrand:2006za}
T.~Sjostrand, S.~Mrenna, and P.~Z. Skands, ``{PYTHIA 6.4 Physics and Manual},''
  \href{http://dx.doi.org/10.1088/1126-6708/2006/05/026}{{\em JHEP} {\bfseries
  0605} (2006) 026},
\href{http://arxiv.org/abs/hep-ph/0603175}{{\ttfamily arXiv:hep-ph/0603175
  [hep-ph]}}.
%%CITATION = HEP-PH/0603175;%%.

\bibitem{Wang:1991hta}
X.-N. Wang and M.~Gyulassy, ``{HIJING: A Monte Carlo model for multiple jet
  production in pp, pA and AA collisions},''
\href{http://dx.doi.org/10.1103/PhysRevD.44.3501}{{\em Phys.Rev.} {\bfseries
  D44} (1991) 3501--3516}.
%%CITATION = PHRVA,D44,3501;%%.

\bibitem{Acharya:2017jgo}
{\bfseries ALICE} Collaboration, S.~Acharya {\em et~al.}, ``{Measurement of
  D-meson production at mid-rapidity in pp collisions at $\mathbf{\sqrt{s}=7}$
  TeV},''
\href{http://arxiv.org/abs/1702.00766}{{\ttfamily arXiv:1702.00766 [hep-ex]}}.
%%CITATION = ARXIV:1702.00766;%%.

\bibitem{Ryd:2005zz}
A.~Ryd, D.~Lange, N.~Kuznetsova, S.~Versille, M.~Rotondo, D.~P. Kirkby, F.~K.
  Wuerthwein, and A.~Ishikawa,
``{EvtGen: A Monte Carlo Generator for B-Physics},''.
%%CITATION = EVTGEN-V00-11-07;%%.

\bibitem{Khachatryan:2015uja}
{\bfseries CMS} Collaboration, V.~Khachatryan {\em et~al.}, ``{Study of B meson
  production in p--Pb collisions at $\sqrt{s_{\rm NN}}=5.02$ TeV using
  exclusive hadronic decays},''
  \href{http://dx.doi.org/10.1103/PhysRevLett.116.032301}{{\em Phys. Rev.
  Lett.} {\bfseries 116} no.~3, (2016) 032301},
\href{http://arxiv.org/abs/1508.06678}{{\ttfamily arXiv:1508.06678 [nucl-ex]}}.
%%CITATION = ARXIV:1508.06678;%%.

\bibitem{ALICE-PUBLIC-2017-005}
{\bfseries ALICE} Collaboration, ``{The ALICE definition of primary
  particles},'' Jun, 2017.
\newblock \url{https://cds.cern.ch/record/2270008}. ALICE-PUBLIC-2017-005.

\bibitem{Gladilin:2014tba}
L.~Gladilin, ``{Fragmentation fractions of $c$ and $b$ quarks into charmed
  hadrons at LEP},''
  \href{http://dx.doi.org/10.1140/epjc/s10052-014-3250-3}{{\em Eur. Phys. J.}
  {\bfseries C75} no.~1, (2015) 19},
\href{http://arxiv.org/abs/1404.3888}{{\ttfamily arXiv:1404.3888 [hep-ex]}}.
%%CITATION = ARXIV:1404.3888;%%.

\bibitem{Acharya2018}
{ALICE Collaboration}, ``{$\Lambda_c^+$ production in pp collisions at
  $\sqrt{s}=7 $ and in p--Pb collisions at $\sqrt{s_{\mathrm{NN}}}=5.02 $
  TeV},'' \href{http://dx.doi.org/10.1007/JHEP04(2018)108}{{\em JHEP}
  {\bfseries 2018} no.~4, (Apr, 2018) 108}.
  \url{https://doi.org/10.1007/JHEP04(2018)108}.

\bibitem{Aaij:2018iyy}
{\bfseries LHCb} Collaboration, R.~Aaij {\em et~al.}, ``{Prompt $\Lambda^+_c$
  production in $p\mathrm{Pb}$ collisions at $\sqrt{s_{NN}} = 5.02$ TeV},''
  \href{http://dx.doi.org/10.1007/JHEP02(2019)102}{{\em JHEP} {\bfseries 02}
  (2019) 102},
\href{http://arxiv.org/abs/1809.01404}{{\ttfamily arXiv:1809.01404 [hep-ex]}}.
%%CITATION = ARXIV:1809.01404;%%.

\bibitem{Fujii:2017rqa}
H.~Fujii and K.~Watanabe, ``{Nuclear modification of forward $D$ production in
  p--Pb collisions at the LHC},''
\href{http://arxiv.org/abs/1706.06728}{{\ttfamily arXiv:1706.06728 [hep-ph]}}.
%%CITATION = ARXIV:1706.06728;%%.

\bibitem{Pumplin:2002vw}
J.~Pumplin, D.~R. Stump, J.~Huston, H.~L. Lai, P.~M. Nadolsky, and W.~K. Tung,
  ``{New generation of parton distributions with uncertainties from global QCD
  analysis},'' \href{http://dx.doi.org/10.1088/1126-6708/2002/07/012}{{\em
  JHEP} {\bfseries 07} (2002) 012},
\href{http://arxiv.org/abs/hep-ph/0201195}{{\ttfamily arXiv:hep-ph/0201195
  [hep-ph]}}.
%%CITATION = HEP-PH/0201195;%%.

\bibitem{Aaij:2017gcy}
{\bfseries LHCb} Collaboration, R.~Aaij {\em et~al.}, ``{Study of prompt
  D$^{0}$ meson production in $p$Pb collisions at $ \sqrt{s_{\mathrm{NN}}}=5 $
  TeV},'' \href{http://dx.doi.org/10.1007/JHEP10(2017)090}{{\em JHEP}
  {\bfseries 10} (2017) 090},
\href{http://arxiv.org/abs/1707.02750}{{\ttfamily arXiv:1707.02750 [hep-ex]}}.
%%CITATION = ARXIV:1707.02750;%%.

\bibitem{Sharma:2009hn}
R.~Sharma, I.~Vitev, and B.-W. Zhang, ``{Light-cone wave function approach to
  open heavy flavor dynamics in QCD matter},''
  \href{http://dx.doi.org/10.1103/PhysRevC.80.054902}{{\em Phys.Rev.}
  {\bfseries C80} (2009) 054902},
\href{http://arxiv.org/abs/0904.0032}{{\ttfamily arXiv:0904.0032 [hep-ph]}}.
%%CITATION = ARXIV:0904.0032;%%.

\bibitem{Kang:2014hha}
Z.-B. Kang, I.~Vitev, E.~Wang, H.~Xing, and C.~Zhang, ``{Multiple scattering
  effects on heavy meson production in p--A collisions at backward rapidity},''
  \href{http://dx.doi.org/10.1016/j.physletb.2014.11.024}{{\em Phys. Lett.}
  {\bfseries B740} (2015) 23--29},
\href{http://arxiv.org/abs/1409.2494}{{\ttfamily arXiv:1409.2494 [hep-ph]}}.
%%CITATION = ARXIV:1409.2494;%%.

\bibitem{Loizides:2295119}
C.~Loizides, J.~Kamin, and D.~d'Enterria, ``{Improved Monte Carlo Glauber
  predictions at present and future nuclear colliders. Precision Monte Carlo
  Glauber predictions at present and future nuclear colliders},'' {\em Phys.
  Rev. C} {\bfseries 97} no.~arXiv:1710.07098, (Oct, 2017) 054910. 23 p.
  \url{http://cds.cern.ch/record/2295119}.

\bibitem{TpACDS}
{\bfseries ALICE} Collaboration, S.~Acharya {\em et~al.},
  ``{Centrality determination in heavy ion collisions},''.
  \url{http://cds.cern.ch/record/2636623}.

\bibitem{Abelev:2012rz}
{\bfseries ALICE} Collaboration, B.~Abelev {\em et~al.}, ``{$\Jpsi$ production
  as a function of charged particle multiplicity in pp collisions at $\sqrt{s}
  = 7$ TeV},'' \href{http://dx.doi.org/10.1016/j.physletb.2012.04.052}{{\em
  Phys.Lett.} {\bfseries B712} (2012) 165},
\href{http://arxiv.org/abs/1202.2816}{{\ttfamily arXiv:1202.2816 [hep-ex]}}.
%%CITATION = ARXIV:1202.2816;%%.

\bibitem{PbPb:2018}
{\bfseries ALICE} Collaboration, S.~Acharya {\em et~al.}, ``{Measurement of
  D$^{0}$, D$^{+}$, D$^{*+}$ and D$_{s}^{+}$ production in Pb--Pb collisions at
  $ \sqrt{s_{\mathrm{NN}}}=5.02 $ TeV},''
  \href{http://dx.doi.org/10.1007/JHEP10(2018)174}{{\em JHEP} {\bfseries 10}
  (2018) 174},
\href{http://arxiv.org/abs/1804.09083}{{\ttfamily arXiv:1804.09083 [nucl-ex]}}.
%%CITATION = ARXIV:1804.09083;%%.

\bibitem{Eskola:1991ec}
K.~J. Eskola, ``{Shadowing effects on quark and gluon production in
  ultrarelativistic heavy ion collisions},''
\href{http://dx.doi.org/10.1007/BF01565590}{{\em Z. Phys.} {\bfseries C51}
  (1991) 633--642}.
%%CITATION = ZEPYA,C51,633;%%.

\bibitem{Helenius:2012wd}
I.~Helenius, K.~J. Eskola, H.~Honkanen, and C.~A. Salgado, ``{Impact-Parameter
  Dependent Nuclear Parton Distribution Functions: EPS09s and EKS98s and Their
  Applications in Nuclear Hard Processes},''
  \href{http://dx.doi.org/10.1007/JHEP07(2012)073}{{\em JHEP} {\bfseries 07}
  (2012) 073},
\href{http://arxiv.org/abs/1205.5359}{{\ttfamily arXiv:1205.5359 [hep-ph]}}.
%%CITATION = ARXIV:1205.5359;%%.

\end{thebibliography}\endgroup

\newpage
%%
%%%----------------------------------------------------------
\appendix

\section{The ALICE Collaboration}
\label{app:collab} 
% Collaboration: CERN-LHC-ALICE
% Generation Date is 2019-May-23

% How to use:
%%%%%%%%% appendix with author list
%\appendix
%\section{The ALICE Collaboration}
%\label{app:collab}
%\input{Alice_Authorslist_XXXX-Axx-XX.tex}
\begingroup
\small
\begin{flushleft}
S.~Acharya\Irefn{org141}\And 
D.~Adamov\'{a}\Irefn{org93}\And 
S.P.~Adhya\Irefn{org141}\And 
A.~Adler\Irefn{org73}\And 
J.~Adolfsson\Irefn{org79}\And 
M.M.~Aggarwal\Irefn{org98}\And 
G.~Aglieri Rinella\Irefn{org34}\And 
M.~Agnello\Irefn{org31}\And 
N.~Agrawal\Irefn{org10}\textsuperscript{,}\Irefn{org48}\textsuperscript{,}\Irefn{org53}\And 
Z.~Ahammed\Irefn{org141}
% ...
\And
\renewcommand\labelenumi{\textsuperscript{\theenumi}~}

\section*{Affiliation notes}
\renewcommand\theenumi{\roman{enumi}}
\begin{Authlist}
\item \Adef{org*}Deceased
\item \Adef{orgI}Dipartimento DET del Politecnico di Torino, Turin, Italy
\item \Adef{orgII}M.V. Lomonosov Moscow State University, D.V. Skobeltsyn Institute of Nuclear, Physics, Moscow, Russia
\item \Adef{orgIII}Department of Applied Physics, Aligarh Muslim University, Aligarh, India
\item \Adef{orgIV}Institute of Theoretical Physics, University of Wroclaw, Poland
\end{Authlist}

\section*{Collaboration Institutes}
\renewcommand\theenumi{\arabic{enumi}~}
\begin{Authlist}
\item \Idef{org1}A.I. Alikhanyan National Science Laboratory (Yerevan Physics Institute) Foundation, Yerevan, Armenia
\item \Idef{org2}Bogolyubov Institute for Theoretical Physics, National Academy of Sciences of Ukraine, Kiev, Ukraine
\item \Idef{org3}Bose Institute, Department of Physics  and Centre for Astroparticle Physics and Space Science (CAPSS), Kolkata, India
\item \Idef{org4}Budker Institute for Nuclear Physics, Novosibirsk, Russia
\item \Idef{org5}California Polytechnic State University, San Luis Obispo, California, United States
\item \Idef{org6}Central China Normal University, Wuhan, China
\item \Idef{org7}Centre de Calcul de l'IN2P3, Villeurbanne, Lyon, France
\item \Idef{org8}Centro de Aplicaciones Tecnol\'{o}gicas y Desarrollo Nuclear (CEADEN), Havana, Cuba
\item \Idef{org9}Centro de Investigaci\'{o}n y de Estudios Avanzados (CINVESTAV), Mexico City and M\'{e}rida, Mexico
\item \Idef{org10}Centro Fermi - Museo Storico della Fisica e Centro Studi e Ricerche ``Enrico Fermi', Rome, Italy
\item \Idef{org11}Chicago State University, Chicago, Illinois, United States
\item \Idef{org12}China Institute of Atomic Energy, Beijing, China
\item \Idef{org13}Chonbuk National University, Jeonju, Republic of Korea
\item \Idef{org14}Comenius University Bratislava, Faculty of Mathematics, Physics and Informatics, Bratislava, Slovakia
\item \Idef{org15}COMSATS University Islamabad, Islamabad, Pakistan
\item \Idef{org16}Creighton University, Omaha, Nebraska, United States
\item \Idef{org17}Department of Physics, Aligarh Muslim University, Aligarh, India
\item \Idef{org18}Department of Physics, Pusan National University, Pusan, Republic of Korea
\item \Idef{org19}Department of Physics, Sejong University, Seoul, Republic of Korea
\item \Idef{org20}Department of Physics, University of California, Berkeley, California, United States
\item \Idef{org21}Department of Physics, University of Oslo, Oslo, Norway
\item \Idef{org22}Department of Physics and Technology, University of Bergen, Bergen, Norway
\item \Idef{org23}Dipartimento di Fisica dell'Universit\`{a} 'La Sapienza' and Sezione INFN, Rome, Italy
\item \Idef{org24}Dipartimento di Fisica dell'Universit\`{a} and Sezione INFN, Cagliari, Italy
\item \Idef{org25}Dipartimento di Fisica dell'Universit\`{a} and Sezione INFN, Trieste, Italy
\item \Idef{org26}Dipartimento di Fisica dell'Universit\`{a} and Sezione INFN, Turin, Italy
\item \Idef{org27}Dipartimento di Fisica e Astronomia dell'Universit\`{a} and Sezione INFN, Bologna, Italy
\item \Idef{org28}Dipartimento di Fisica e Astronomia dell'Universit\`{a} and Sezione INFN, Catania, Italy
\item \Idef{org29}Dipartimento di Fisica e Astronomia dell'Universit\`{a} and Sezione INFN, Padova, Italy
\item \Idef{org30}Dipartimento di Fisica `E.R.~Caianiello' dell'Universit\`{a} and Gruppo Collegato INFN, Salerno, Italy
\item \Idef{org31}Dipartimento DISAT del Politecnico and Sezione INFN, Turin, Italy
\item \Idef{org32}Dipartimento di Scienze e Innovazione Tecnologica dell'Universit\`{a} del Piemonte Orientale and INFN Sezione di Torino, Alessandria, Italy
\item \Idef{org33}Dipartimento Interateneo di Fisica `M.~Merlin' and Sezione INFN, Bari, Italy
\item \Idef{org34}European Organization for Nuclear Research (CERN), Geneva, Switzerland
\item \Idef{org35}Faculty of Electrical Engineering, Mechanical Engineering and Naval Architecture, University of Split, Split, Croatia
\item \Idef{org36}Faculty of Engineering and Science, Western Norway University of Applied Sciences, Bergen, Norway
\item \Idef{org37}Faculty of Nuclear Sciences and Physical Engineering, Czech Technical University in Prague, Prague, Czech Republic
\item \Idef{org38}Faculty of Science, P.J.~\v{S}af\'{a}rik University, Ko\v{s}ice, Slovakia
\item \Idef{org39}Frankfurt Institute for Advanced Studies, Johann Wolfgang Goethe-Universit\"{a}t Frankfurt, Frankfurt, Germany
\item \Idef{org40}Gangneung-Wonju National University, Gangneung, Republic of Korea
\item \Idef{org41}Gauhati University, Department of Physics, Guwahati, India
\item \Idef{org42}Helmholtz-Institut f\"{u}r Strahlen- und Kernphysik, Rheinische Friedrich-Wilhelms-Universit\"{a}t Bonn, Bonn, Germany
\item \Idef{org43}Helsinki Institute of Physics (HIP), Helsinki, Finland
\item \Idef{org44}High Energy Physics Group,  Universidad Aut\'{o}noma de Puebla, Puebla, Mexico
\item \Idef{org45}Hiroshima University, Hiroshima, Japan
\item \Idef{org46}Hochschule Worms, Zentrum  f\"{u}r Technologietransfer und Telekommunikation (ZTT), Worms, Germany
\item \Idef{org47}Horia Hulubei National Institute of Physics and Nuclear Engineering, Bucharest, Romania
\item \Idef{org48}Indian Institute of Technology Bombay (IIT), Mumbai, India
\item \Idef{org49}Indian Institute of Technology Indore, Indore, India
\item \Idef{org50}Indonesian Institute of Sciences, Jakarta, Indonesia
\item \Idef{org51}INFN, Laboratori Nazionali di Frascati, Frascati, Italy
\item \Idef{org52}INFN, Sezione di Bari, Bari, Italy
\item \Idef{org53}INFN, Sezione di Bologna, Bologna, Italy
\item \Idef{org54}INFN, Sezione di Cagliari, Cagliari, Italy
\item \Idef{org55}INFN, Sezione di Catania, Catania, Italy
\item \Idef{org56}INFN, Sezione di Padova, Padova, Italy
\item \Idef{org57}INFN, Sezione di Roma, Rome, Italy
\item \Idef{org58}INFN, Sezione di Torino, Turin, Italy
\item \Idef{org59}INFN, Sezione di Trieste, Trieste, Italy
\item \Idef{org60}Inha University, Incheon, Republic of Korea
\item \Idef{org61}Institut de Physique Nucl\'{e}aire d'Orsay (IPNO), Institut National de Physique Nucl\'{e}aire et de Physique des Particules (IN2P3/CNRS), Universit\'{e} de Paris-Sud, Universit\'{e} Paris-Saclay, Orsay, France
\item \Idef{org62}Institute for Nuclear Research, Academy of Sciences, Moscow, Russia
\item \Idef{org63}Institute for Subatomic Physics, Utrecht University/Nikhef, Utrecht, Netherlands
\item \Idef{org64}Institute of Experimental Physics, Slovak Academy of Sciences, Ko\v{s}ice, Slovakia
\item \Idef{org65}Institute of Physics, Homi Bhabha National Institute, Bhubaneswar, India
\item \Idef{org66}Institute of Physics of the Czech Academy of Sciences, Prague, Czech Republic
\item \Idef{org67}Institute of Space Science (ISS), Bucharest, Romania
\item \Idef{org68}Institut f\"{u}r Kernphysik, Johann Wolfgang Goethe-Universit\"{a}t Frankfurt, Frankfurt, Germany
\item \Idef{org69}Instituto de Ciencias Nucleares, Universidad Nacional Aut\'{o}noma de M\'{e}xico, Mexico City, Mexico
\item \Idef{org70}Instituto de F\'{i}sica, Universidade Federal do Rio Grande do Sul (UFRGS), Porto Alegre, Brazil
\item \Idef{org71}Instituto de F\'{\i}sica, Universidad Nacional Aut\'{o}noma de M\'{e}xico, Mexico City, Mexico
\item \Idef{org72}iThemba LABS, National Research Foundation, Somerset West, South Africa
\item \Idef{org73}Johann-Wolfgang-Goethe Universit\"{a}t Frankfurt Institut f\"{u}r Informatik, Fachbereich Informatik und Mathematik, Frankfurt, Germany
\item \Idef{org74}Joint Institute for Nuclear Research (JINR), Dubna, Russia
\item \Idef{org75}Korea Institute of Science and Technology Information, Daejeon, Republic of Korea
\item \Idef{org76}KTO Karatay University, Konya, Turkey
\item \Idef{org77}Laboratoire de Physique Subatomique et de Cosmologie, Universit\'{e} Grenoble-Alpes, CNRS-IN2P3, Grenoble, France
\item \Idef{org78}Lawrence Berkeley National Laboratory, Berkeley, California, United States
\item \Idef{org79}Lund University Department of Physics, Division of Particle Physics, Lund, Sweden
\item \Idef{org80}Nagasaki Institute of Applied Science, Nagasaki, Japan
\item \Idef{org81}Nara Women{'}s University (NWU), Nara, Japan
\item \Idef{org82}National and Kapodistrian University of Athens, School of Science, Department of Physics , Athens, Greece
\item \Idef{org83}National Centre for Nuclear Research, Warsaw, Poland
\item \Idef{org84}National Institute of Science Education and Research, Homi Bhabha National Institute, Jatni, India
\item \Idef{org85}National Nuclear Research Center, Baku, Azerbaijan
\item \Idef{org86}National Research Centre Kurchatov Institute, Moscow, Russia
\item \Idef{org87}Niels Bohr Institute, University of Copenhagen, Copenhagen, Denmark
\item \Idef{org88}Nikhef, National institute for subatomic physics, Amsterdam, Netherlands
\item \Idef{org89}NRC Kurchatov Institute IHEP, Protvino, Russia
\item \Idef{org90}NRC << Kurchatov Institute >>  - ITEP, Moscow, Russia
\item \Idef{org91}NRNU Moscow Engineering Physics Institute, Moscow, Russia
\item \Idef{org92}Nuclear Physics Group, STFC Daresbury Laboratory, Daresbury, United Kingdom
\item \Idef{org93}Nuclear Physics Institute of the Czech Academy of Sciences, \v{R}e\v{z} u Prahy, Czech Republic
\item \Idef{org94}Oak Ridge National Laboratory, Oak Ridge, Tennessee, United States
\item \Idef{org95}Ohio State University, Columbus, Ohio, United States
\item \Idef{org96}Petersburg Nuclear Physics Institute, Gatchina, Russia
\item \Idef{org97}Physics department, Faculty of science, University of Zagreb, Zagreb, Croatia
\item \Idef{org98}Physics Department, Panjab University, Chandigarh, India
\item \Idef{org99}Physics Department, University of Jammu, Jammu, India
\item \Idef{org100}Physics Department, University of Rajasthan, Jaipur, India
\item \Idef{org101}Physikalisches Institut, Eberhard-Karls-Universit\"{a}t T\"{u}bingen, T\"{u}bingen, Germany
\item \Idef{org102}Physikalisches Institut, Ruprecht-Karls-Universit\"{a}t Heidelberg, Heidelberg, Germany
\item \Idef{org103}Physik Department, Technische Universit\"{a}t M\"{u}nchen, Munich, Germany
\item \Idef{org104}Politecnico di Bari, Bari, Italy
\item \Idef{org105}Research Division and ExtreMe Matter Institute EMMI, GSI Helmholtzzentrum f\"ur Schwerionenforschung GmbH, Darmstadt, Germany
\item \Idef{org106}Rudjer Bo\v{s}kovi\'{c} Institute, Zagreb, Croatia
\item \Idef{org107}Russian Federal Nuclear Center (VNIIEF), Sarov, Russia
\item \Idef{org108}Saha Institute of Nuclear Physics, Homi Bhabha National Institute, Kolkata, India
\item \Idef{org109}School of Physics and Astronomy, University of Birmingham, Birmingham, United Kingdom
\item \Idef{org110}Secci\'{o}n F\'{\i}sica, Departamento de Ciencias, Pontificia Universidad Cat\'{o}lica del Per\'{u}, Lima, Peru
\item \Idef{org111}Shanghai Institute of Applied Physics, Shanghai, China
\item \Idef{org112}St. Petersburg State University, St. Petersburg, Russia
\item \Idef{org113}Stefan Meyer Institut f\"{u}r Subatomare Physik (SMI), Vienna, Austria
\item \Idef{org114}SUBATECH, IMT Atlantique, Universit\'{e} de Nantes, CNRS-IN2P3, Nantes, France
\item \Idef{org115}Suranaree University of Technology, Nakhon Ratchasima, Thailand
\item \Idef{org116}Technical University of Ko\v{s}ice, Ko\v{s}ice, Slovakia
\item \Idef{org117}Technische Universit\"{a}t M\"{u}nchen, Excellence Cluster 'Universe', Munich, Germany
\item \Idef{org118}The Henryk Niewodniczanski Institute of Nuclear Physics, Polish Academy of Sciences, Cracow, Poland
\item \Idef{org119}The University of Texas at Austin, Austin, Texas, United States
\item \Idef{org120}Universidad Aut\'{o}noma de Sinaloa, Culiac\'{a}n, Mexico
\item \Idef{org121}Universidade de S\~{a}o Paulo (USP), S\~{a}o Paulo, Brazil
\item \Idef{org122}Universidade Estadual de Campinas (UNICAMP), Campinas, Brazil
\item \Idef{org123}Universidade Federal do ABC, Santo Andre, Brazil
\item \Idef{org124}University of Cape Town, Cape Town, South Africa
\item \Idef{org125}University of Houston, Houston, Texas, United States
\item \Idef{org126}University of Jyv\"{a}skyl\"{a}, Jyv\"{a}skyl\"{a}, Finland
\item \Idef{org127}University of Liverpool, Liverpool, United Kingdom
\item \Idef{org128}University of Science and Techonology of China, Hefei, China
\item \Idef{org129}University of South-Eastern Norway, Tonsberg, Norway
\item \Idef{org130}University of Tennessee, Knoxville, Tennessee, United States
\item \Idef{org131}University of the Witwatersrand, Johannesburg, South Africa
\item \Idef{org132}University of Tokyo, Tokyo, Japan
\item \Idef{org133}University of Tsukuba, Tsukuba, Japan
\item \Idef{org134}Universit\'{e} Clermont Auvergne, CNRS/IN2P3, LPC, Clermont-Ferrand, France
\item \Idef{org135}Universit\'{e} de Lyon, Universit\'{e} Lyon 1, CNRS/IN2P3, IPN-Lyon, Villeurbanne, Lyon, France
\item \Idef{org136}Universit\'{e} de Strasbourg, CNRS, IPHC UMR 7178, F-67000 Strasbourg, France, Strasbourg, France
\item \Idef{org137}Universit\'{e} Paris-Saclay Centre d'Etudes de Saclay (CEA), IRFU, D\'{e}partment de Physique Nucl\'{e}aire (DPhN), Saclay, France
\item \Idef{org138}Universit\`{a} degli Studi di Foggia, Foggia, Italy
\item \Idef{org139}Universit\`{a} degli Studi di Pavia, Pavia, Italy
\item \Idef{org140}Universit\`{a} di Brescia, Brescia, Italy
\item \Idef{org141}Variable Energy Cyclotron Centre, Homi Bhabha National Institute, Kolkata, India
\item \Idef{org142}Warsaw University of Technology, Warsaw, Poland
\item \Idef{org143}Wayne State University, Detroit, Michigan, United States
\item \Idef{org144}Westf\"{a}lische Wilhelms-Universit\"{a}t M\"{u}nster, Institut f\"{u}r Kernphysik, M\"{u}nster, Germany
\item \Idef{org145}Wigner Research Centre for Physics, Hungarian Academy of Sciences, Budapest, Hungary
\item \Idef{org146}Yale University, New Haven, Connecticut, United States
\item \Idef{org147}Yonsei University, Seoul, Republic of Korea
\end{Authlist}
\endgroup

\end{document}